\documentclass[]{aastex631}

\usepackage{hyperref}
\usepackage{xcolor} 
\usepackage{natbib}

\newcommand\cmt[1]{{\color{green!50!black}[]}} 
\shorttitle{Radio quasars at $z>4$}
\shortauthors{Krezinger et al.}
\graphicspath{{./}{figures/}}

\begin{document}

\title{Radio-loud Quasars above Redshift 4: VLBI Imaging of an Extended Sample}

\author[0000-0002-8813-4884]{M\'at\'e Krezinger}
\affiliation{Department of Astronomy, Institute of Geography and Earth Sciences, ELTE E\"otv\"os Lor\'and University, P\'azm\'any P\'eter s\'et\'any 1/A,
H-1117 Budapest, Hungary}
\affiliation{Konkoly Observatory, ELKH Research Centre for Astronomy and Earth Sciences, Konkoly Thege Mikl\'os \'ut 15-17, H-1121 Budapest, Hungary}

\author[0000-0002-6044-6069]{Krisztina Perger}
\affiliation{Konkoly Observatory, ELKH Research Centre for Astronomy and Earth Sciences, Konkoly Thege Mikl\'os \'ut 15-17, H-1121 Budapest, Hungary}

\author[0000-0003-1020-1597]{Krisztina \'Eva Gab\'anyi}
\affiliation{Department of Astronomy, Institute of Geography and Earth Sciences, ELTE E\"otv\"os Lor\'and University, P\'azm\'any P\'eter s\'et\'any 1/A,
H-1117 Budapest, Hungary}
\affiliation{ELKH-ELTE Extragalactic Astrophysics Research Group, E\"otv\"os Lor\'and University, P\'azm\'any P\'eter s\'et\'any 1/A, H-1117 Budapest, Hungary}
\affiliation{Konkoly Observatory, ELKH Research Centre for Astronomy and Earth Sciences, Konkoly Thege Mikl\'os \'ut 15-17, H-1121 Budapest, Hungary}

\author[0000-0003-3079-1889]{S\'andor Frey}
\affiliation{Konkoly Observatory, ELKH Research Centre for Astronomy and Earth Sciences, Konkoly Thege Mikl\'os \'ut 15-17, H-1121 Budapest, Hungary}
\affiliation{Institute of Physics, ELTE E\"otv\"os Lor\'and University, P\'azm\'any P\'eter s\'et\'any 1/A,
H-1117 Budapest, Hungary}

\author[0000-0002-0694-2459]{Leonid I. Gurvits}
\affiliation{Joint Institute for VLBI ERIC, Oude Hoogeveensedijk 4, 7991 PD Dwingeloo, The Netherlands}
\affiliation{Department of Astrodynamics and Space Missions, Delft University of Technology, Kluyverweg 1, 2629 HS Delft, The Netherlands}

\author[0000-0002-5195-335X]{Zsolt Paragi}
\affiliation{Joint Institute for VLBI ERIC, Oude Hoogeveensedijk 4, 7991 PD Dwingeloo, The Netherlands}

\author[0000-0003-4341-0029]{Tao An}
\affiliation{Shanghai Astronomical Observatory, Key Laboratory of Radio Astronomy, Chinese Academy of Sciences, 80 Nandan Road, Shanghai 200030, China}

\author[0000-0001-8256-8887]{Yingkang Zhang}
\affiliation{Shanghai Astronomical Observatory, Key Laboratory of Radio Astronomy, Chinese Academy of Sciences, 80 Nandan Road, Shanghai 200030, China}

\author[0000-0003-1514-881X]{Hongmin Cao}
\affiliation{School of Physics and Electrical Information, Shangqiu Normal University, 298 Wenhua Road, Shangqiu, Henan 476000, China}

\author[0000-0002-3069-9399]{Tullia Sbarrato}
\affiliation{INAF -- Osservatorio Astronomico di Brera, via E. Bianchi 46, I-23807 Merate, Italy}

\correspondingauthor{M\'at\'e Krezinger}
\email{krezinger.mate@csfk.org}



\begin{abstract}

High-redshift radio sources provide plentiful opportunities for studying the formation and evolution of early galaxies and supermassive black holes. However, the number of known radio-loud active galactic nuclei (AGN) above redshift 4 is rather limited. At high redshifts, it appears that blazars, with relativistically beamed jets pointing towards the observer, are in majority compared to radio-loud sources with jets misaligned with respect to the line of sight. To find more of these misaligned AGN, milliarcsec-scale imaging studies carried out with very long baseline interferometry (VLBI) are needed, as they allow us to distinguish between compact core--jet radio sources and those with more extended emission. Previous high-resolution VLBI studies revealed that some of the radio sources among blazar candidates in fact show unbeamed radio emission on milliarcsecond scales. The most accurate optical coordinates determined with the Gaia astrometric space mission are also useful in the classification process. Here, we report on dual-frequency imaging observations of 13 high-redshift ($4 < z < 4.5$) quasars at 1.7 and 5~GHz with the European VLBI Network. This sample increases the number of $z>4$ radio sources for which VLBI observations are available by about a quarter. Using structural and physical properties, such as radio morphology, spectral index, variability, brightness temperature, as well as optical coordinates, we identified six blazars and six misaligned radio AGNs, with the remaining one tentatively identified as blazar.
\end{abstract}

\keywords{Extragalactic radio sources (508) --- Radio active galactic nuclei (2134) --- Quasars (1319) --- High-redshift galaxies (734) --- Interferometry (808)}


\section{Introduction} 
\label{sec:intro}

Studying the high-redshift Universe can help better understand how are galaxies formed and evolved. Active galactic nuclei (AGN) are of special value for this purpose because they are luminous and already present in the early Universe, less than about a billion years (redshift $z \ga 6$) after the Big Bang. AGN are powered by accreting supermassive black holes (SMBHs) with masses of $\sim 10^6-10^{10}$~M$_{\odot}$ and located in the center of the galaxies. In some objects, two-sided relativistic plasma jets are launched along the axis perpendicular to the accretion disk. High-redshift AGN offer a key to understand the evolution of their host galaxies \citep[e.g.][]{2005MNRAS.362...25B, 2012ARA&A..50..455F,2013Sci...341.1082M} as they set constrains on the properties of the accretion process and the black hole growth \citep[e.g.][]{2012MNRAS.425.2892W,2014MNRAS.440L..91P}. There is still much to learn about these objects, and the sample of high-redshift AGN is limited.  

Only about $10\%$ of the AGN population are radio-loud with (highly energetic) synchrotron-emitting jets \citep{2002AJ....124.2364I}. These are the radio quasars that are the easiest to observe from extreme distances. They are called blazars when their jet points nearly towards the observer with $\theta \la 10\degr$ \citep{1995PASP..107..803U}, where $\theta$ is the inclination angle of the jet axis with respect to the line of sight. Because of relativistic beaming effects, the blazar jets are Doppler-boosted, which could make their emission dominate most of the AGN spectrum. Given limited sensitivity of the instruments, blazars with enhanced strong synchrotron emission are expected to be the most easily detected in the radio AGN population at a given redshift. Blazars have characteristic properties in their X-ray and $\gamma$-ray emission \citep[e.g.][]{2013MNRAS.433.2182S,2015MNRAS.452.3457G}, flat radio spectrum, and compact radio structure with brightness temperatures exceeding the $T_\mathrm{b,eq} \approx 5 \times 10^{10}$\,K equipartition limit \citep{1994ApJ...426...51R}. Blazars are traditionally divided into two main sub-classes, flat-spectrum radio quasars (FSRQs) that show prominent optical and ultraviolet emission lines in their spectra, and BL Lac objects with weak or even no emission lines \citep[e.g.][]{1999ASPC..159..351F,2015Ap&SS.357...75M}. Since FSRQs are quite luminous in the optical, they can be detected from large cosmological distances. Due to the lack of emission lines, the redshifts of BL Lacs are often hard to determine, thus most of them with known redshifts are at $z<2$ \citep{2008AJ....135.2453P}.

As the AGN jets propagate through the interstellar or intergalactic space, they interact with the ambient medium, creating radio-emitting regions, i.e. hotspots and lobes. Quasars with jet inclination angles $\theta > 10\degr$ are often referred to as misaligned sources. These include objects like the unbeamed versions of gigahertz-peaked spectrum (GPS) and compact steep-spectrum (CSS) sources, and the symmetric ones are called compact symmetric objects (CSOs) or medium symmetric objects (MSOs). They are found at $z>4$ as well \citep[e.g.][]{2017MNRAS.467.2039C,2022A&A...659A.159S}. In contrast to blazars, the radio emission of these types of AGN is dominated by the outer regions of the jet, while the `core' emission coinciding with the optical position, due to the absence of Doppler-boosting, often remains undetected \citep[e.g.,][and references therein]{2012ApJ...760...77A, 2021AARv..29....3O}. Characteristic features of young misaligned sources are the extended kpc-scale radio emission, symmetric radio structure, and steep radio spectrum. Non-variable and non-Doppler-boosted emission is also a common feature. Young radio sources like CSOs have concave-shape (peaked) continuum spectra, in addition to the symmetric radio structure \citep[e.g.][]{2016MNRAS.459..820T,2020MNRAS.496.1811K}.         

To study the family of jetted high-redshift radio sources in detail, or to use them for classical cosmological tests \citep[e.g.][]{1999A&A...342..378G,1999NewAR..43..757K}, a comprehensive sample of radio quasars is needed. Furthermore, it also has a critical importance in the building of a quasar luminosity function \citep{2004ApJ...612..698H}. As \citet{2011MNRAS.416..216V} showed, at a given redshift, for every blazar there must be hundreds of misaligned sources whose jets point in a different direction. This number can be estimated as about $2\Gamma^2$ \citep{2016MNRAS.461L..21G}, where $\Gamma$ denotes the bulk Lorentz factor of the jet plasma, and blazars are defined as jetted sources with viewing angles of $\theta< 1/\Gamma$. \citet{2011MNRAS.416..216V} compared the number of high-redshift ($z>3$) radio-loud quasars derived by cross-matching the Sloan Digital Sky Survey \citep[SDSS,][]{2010AJ....139.2360S} and the Faint Images of the Radio Sky at Twenty-Centimeters \citep[FIRST,][]{1997ApJ...475..479W} surveys to the number of radio-loud quasars expected from the blazar luminosity function at different redshifts. They found the observed number of misaligned radio-loud  quasars much smaller than that estimated based on the blazar luminosity function. Every new high-redshift blazar discovered further increases this apparent deficit, as each of them implies the existence of hundreds of more sources with misaligned jets. Several scenarios were proposed as a solution for this problem \citep[e.g.][]{2011MNRAS.416..216V,2016MNRAS.461L..21G}. \citet{2011MNRAS.416..216V} suggested that the difference might be due to internal and external absorption mechanisms, lower bulk Lorentz factors in jets in the early Universe, or observational effects arising from survey sensitivity.
At high redshift, the radio lobes are dimmed due to the interaction of their electrons with photons of the cosmic microwave background \citep{2015MNRAS.452.3457G}. Compact hotspots would be less affected and thus more easily detectable.
Also, high-redshift galaxies may be heavily obscured due to a dense bubble surrounding the central region of the AGN \citep{2016MNRAS.461L..21G}, effectively reducing the number of misaligned jetted quasars that can be identified optically. Only the most powerful jets are able to penetrate this bubble, sweeping away most of the material along their path, revealing the central engine to our telescopes. 

In contrast to the arguments outlined above, flux density-limited samples \citep[e.g.][]{2019MNRAS.484..204C,2019ApJ...874...43L} indicate that there is no inconsistency between the number of blazars and misaligned radio AGN at all. Their results suggest that the most luminous radio-loud quasars have a density peak at higher redshifts ($z \sim 4$), while the less luminous population has similar cosmological evolution to radio-quiet quasars with space density peak at $z \approx 2$. Using a sample of $z < 4$ quasars, \citet{2017ApJ...842...87M} arrived at a similar conclusion. Both arguments above are in conflict with what \citet{2009ApJ...699..603A} found while investigating an X-ray selected sample of blazars. It seems that radio and X-ray selections are sampling different classes of objects with different cosmological behavior. However, \citet{2021MNRAS.505.4120I} found that the different blazar evolutions observed in radio and X-rays can be explained by the inverse-Compton interaction of the relativistic electrons in the jet and with the cosmic microwave background photons. This effect becomes more important at higher redshifts, leading to the increase of the observed X-ray-to-radio luminosity ratios with redshift.

\citet{2019ApJ...874...43L} argue that estimating the number of misaligned sources with $\sim \Gamma^2$ is oversimplified because it does not account for the observational biases caused by flux density-limited sampling. The distribution of jet Lorentz factors has a broad peak between 5 and 15, then falls off rapidly until $\Gamma \approx 50$. As a consequence, there is only a shallow increase in the misaligned (parent) population of jetted sources above $\Gamma \approx 15$. Low-$\Gamma$ jets require strong Doppler boosting and thus very small viewing angles to be detectable. The underlying misaligned population is therefore large. On the other hand, the relatively rare high-$\Gamma$ jets enter more easily in the flux density-limited samples, allowing for comparatively larger jet inclinations. Their parent population is therefore smaller. The general trend with $\Gamma$ is in fact opposite to what is expected from the $\sim \Gamma^2$ rule \citep{2019ApJ...874...43L}.

Furthermore, e.g. \citet{2016MNRAS.463.3260C} and \citet{2017MNRAS.467..950C} found that AGN may in some cases be mistakenly classified as blazar candidates based on X-ray and low-resolution radio observations only. High-resolution imaging with very long baseline interferometry (VLBI) revealed that some of these sources show low-brightness-temperature radio cores, and extended structures with steep radio spectrum on scales of $\sim 10-100$ milliarcsecond (mas) which are at odds with the blazar classification.

Correctly identifying the blazars is critical for constructing the blazar luminosity function at high redshifts. By offering the finest, mas-scale angular resolution at cm wavelengths, observations of the known high-redshift radio sources with VLBI are uniquely suited for distinguishing between compact, high brightness temperature radio emission of blazars and the more extended structures of misaligned jetted AGN. Multi-frequency radio interferometric observations can distinguish between compact cores from hotspots and more extended radio lobes, through the observed spectral characteristics, morphological information, and and brightness temperatures. Comparing the results of VLBI observations with lower-resolution radio images and data obtained at different wavebands is also essential in the classification process \citep{2021Galax...9...23S}. 

Phase-referenced VLBI observations using well-known, nearby calibrator sources also provide relative astrometric positions for the radio-emitting features, accurate to mas level. The comparison of VLBI positions with the precise optical astrometric data from the recent Early Data Release 3 \citep[EDR3,][]{2021A&A...649A...1G} of the Gaia space mission \citep{2016A&A...595A...1G}, if available, adds further relevant astrophysical information to aid the source classification. While the optical position of an AGN marks the location of the accretion disk in combination with the optical synchrotron emission of the innermost sub-mas scale jet \citep{2019ApJ...871..143P}, the VLBI intensity peak pinpoints the brightest and most compact emission feature, usually the self-absorbed base or the jet (i.e. the core), or maybe a shock front (i.e. a hotspot in a lobe) in an extended radio source. Hence, a significant offset between the radio and optical positions could be related to the misaligned nature of the jet, while for blazars, no large positional difference is expected.
Long-term VLBI monitoring of resolved radio jets can be used to measure component proper motion and to determine their physical and geometric properties \citep[e.g.][]{2015MNRAS.446.2921F,2018MNRAS.477.1065P,2020SciBu..65..525Z}.

Despite the high value of VLBI imaging investigations, only a few dozens of the known radio quasars at $z \geq 4$ have been studied with this technique to date (see their list compiled in Table~\ref{vlbilit}). Jets of distant radio quasars are harder to detect, because the observed frequencies correspond to $(1+z)$ times higher emitted frequencies in the rest frame of the sources, where the steep-spectrum jets become intrinsically fainter. This way, prominent extended features are less likely to be detected in high-redshift radio AGN \citep[e.g.][]{2000pras.conf..183G,2015IAUS..313..327G}. 

In this paper, we present VLBI observations of 13 high-redshift radio quasars at $z > 4$, which were carried out with the European VLBI Network (EVN) combined with e-MERLIN (Multi-Element Radio-Linked Interferometer Network) antennas at 1.7 and 5~GHz frequencies from 2017 to 2020. We investigate the nature of these sources using the newly-obtained VLBI data, combined with the available radio spectral information and the recent Gaia EDR3 optical positions, where available. In Section~\ref{sec:sample}, we present the list of $z \geq 4$ radio quasars previously imaged with VLBI, and describe our sample and the selection process. We give detailed description on the EVN observations and data reduction in Section~\ref{sec:obs}. In Section~\ref{sec:results}, we present the results and the properties derived for the new sample, which we discuss in Section~\ref{sec:discussion}.
Conclusions are given in Section~\ref{sec:conclusion}. Throughout this paper, we assume a standard flat $\Lambda$ Cold Dark Matter cosmology with $\Omega_{\rm m} = 0.3$, $\Omega_{\rm \Lambda} = 0.7$, and $H_0 = 70$~km\,s$^{-1}$\,Mpc$^{-1}$. We used the cosmology calculator of \citet{2006PASP..118.1711W} for determining projected linear sizes and luminosity distances.             

\section{Sample Selection} 
\label{sec:sample}

According to the updated catalog of \citet{2017FrASS...4....9P}, there are 56 sources at $z \geq 4$ that were imaged with VLBI at least at one frequency band. Table~\ref{vlbilit} lists these high-redshift sources with literature references, and can be considered as an update to table~5 of \citet{2016MNRAS.463.3260C}. The observed frequencies range from 1.4 to 43~GHz. We inspected the published radio images to classify the mas-scale structure of the sources visually. The notations of the simple morphological classes: C refers to sources with a single compact component, J to sources having a core--jet type morphology, and E to sources with extended radio structure on angular scales of a few tens of mas. In addition to the morphological classification, sources are also categorized by their total flux density radio continuum spectra, based on low-resolution observations when sufficient data are available in the literature. The range of available frequencies varies from source to source, but typically covers $\sim 0.1-10$~GHz. We define three classes containing flat (f), steep (s) and peaked (p) spectrum sources. However, we note that the multi-frequency measurements may not have been necessarily simultaneous, and the frequency coverage is inhomogeneous in the sample. Therefore this simple classification is indicative only and not used in any quantitative analysis.

\begin{deluxetable*}{llccc}
\tablenum{1}
\tablecaption{List of the VLBI imaged radio quasars at $z \geq 4$ from the literature.
}
\label{vlbilit}
\tablewidth{0pt}
\tablehead{
\colhead{Name} & \colhead{$z$} & \colhead{$\nu$ (GHz)}  & \colhead{Classification} & \colhead{VLBI reference}
}
\decimalcolnumbers
\startdata
J001115.2$+$144601   &   4.96   &   1.7, 5   &   C-f   &   32   \\   
J003126.8$+$150740   &   4.29   &   4.3, 7.6   &  C-f     &   52 \\   
J010013.0$+$280225   &   6.33   &   1.5   &  C-s    &   36, 51   \\   
J012126.1$+$034706   &   4.13   &   2.3, 8.4   &  C-s    &   25   \\   
J012201.8$+$031002   &   4.00   &   4.3, 7.6   &  C-f    &   46  \\   
J013127.3$-$032100   &   5.18   &   1.7, 22, 43   &   C-f   &   29, 35  \\   
J021043.1$-$001818   &   4.65   &   1.7, 5   &  C-f    &   32   \\   
J025759.1$+$433836   &   4.07   &   2.3, 8.4, 8.6, 8.7   &  J-f    &   31   \\   
J030947.0$+$271757   &   6.10   &   1.5, 5, 8.4   &  J-f    &   42   \\   
J031147.0$+$050802   &   4.51   &   1.7, 2.3, 5   &   E-s   &  9, 27  \\   
J032444.3$-$291821   &   4.62   &   2.3, 8.6, 8.7, 22   &  J-p    &   12,  24,  32,  31   \\   
J052506.1$-$334306   &   4.42   &   4.3, 7.6     &  J-f    &   46  \\   
J081333.2$+$350811   &   4.92   &   1.6, 5   &   E-s   &   17   \\   
J083643.8$+$005453   &   5.80   &   1.6, 5   &   C-s   &   5,  11   \\   
J083946.2$+$511201   &   4.40   &   1.7, 5   &   J-f   &   33   \\   
J090630.7$+$693031   &   5.47   &   2.3, 8.4, 8.6, 8.7, 15, 22, 43   &   C-p   &   7,  9, 22,  32,  35,  39,  44   \\   
J091316.5$+$591921   &   5.12   &   1.4   &   C-p   &   8   \\   
J094004.8$+$052630   &   4.50   &   1.7, 5   &   C-s   &   32   \\   
J101335.0$+$281181   &   4.75   &   1.7, 5   &   C-f   &   32   \\   
J102107.5$+$220921   &   4.26   &   4.3, 7.6   &  J-f    &   46  \\   
J102623.6$+$254259   &   5.25   &   1.7, 4.8, 4.9, 5, 22, 43   &   J-f   &   12, 13, 26, 30, 35   \\  
J102838.8$-$084438   &   4.28   &   2.3, 8.7   &   J-f   &   9   \\   
J105320.4$-$001649   &   4.30   &   1.4   &  C-f    &   8   \\   
J112925.4$+$184624   &   6.82   &   1.5   &  C-s    &   48  \\   
J114657.8$+$403708   &   4.98   &   1.6, 5   &   C-s   &   17   \\   
J115502.9$-$310758   &   4.30   &   2.3, 8.6, 8.7, 22   &  C-p    &   4   \\   
J120523.1$-$074232   &   4.69   &   1.4   &  E-f    &   10   \\   
J123503.0$-$000331   &   4.69   &   1.4   &  c-p    &   8   \\   
J124230.6$+$542257   &   4.73   &   1.6, 5   &   E-p   &   17   \\   
J125359.3$-$405932   &   4.46   &   2.3, 4.3, 7.6, 8.6, 8.7   &   C-s   &   4   \\  
J131121.3$+$222738   &   4.61   &   1.7, 5   &  C-f    &   32   \\   
J140025.4$+$314910   &   4.64   &   1.7, 5   &  C-f    &   32   \\ 
J142048.0$+$120546   &   4.03   &   1.5, 1.7, 5   &  E-s    &   33, 46, 49  \\   
J142738.5$+$331241   &   6.12   &   1.4, 1.6, 5   &   E-s   &   15,  16   \\   
J142952.1$+$544717   &   6.21   &   1.6, 5   &   E-s   &   19   \\   
J143023.7$+$420436   &   4.71   &   2.3, 4.8, 5, 8.3, 8.4, 8.6, 8.7, 15, 22   &  J-f    &    2 3,  6, 13, 18, 21  23,  24,  32,  43,  52  \\ 
J145147.1$-$151220   &   4.76   &   4.3, 7.6   &  C-s  &   37   \\   
J145459.0$+$110928   &   4.93   &   1.7, 5   &   C-f   &   32   \\   
J151002.9$+$570243   &   4.31   &   2.3, 4.8, 5, 7.6, 8.3, 8.6, 8.7, 15   & J-p &   1, 3, 9, 13, 20, 21, 31, 46, 52  \\   
J153049.9$+$104931   &   5.72   &   1.7   &   E-s    &   41   \\   
J153533.9$+$025423   &   4.39   &   4.3, 7.6   &  C-f     &   52  \\   
J154823.9$+$333459   &   4.68   &   1.7, 5   &  E-s    &   32   \\   
\enddata
\end{deluxetable*}
\begin{deluxetable*}{llccc}
\tablenum{1}
\tablecaption{Continued}
\tablewidth{0pt}
\tablehead{
\colhead{Name} & \colhead{$z$} & \colhead{$\nu$ (GHz)} & \colhead{Classification} & \colhead{VLBI reference}
}
\decimalcolnumbers
\startdata
J160608.5$+$312446   &   4.56   &   2.3, 4.8, 5, 8.3, 8.4, 8.7, 22   &  J-p    &   3,  13, 31, 32,  34,  50\\ 
J161105.6$+$084435   &   4.55   &   1.6, 5   &  C-f    &   17   \\   
J162830.0$+$115403   &   4.47   &   1.7, 5   &   C-s   &   32   \\   
J165913.2$+$210115   &   4.83   &   1.6, 5   &   J-p   &   17   \\   
J171521.1$+$214534   &   4.01   &   2.3, 4.3, 7.6, 8.6, 8.7   &  J-s    &  21, 46  \\   
J172026.6$+$310431   &   4.62   &   1.7, 5   &  C-f    &   32   \\   
J195136.0$+$013442   &   4.11   &   2.3, 4.3, 7.6, 8.4, 8.6, 8.7   &  C-p    &   9, 31, 37  \\ 
J210240.3$+$601510   &   4.58   &   2.3, 8.3, 8.6, 8.7   &   J-p   &   3,  14,  32, 39   \\   
J213412.0$-$041909   &   4.33   &   1.7, 4.3, 5, 7.6   &  J-p    &   33,  38,  46  \\   
J222032.6$+$002536   &   4.20   &   1.7, 5   &   E-s   &   33   \\   
J222843.5$+$011032   &   5.95   &   1.6   &   J-p   &   28   \\   
J224607.6$-$052635   &   4.60   &   1.4, 1.6  &  C-s       &  45      \\
J231448.7$+$020151   &   4.11   &   4.3, 7.6   &  J-f    &  52  \\   
J232936.8$-$152014   &   5.84   &   1.6   &  J-s     &   40   \\    
J235758.6$+$140202   &   4.33   &   4.3, 7.6   &  C-f    &   52  \\    
\enddata
\tablecomments{Col.~1 -- radio source name derived from J2000 right ascension and declination; Col.~2 -- redshift; Col.~3 -- VLBI observing frequencies; Col.~4 -- mas-scale morphological classification (capital letters) and spectral classification (small letters) based on low-resolution observations (see text in Section \ref{sec:sample}); Col.~5 -- VLBI references:
1: \citet{1997AA...325..511F};
2: \citet{1999AA...344...51P}; 
3: \citet{2002ApJS..141...13B}; 
4: \citet{2003AJ....126.2562F}; 
5: \citet{2003MNRAS.343L..20F}; 
6: \citet{2004AJ....127.3587F}; 
7: \citet{2004ApJ...610L...9R}; 
8: \citet{2004AJ....127..587M}; 
9: \citet{2005AJ....129.1163P}; 
10: \citet{2005AJ....129.1809M}; 
11: \citet{2005AA...436L..13F}; 
12: \citet{2006AJ....131.1872P}; 
13: \citet{2007ApJ...658..203H}; 
14: \citet{2008AJ....136..580P}; 
15: \citet{2008AJ....136..344M}; 
16: \citet{2008AA...484L..39F}; 
17: \citet{2010AA...524A..83F}; 
18: \citet{2010AA...521A...6V}; 
19: \citet{2011AA...531L...5F}; 
20: \citet{2011MNRAS.415.3049O};
21: \citet{2012AA...544A..34P}; 
22: \citet{2012AJ....144..150P}; 
23: \citet{2012ApJ...756L..20C}; 
24: \citet{2012AJ....143...35P}; 
25: \citet{2013AJ....146....5P}; 
26: \citet{2013MNRAS.431.1314F}; 
27: \citet{2014MNRAS.439.2314P}; 
28: \citet{2014AA...563A.111C}; 
29: \citet{2015MNRAS.450L..57G}; 
30: \citet{2015MNRAS.446.2921F}; 
31: \citet{2016AJ....151..154G}; 
32: \citet{2016MNRAS.463.3260C}; 
33: \citet{2017MNRAS.467..950C}; 
34: \citet{2017ApJS..228...22L}; 
35: \citet{2017MNRAS.468...69Z}; 
36: \citet{2017ApJ...835L..20W}; 
37: \citet{2017ApJ...838..139S}; 
38: \citet{2018MNRAS.477.1065P}; 
39: \citet{2018AA...618A..68F}; 
40: \citet{2018ApJ...861...86M}; 
41: \citet{2018RNAAS...2..200G}; 
42: \citet{2020AA...643L..12S}; 
43: \citet{2020SciBu..65..525Z}; 
44: \citet{2020NatCo..11..143A}; 
45: \citet{2020ApJ...905L..32F}
46: \citet{2021AJ....161...14P};
47: \citet{2021AJ....162..121H}; 
48: \citet{2021AJ....161..207M}; 
49: \citet{2021AN....342.1092G}; 
50: \citet{2022MNRAS.tmp..166A}; 
51: \citet{2022arXiv220302922L}
52: Astrogeo database (\url{http://astrogeo.org/})}
\label{vlbilit2}
\end{deluxetable*}

To increase the sample compiled in Table~\ref{vlbilit}, we chose 13 high-redshift radio-loud quasars from the list of \citet{2013MNRAS.433.2182S} which contains 31 $z>4$ blazar candidates with radio loudness $R > 100$. Their selection was based on the catalog of \citet{2011ApJS..194...45S} who matched the SDSS Data Release 7 (DR7) quasar list \citep{2010AJ....139.2360S} with radio sources from the FIRST survey \citep{1997ApJ...475..479W}. Out of the 31 sources from the list of \citet{2013MNRAS.433.2182S}, 15 were already imaged with VLBI. A careful inspection of the original survey catalogs revealed that as many as 3 SDSS DR7 optical quasars (i.e. about $10\%$) were mistakenly identified by \citet{2011ApJS..194...45S} with apparently unrelated radio sources, due to the large matching radius they applied, $30\arcsec$. In contrast, \citet{2002AJ....124.2364I} found that with just $3\arcsec$ radius, practically all true SDSS--FIRST matches are included. The three misidentified objects in \citet{2011ApJS..194...45S} (SDSS~J111856.15+370255.9, SDSS~J143003.96+144354.8, and SDSS~J145212.86+023526.3) are in fact not radio-loud high-redshift quasars.
 
At the time of initiating our VLBI observing project, the sample of these 13 new sources was as large as about one-third of all VLBI-imaged radio sources above redshift 4. Even now, with new results in the literature published after the start of our project, the increase due to our results is by more than $25\%$ (cf. Table~\ref{vlbilit}). Our target sources and their parameters are listed in Table~\ref{tabsample}. Each of the targets can be found in the FIRST \citep{1997ApJ...475..479W} 
and the U.S. National Radio Astronomy Observatory (NRAO) VLA Sky Survey \citep[NVSS,][]{1998AJ....115.1693C} 
catalogs. Each of them is detected also
in the ongoing Karl G. Jansky Very Large Array All Sky Survey \citep[VLASS,][]{2020PASP..132c5001L,2020RNAAS...4..175G}. Their redshifts are within a narrow range of $4 < z < 4.5$, thus they represent radio AGN that are at similar cosmological distances. Note that J1307$+$1507 is blended in NVSS with a bright radio source seen within $\sim 13\arcsec$ separation. Therefore we have to consider its flux density as an upper limit in Table~\ref{tabsample}.

Out of the 13 sources chosen, we found 6 having X-ray measurements available in the literature. All 6 are observed with Chandra \citep{2002PASP..114....1W} in various surveys \citep{2004AJ....128..523B, 2005AJ....129.2519V, 2013ApJ...763..109W, 2019MNRAS.482.2016Z}. One of them (J1309$+$5733) is also detected in the XMM-Newton Slew Survey \citep{2008AA...480..611S}.

\begin{deluxetable*}{llcccc}
\tablenum{2}
\tablecaption{The $z > 4$ radio quasar sample presented in this study.}
\tablewidth{0pt}
\tablehead{
\colhead{Name} & \colhead{Source ID} & \colhead{$z$} & \colhead{$S_\mathrm{FIRST,1.4 GHz}$} & \colhead{$S_\mathrm{NVSS,1.4 GHz}$} & \colhead{$S_\mathrm{VLASS,2.7 GHz}$} \\
\nocolhead{} & \nocolhead{} & \nocolhead{} & (mJy) & (mJy) & (mJy)
}
\decimalcolnumbers
\startdata
J030437.21$+$004653.5 & J0304$+$0046 & 4.31 & 21.0 (0.1) & 24.6 (0.8)      & 14.63 (0.16)    \\
J085111.59$+$142337.7 & J0851$+$1423 & 4.31 & 16.2 (0.1) & 12.3 (0.5)      & 6.75 (0.21)    \\
J091824.38$+$063653.3 & J0918$+$0636 & 4.19 & 26.5 (0.1) & 30.9 (1.0)	    & 41.18 (0.24)    \\
J100645.58$+$462717.2 & J1006$+$4627 & 4.44 & 6.3 (0.4)  & 6.3 (0.1)	    & 6.73 (0.24)   \\
J103717.72$+$182303.0 & J1037$+$1823 & 4.05 & 13.7 (0.2) & 11.4 (0.5)	    & 7.89 (0.23)  \\
J123142.17$+$381658.9 & J1231$+$3816 & 4.14 & 24.0 (0.1) & 25.7 (0.9)      & 11.09 (0.20)  \\
J130738.83$+$150752.0 & J1307$+$1507 & 4.11 & 3.9 (0.1)  & $<$16.2 (0.6)	& 2.07 (0.22)  \\
J130940.70$+$573309.9 & J1309$+$5733 & 4.27 & 11.3 (0.1) & 11.2 (0.9)	    & 11.07 (0.24) \\
J132512.49$+$112329.7 & J1325$+$1123 & 4.41 & 71.1 (0.1) & 81.4 (2.5)	    & 51.03 (0.26) \\
J141209.96$+$062406.9 & J1412$+$0624 & 4.47 & 43.5 (0.1) & 47.2 (1.5)  	& 25.98 (0.25)  \\
J143413.05$+$162852.7 & J1434$+$1628 & 4.20 & 4.2 (0.1)  & 5.0 (0.5)	    & 2.30 (0.29)  \\
J152028.14$+$183556.1 & J1520$+$1835 & 4.12 & 6.9 (0.2)  & 8.8 (0.5)	    & 2.54 (0.25)  \\
J172007.19$+$602824.0 & J1720$+$6028 & 4.42 & 5.1 (0.2)  & 6.6 (0.4)	    & 5.15 (0.17)  \\
\enddata
\tablecomments{Col.~1 -- radio source name derived from J2000 right ascension and declination; Col.~2 -- shortened source name used throughout this paper; Col.~3 -- redshift; Col.~4 -- FIRST 1.4~GHz flux density and its uncertainty; Col.~5 -- NVSS 1.4~GHz flux density and its uncertainty; Col.~6 -- VLASS 2.7~GHz flux density and its uncertainty.}
\label{tabsample}
\label{targetstable}
\end{deluxetable*}

\section{Observations and Data Reduction} 
\label{sec:obs}
\subsection{EVN Observations}

The selected target sources were observed with the EVN in e-VLBI mode \citep[e-EVN,][]{2004evn..conf..257S} at two central frequencies, 1.66 and 4.99~GHz. In the e-VLBI experiments, data from the radio telescopes were streamed to the correlator through an optical fiber network in real time. The data were processed at the SFXC software correlator \citep{2015ExA....39..259K} at the Joint Institute for VLBI European Research Infrastructure Consortium (JIVE) in Dwingeloo, The Netherlands.

The series of experiments started in 2017 December and ended in 2020 November, under the project code EG102 (PI: K. Gab\'{a}nyi). A total of 12 observing sessions (EG102A to EG102L) were scheduled. In addition to the elements of the EVN, antennas of the e-MERLIN were occasionally also included in the observing network. Table~\ref{observations} contains the details of the EVN observations, including the date, the frequency, the target sources with their corresponding phase-reference calibrators, and the participating telescopes in each project segment. The following radio telescopes participated in the various segments: e-EVN: Jodrell Bank Mk2 (Jb, United Kingdom), Westerbork (Wb, The Netherlands), Effelsberg (Ef, Germany), Medicina (Mc, Italy), Noto (Nt, Italy), Sardinia (Sr, Italy), Onsala 25-m (O8, Sweden), Toru\'{n} (Tr, Poland), Irbene 32-m (Ir, Latvia), Irbene 16-m (Ib, Latvia), Yebes (Ys, Spain), Svetloe (Sv, Russia), Zelenchukskaya (Zc, Russia), Badary (Bd, Russia), Tianma (T6, China), Sheshan (Sh, China), Kunming (Km, China), Hartebeesthoek (Hh, South Africa); e-MERLIN (United Kingdom): Cambridge (Cm), Darnhall (Da), Defford (De), Knockin (Kn), Pickmere (Pi).

The observations were performed in phase-reference mode \citep{1995ASPC...82..327B}, by regularly nodding between the target sources and the corresponding bright nearby calibrators (all within $\approx 2\degr$ angular separation). The same phase-reference calibrator was used at both frequencies for each target source (Table~\ref{observations}).

\begin{longrotatetable}
\begin{deluxetable*}{ccclccccc}
\tablenum{3}
\tablecaption{Details of the observations in the EVN project EG102.}
\tablewidth{0pt}
\tablehead{
\colhead{Project} & \colhead{Frequency} & \colhead{Observing date} & \colhead{Participating radio telescopes} & \colhead{Source ID} & \colhead{Total on-source } & \colhead{Datarate} & \colhead{Phase} & \colhead{Separation}
\\ \colhead{segment} & \colhead{$\nu$ (GHz)} & \nocolhead{} & \nocolhead{} & \nocolhead{} & \colhead{time (min)} & \colhead{(Mbps)} & \colhead{calibrator}  & \colhead{($\degr$)}
}
\decimalcolnumbers
\startdata
A & 1.7 & 2017 Dec 14 & Jb, Wb, Ef, Mc, O8, T6, Tr, Hh, Ir & J0304$+$0046 & 198 & 1024 & J0301$+$0118 & 0.96 \\
B & 5 & 2018 Jan 16 & Jb, Ef, Mc, Nt, O8, Sh, Tr, Hh, Ib & J0304$+$0046 & 108 & 2048 & J0301$+$0118 & 0.96  \\
C & 5 & 2018 Apr 10 & Jb, Wb, Ef, Mc, Nt, O8, T6, Tr, Ys, Hh, Ib & J0918$+$0636 & 222 & 2048 & J0915$+$0745 & 1.31 \\
D & 1.7 & 2018 May 17 & Jb, Wb, Ef, Mc, O8, T6, Tr, Hh, Ib, Sr & J0918$+$0636 & 270 & 1024 & J0915$+$0745 &  1.31 \\
 & & &                                                            & J1309$+$5733 & 144 & 1024 & J1302$+$5748 & 0.95 \\
E & 5 & 2018 Jun 19 & Jb, Wb, Ef, Mc, Nt, Tr, Ys, Hh, Ib, T6 & J1309$+$5733 & 198 & 1024 & J1302$+$5748 & 0.95 \\
F & 1.7 & 2019 Jan 23 & Jb, Wb, Ef, Mc, O8, T6, Tr, Hh, Ir, Sr & J1307$+$1507 & 108 & 1024 & J1300$+$1417 & 1.88 \\
& & &                                                            & J1520$+$1835 & 138 & 1024 & J1521$+$1756 & 0.69 \\
G & 5 & 2019 Feb 14 & Jb, Wb, Ef, Mc, Nt, O8, T6, Tr, Ys, Hh, Ib & J1307$+$1507 & 150 & 2048 & J1300$+$1417 & 1.88  \\
& & &                                                            & J1520$+$1835 & 126 & 2048 & J1521$+$1756 & 0.69 \\
H & 5 & 2019 Mar 19 & Jb, Wb, Ef, Nt, Mc, O8, T6, Tr, Ys, Hh, Ib, & J1037$+$1823 & 102 & 2048 & J1045$+$1735 & 2.05 \\
&&&                       Cm, Da, De, Kn, Pi                      & J1231$+$3816 & 132 & 2048 & J1228$+$370 & 1.31 \\
&&&                                                               & J1412$+$0624 & 132 & 2048 & J1410$+$0731 & 1.19 \\
I & 5 & 2019 May 14 & Jb, Wb, Ef, Mc, Nt, O8, T6, Tr, Ys, Hh, Ir,  & J0851$+$1423 & 96 & 2048 & J0858$+$1409 & 1.83 \\
&&&              Sv, Bd, Zc, Cm, Da, De, Kn, Pi                   & J1006$+$4627 & 144 & 2048 & J0958$+$4725 & 1.73 \\
&&&                                                               & J1325$+$1123 & 108 & 2048 & J1327$+$1223 & 1.19 \\
J & 1.7 & 2020 Jun 23 & Jb, Wb, Ef, O8, T6, Hh, Ir,              & J0851$+$1423 & 102 & 1024 &  J0858$+$1409 & 1.83 \\
&&&              Cm, Da, De, Kn, Pi                              & J1037$+$1823 & 150 & 1024 & J1045$+$1735 & 2.05 \\
&&&                                                               & J1231$+$3816 & 108 & 1024 & J1228$+$3706 & 1.31 \\
&&&                                                               & J1412$+$0624 & 126 & 1024 & J1410$+$0731 & 1.19 \\
&&&                                                               & J1434$+$1628 & 192 & 1024 & J1428$+$1628 & 1.45 \\
K & 1.7 & 2020 Oct 07 & Jb, Wb, Ef, Mc, Nt, O8, T6, Hh, Ir,     & J1006$+$4627 & 132 & 1024 & J0958$+$4725 & 1.73 \\
&&&                Sr, Cm, Da, De, Kn, Pi                       & J1325$+$1123 & 180 & 1024 & J1327$+$1223 & 1.19 \\
&&&                                                               & J1720$+$6028 & 126 & 1024 & J1722$+$6105 & 0.70 \\
L & 5 & 2020 Nov 18 & Jb, Wb, Ef, Mc, O8, T6, Tr, Ys, Hh, Ir,  & J1434$+$1628 & 144 & 2048 & J1428$+$1628 & 1.45 \\
&&&       Sv, Zc, Bd, Km, Cm, Da, De, Kn, Pi                   & J1720$+$6028 & 144 & 2048 & J1722$+$6105 & 0.70 \\
\enddata
\label{observations}
\end{deluxetable*}
\end{longrotatetable}

\subsection{Data Reduction}
\label{datared}

The VLBI data were calibrated with the NRAO Astronomical Image Processing System (\textsc{aips}) software package \citep{2003ASSL..285..109G}, following the standard procedure. After loading the raw correlated data, interferometric visibility amplitudes were calibrated using the antenna gain curves and the system temperatures measured at the telescopes (or nominal values if system temperature measurements were unavailable). The data were corrected for the dispersive ionospheric delay using total electron content maps derived from global navigation satellite systems data. Phase changes due to the time variation of the source parallactic angle were corrected for radio telescopes with azimuth--elevation mount. Then global fringe-fitting \citep{1983AJ.....88..688S} was performed on the phase-reference calibrators and bright fringe-finder sources also scheduled in the experiments. The calibrated visibility data were exported to \textsc{Difmap} \citep{1994BAAS...26..987S}, where we carried out hybrid mapping, including several iterations of the \textsc{clean} algorithm \citep{1974A&AS...15..417H} and phase-only self-calibration \citep{1986A&A...168..365A}.
Then antenna-based gain correction factors were determined for the different calibrator and fringe-finder sources. If exceeding $\pm5\%$, the median correction factors were applied to the visibility amplitudes in \textsc{aips}, for all the calibrator and the target sources. The \textsc{clean} model components of the calibrators produced in \textsc{Difmap} were transferred to \textsc{aips}, as inputs for the repeated fringe-fitting of the calibrator data, to improve phase solutions by taking the calibrator source structure into account. Finally, the fringe-fit solutions obtained for the phase-reference calibrators were interpolated to the target source data. The calibrated visibility data of the target sources were then exported from \textsc{aips}.

\textsc{Difmap} was used to produce the images of the target sources. We applied natural weighting, with amplitude errors raised to the power of $-1$ (\textsc{uvweight $0,-1$}) to reduce the image noise. For the typically weak (mJy-level) targets, up to a few rounds of \textsc{clean} iterations were performed, without self-calibration. Phase-only self-calibration was attempted for four sources only (J0918+0636, J1325+1123, and J1412+0624 at both frequencies, J1309+5733 at 1.7~GHz only) which had flux densities exceeding 15~mJy in their compact components. After reaching an insignificant peak-to-noise level (below $\sim 6$) in the residual images, a \textsc{clean} iteration involving $1000$ steps with a small loop gain $(0.01)$ was applied to smooth the noise features in the final images displayed in Figure~\ref{evnimages}. The image parameters are listed in Table~\ref{imgparams}. Circular Gaussian brightness distribution model components were also fitted to the visibility data \citep{1995ASPC...82..267P} in \textsc{Difmap}, to quantitatively characterize sizes and flux densities (Table~\ref{phyparams}).

We note that the project segments EG102B (J0304$+$0046 at 5~GHz) and EG102C (J0918$+$0636 at 5~GHz) observed in early 2018 were affected by a temporary technical problem with the calibration control in the digital baseband converters at the EVN stations \citep[see also][]{2019Sci...363..968G}. The resulting loss of coherence prevented us from properly calibrating the amplitudes and making reliable images and brightness distribution models of these two sources at 5~GHz. However, they were clearly detected and their phase-referenced positions could still be estimated (Table~\ref{imgparams}). Among the two sources above, J0918$+$0636 was observed previously with the EVN at 5~GHz on 1999 Sep 23 (experiment code ES034A, PI: I.A.G. Snellen). The participating seven radio observatories were Ef, Jb, Mc, Nt, O8, Hh, and the phased array of the Westerbork Synthesis Radio Telescope with 13 antenna elements. The observing time spent on J0918$+$0636 was 1.8~h. 
The observing setup and the calibration process were similar to those reported for the experiment ES034B \citep{2018MNRAS.477.1065P}. Here in Table~\ref{phyparams} we present fitted Gaussian model parameters derived for J0918$+$0636 from this archival EVN observation, to supplement our EG102 data.

For weaker sources below 15~mJy flux density (i.e. those that were not self-calibrated), the effect of coherence loss on the peak brightnesses and flux densities \citep{2010A&A...515A..53M} has to be considered. To estimate the level of coherence loss, we imaged and modeled the stronger sources with and without performing phase self-calibration and found an average model flux density difference of $25\%$. While the actual value may vary from source to source because of different target--calibrator separations, network stations and signal-to-noise ratios, this correction factor is consistent with estimates found for other EVN experiments \citep[e.g.][]{2006A&A...445..413M,2019A&A...630L...5G}. Therefore, we multiplied the fitted flux densities for the weak sources by a factor of 1.25 and marked those values with an asterisk in Table~\ref{phyparams}.

\begin{figure*}
\centering
\gridline{  \includegraphics[width=0.37\textwidth,angle=-90]{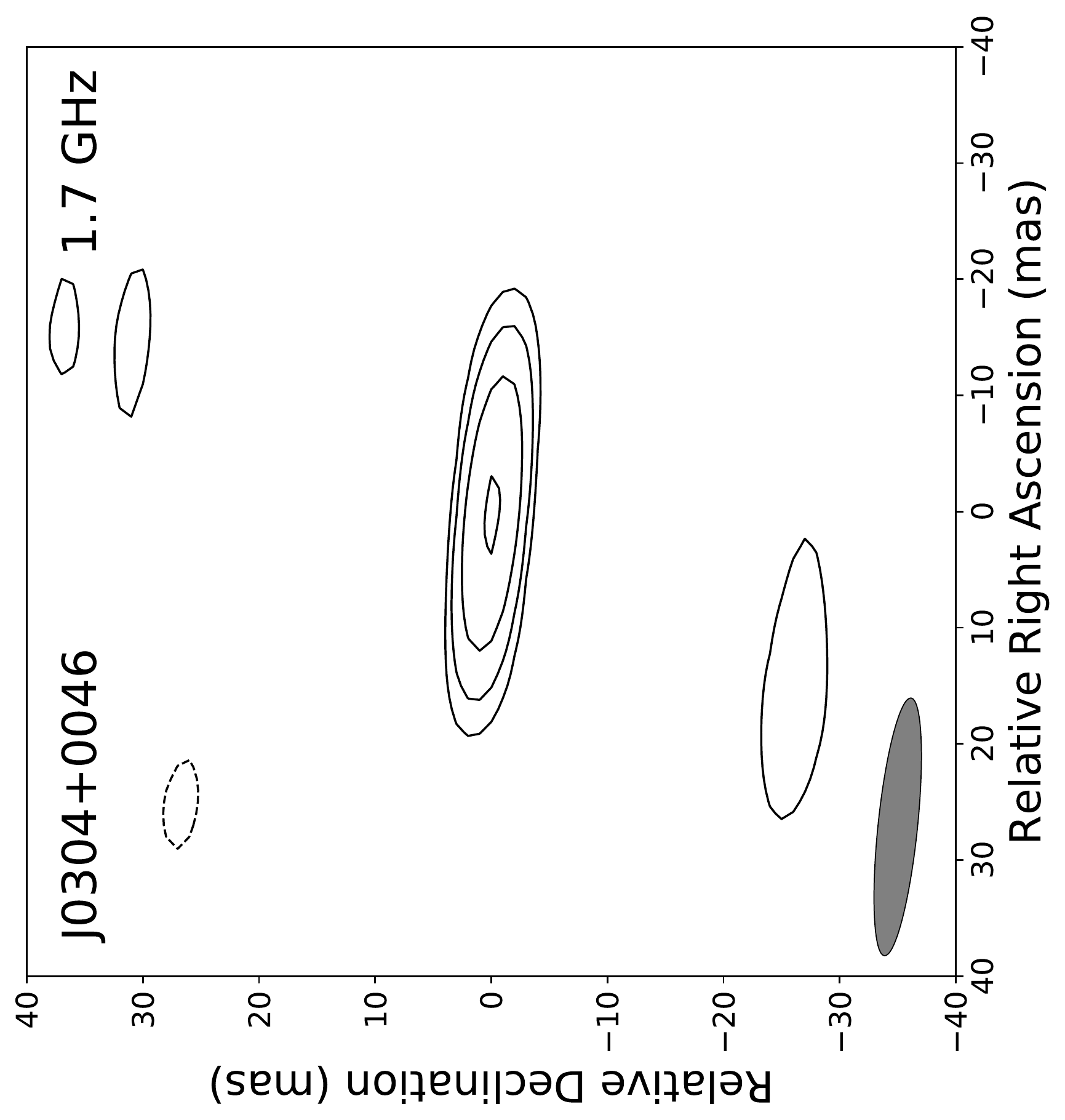}
            \includegraphics[width=0.37\textwidth,angle=-90]{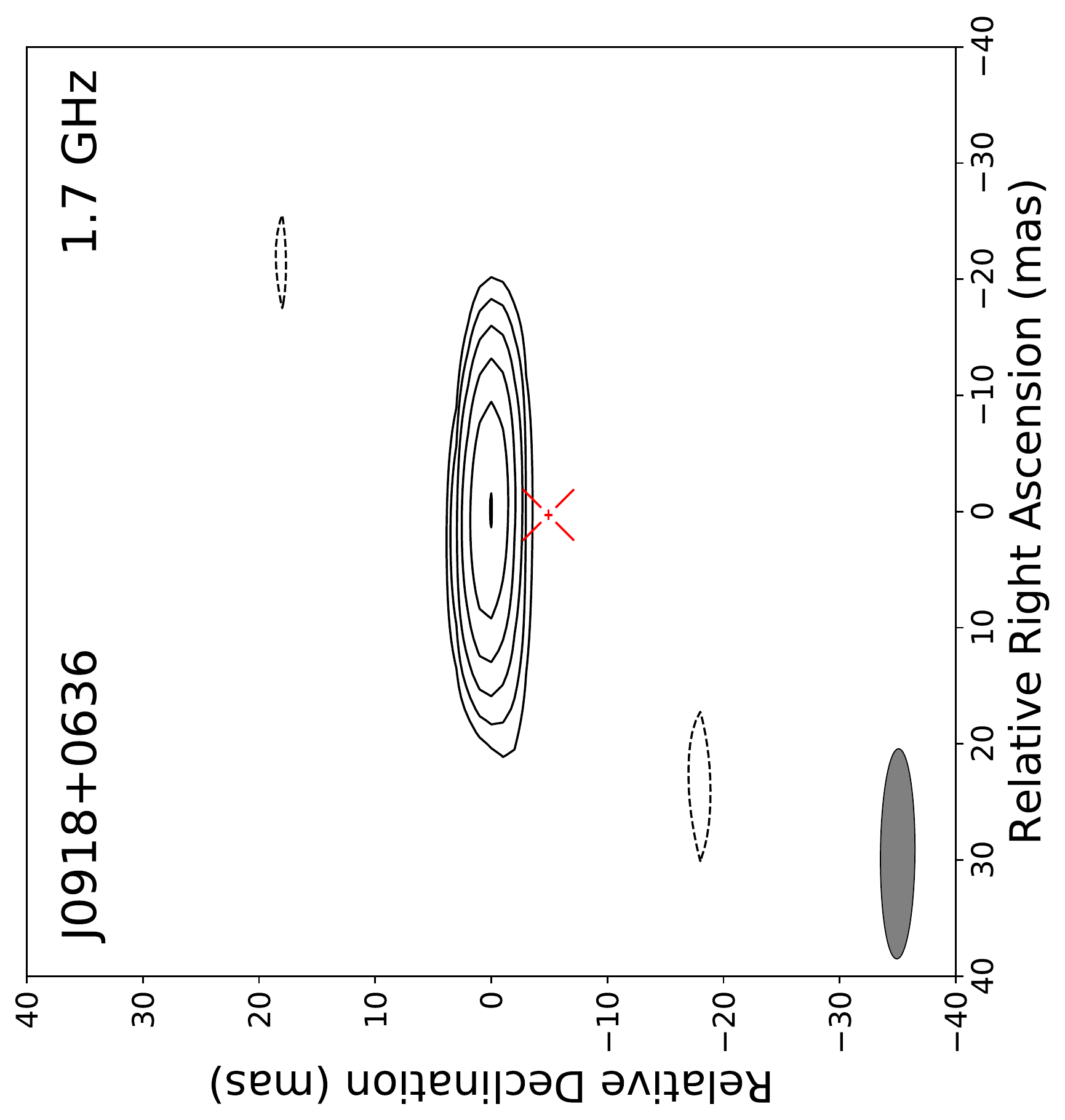}
            }
\gridline{  \includegraphics[width=0.37\textwidth,angle=-90]{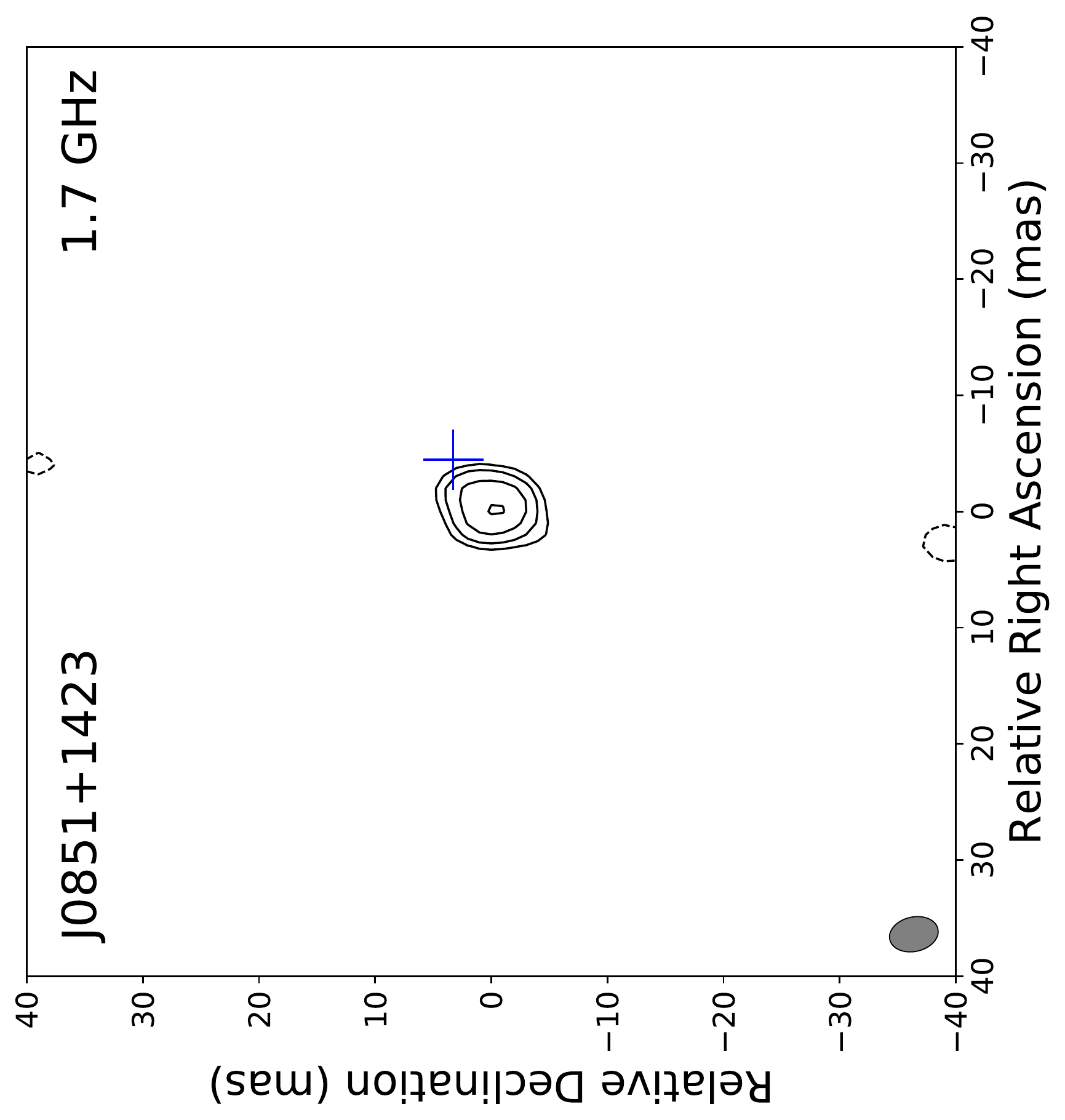}
            \includegraphics[width=0.37\textwidth,angle=-90]{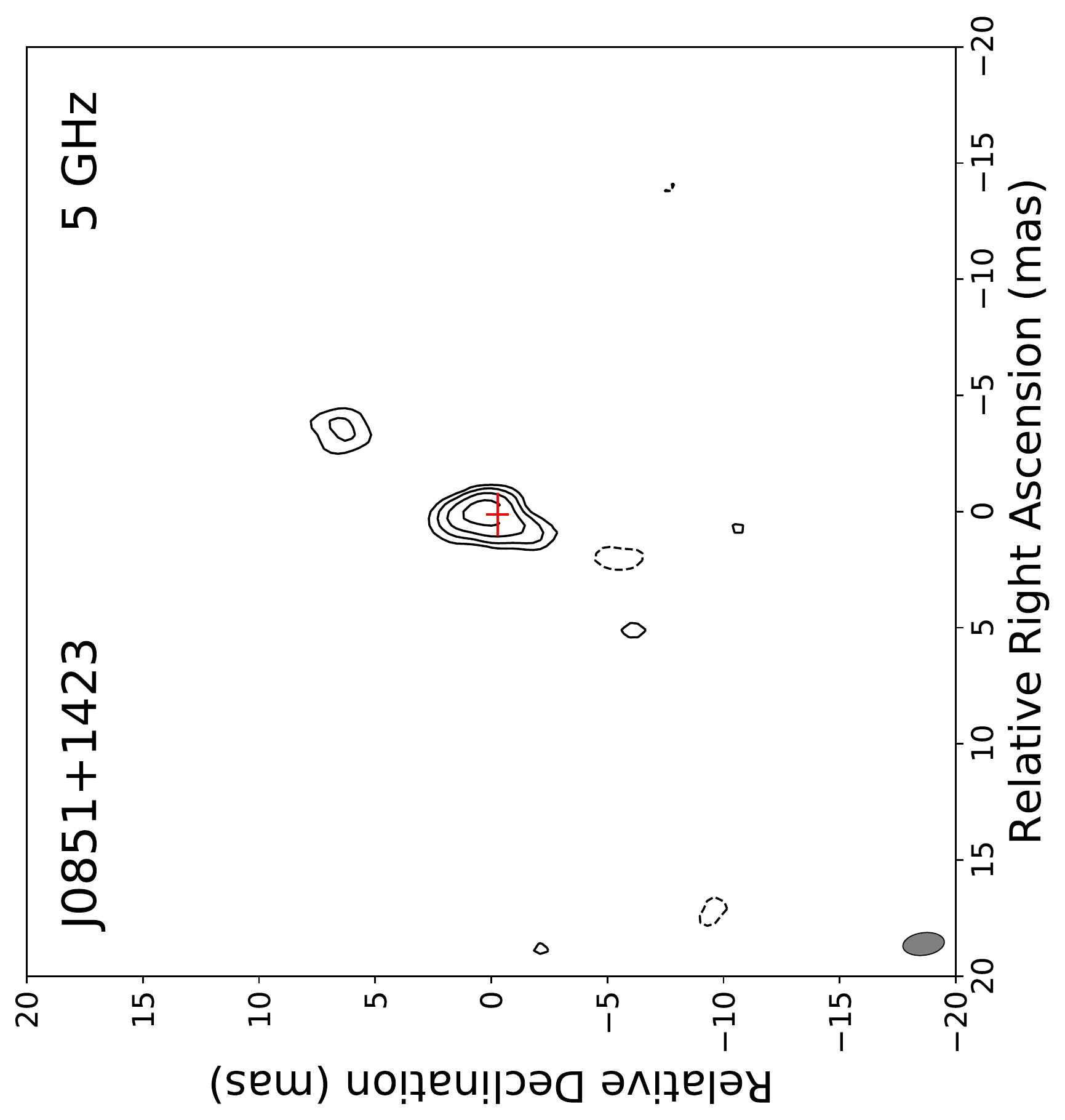}
            }
\gridline{  \includegraphics[width=0.37\textwidth,angle=-90]{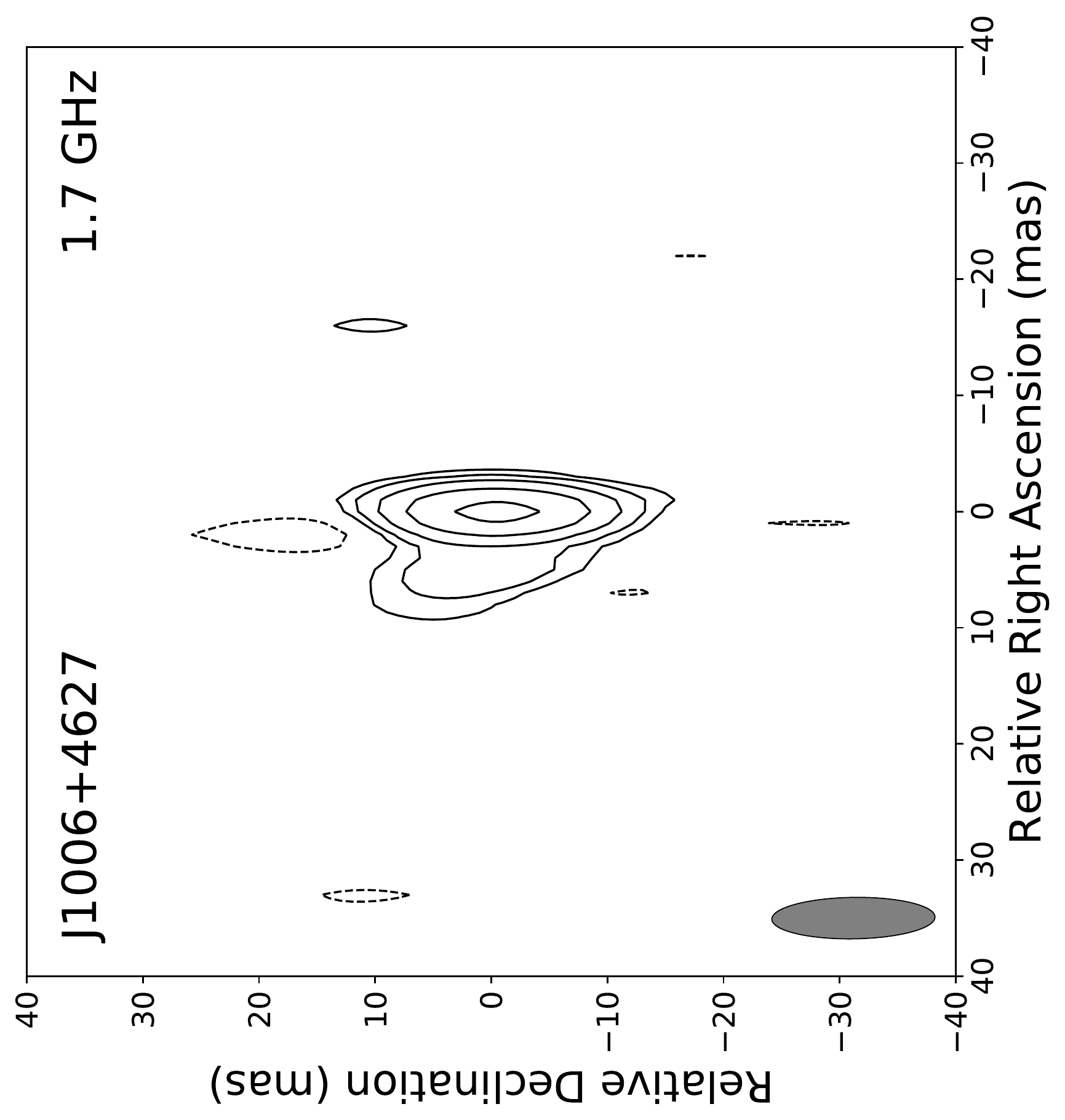}
            \includegraphics[width=0.37\textwidth,angle=-90]{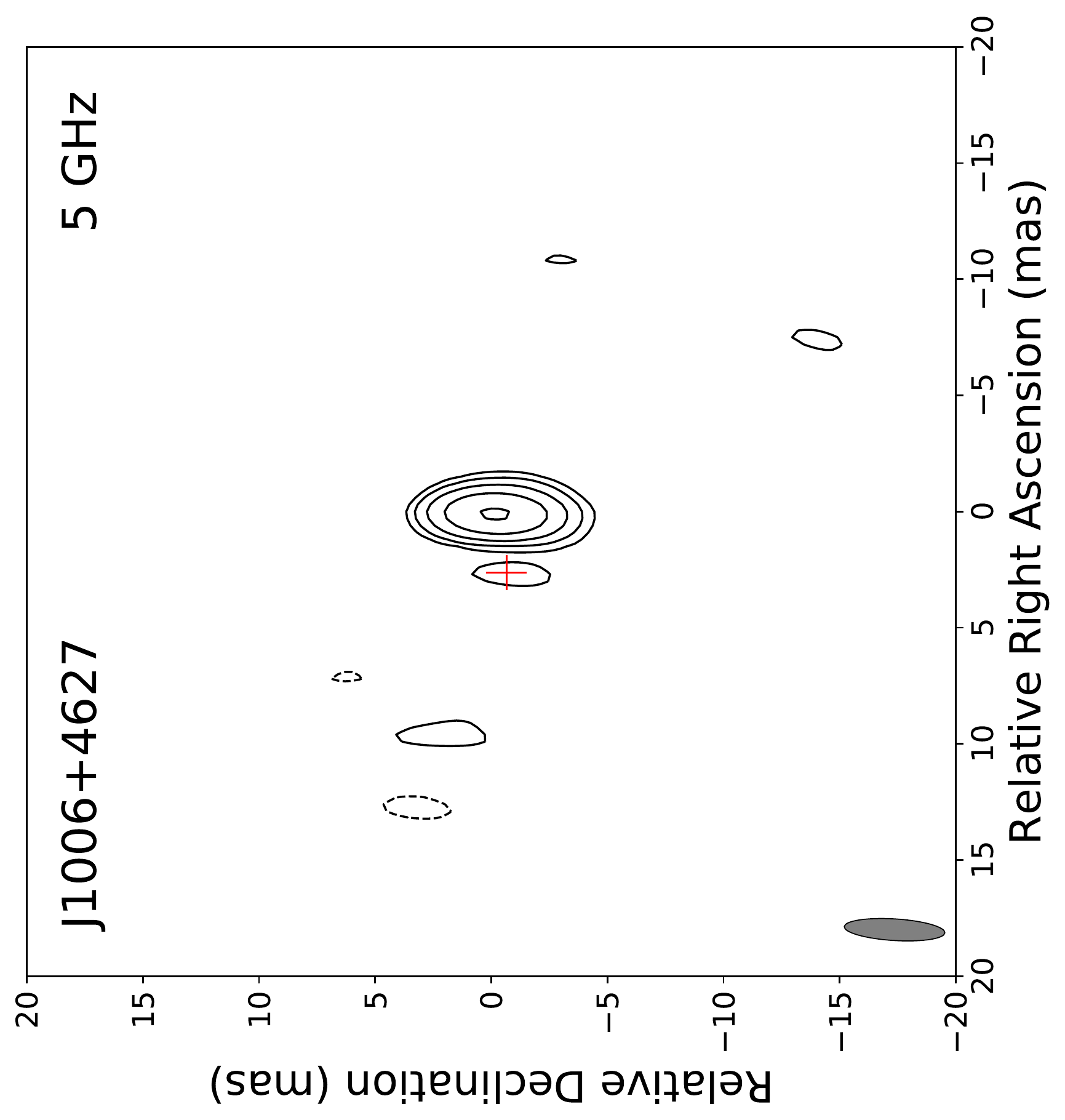}
            }
    \caption{EVN images at 1.7 and 5 GHz. The red cross indicates the Gaia optical position with its uncertainty, where available. In cases where the red crosses are too small, additional large red $\times$ symbol marks the position. Its purpose is simply to guide the eye and the size has no special meaning. The blue cross in the 1.7-GHz image denotes the 5-GHz VLBI position with the uncertainties of the 1.7- and 5-GHz positional difference for  sources where the offset between the two VLBI radio positions is non-negligible. The lowest contours are drawn at $\pm3$ times the image noise. The positive contours increase by a factor of 2. The restoring beam is shown in the bottom-left corner. Table~\ref{imgparams} contains the image parameters.} 
\end{figure*}
\begin{figure*}
    \setcounter{figure}{0}
\gridline{  \includegraphics[width=0.37\textwidth,angle=-90]{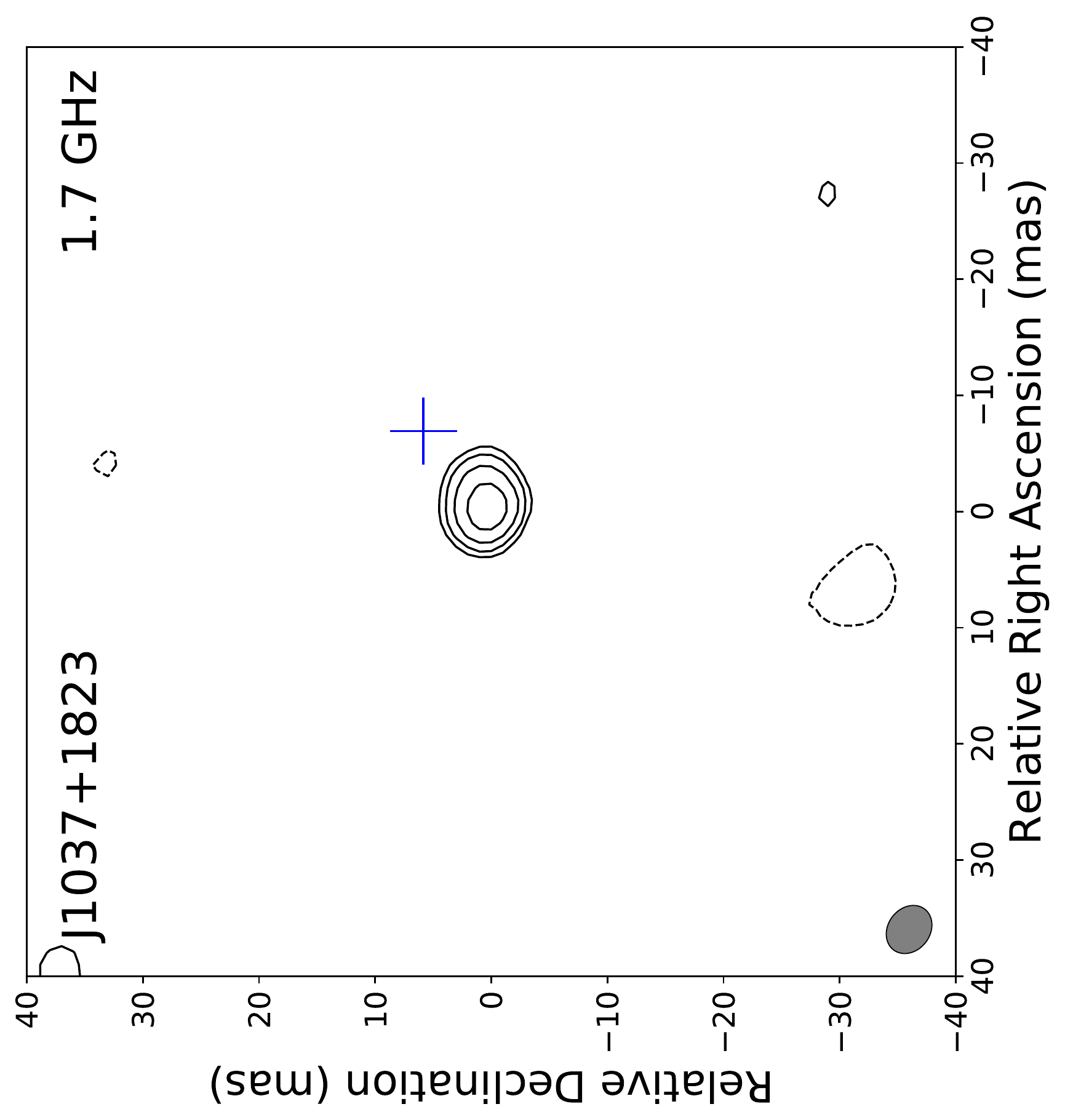}
            \includegraphics[width=0.37\textwidth,angle=-90]{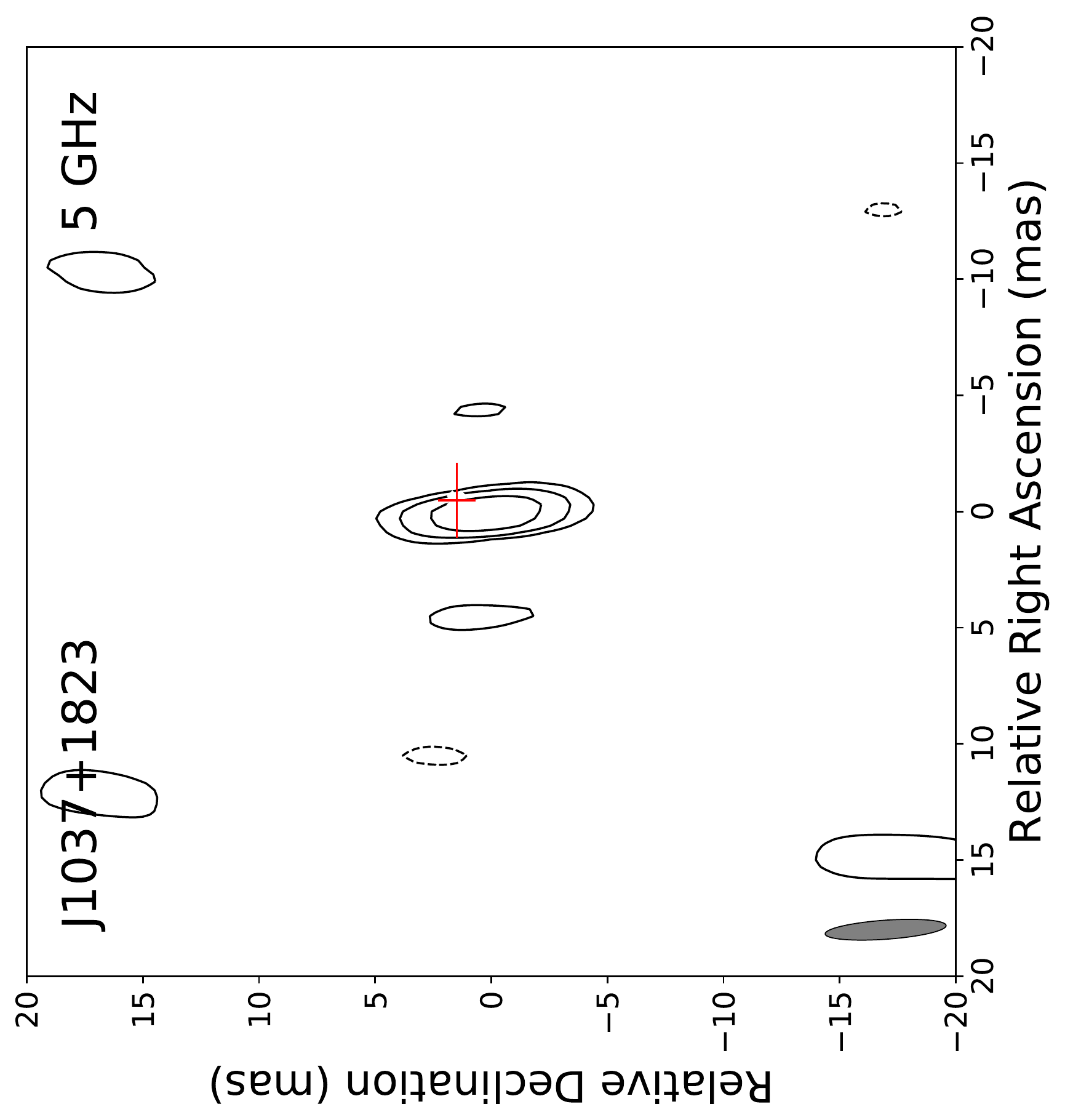}
            }
\gridline{  \includegraphics[width=0.37\textwidth,angle=-90]{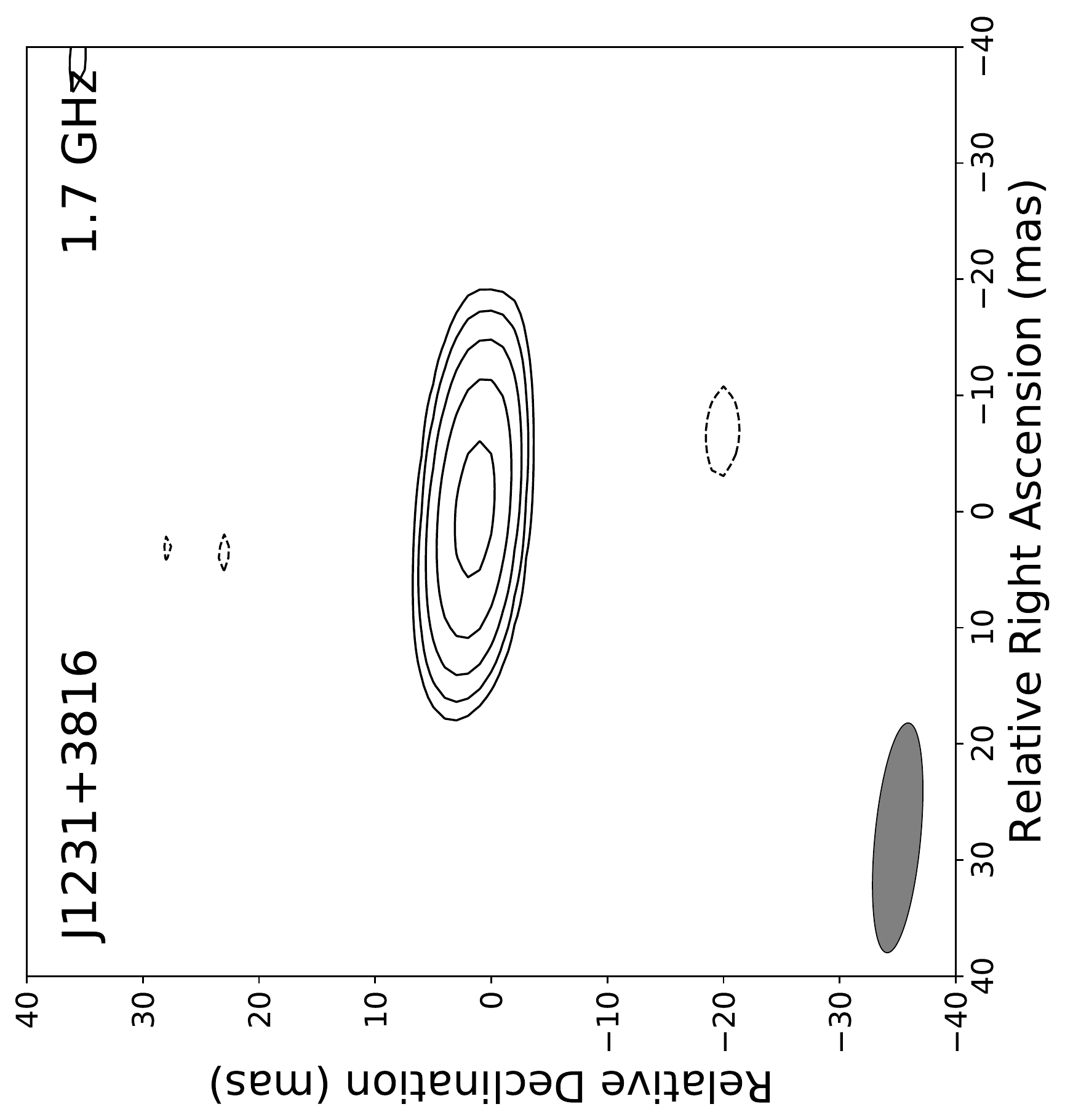}
            \includegraphics[width=0.37\textwidth,angle=-90]{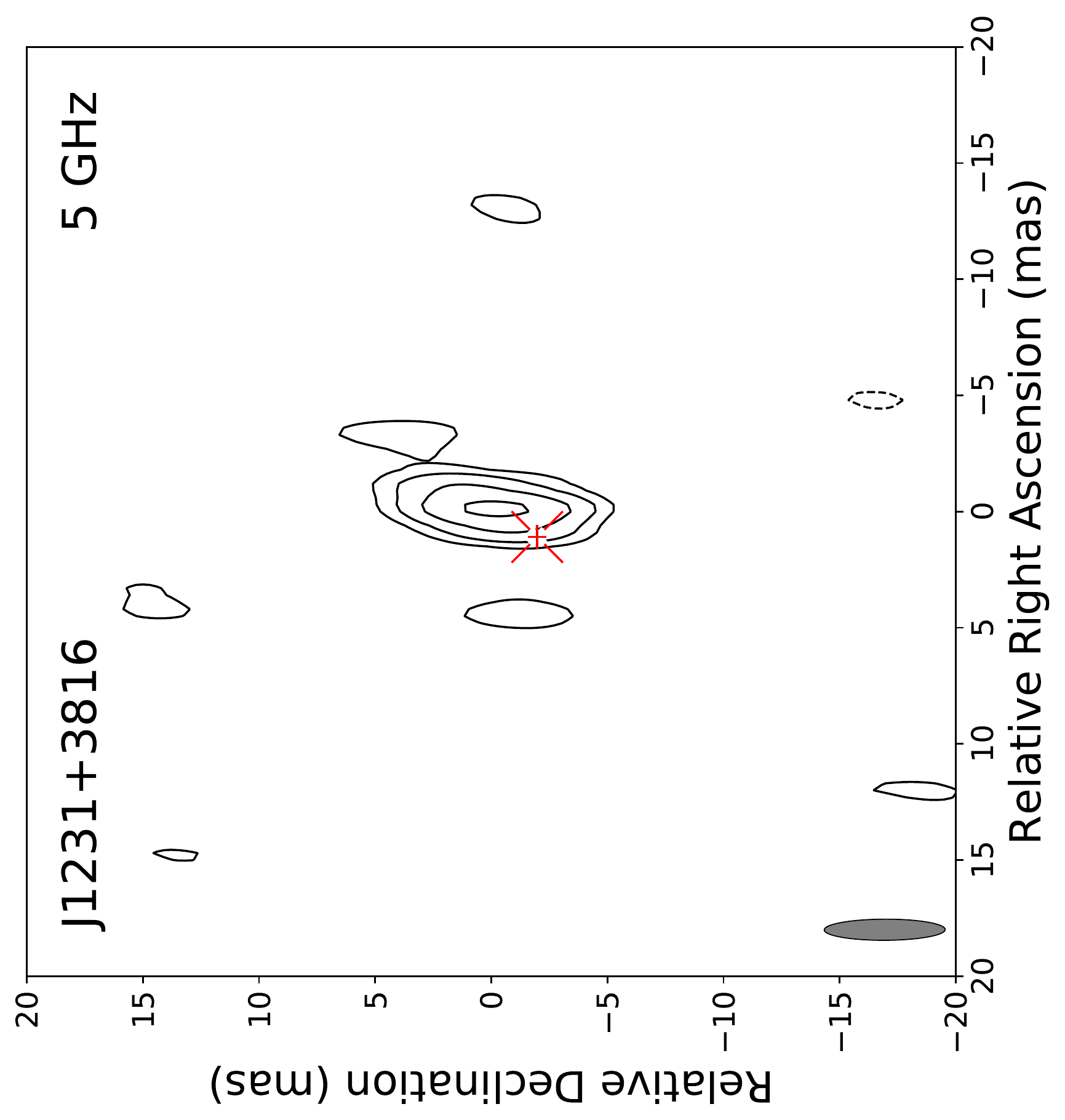}
            }
\gridline{  \includegraphics[width=0.37\textwidth,angle=-90]{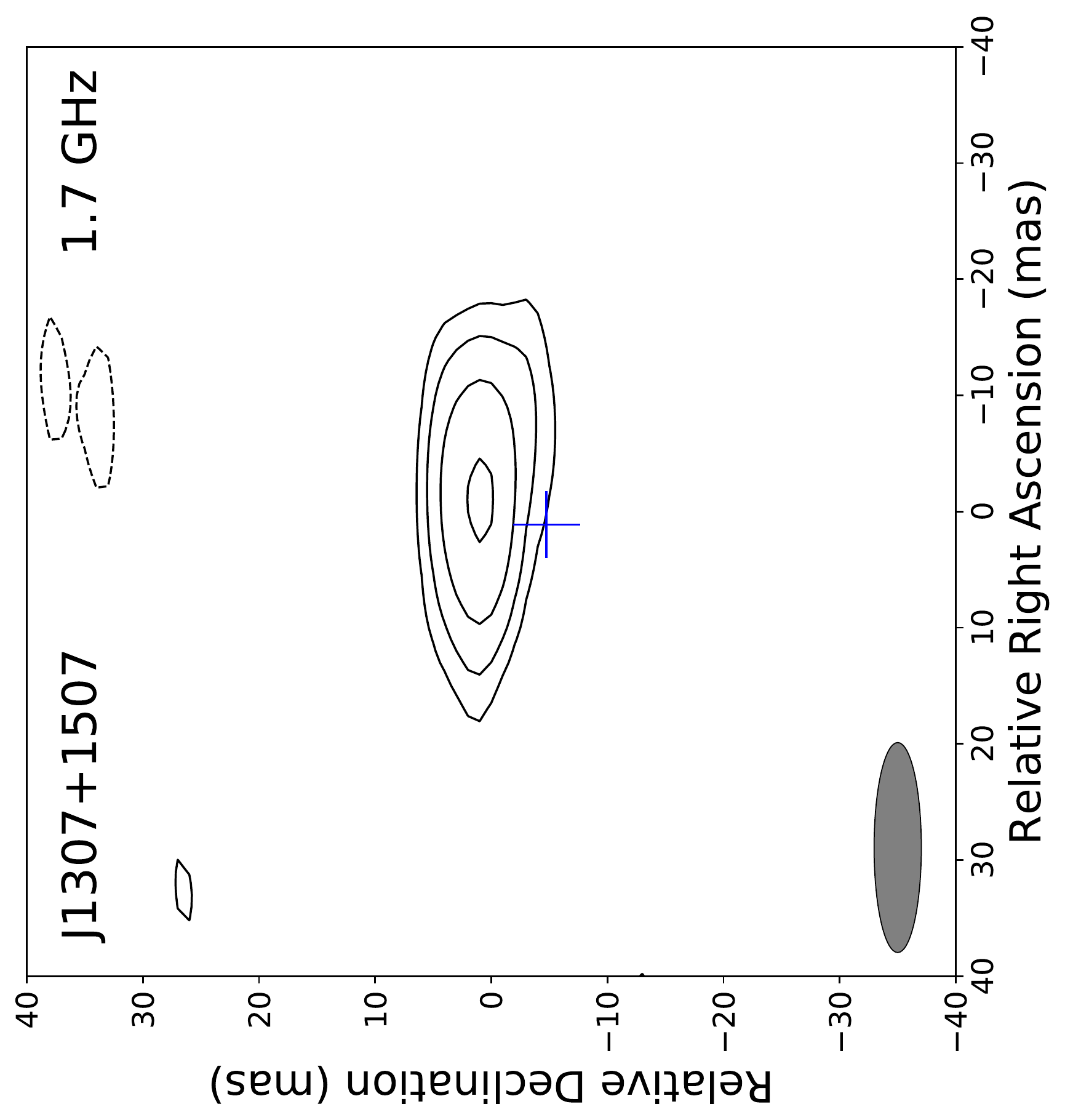}
            \includegraphics[width=0.37\textwidth,angle=-90]{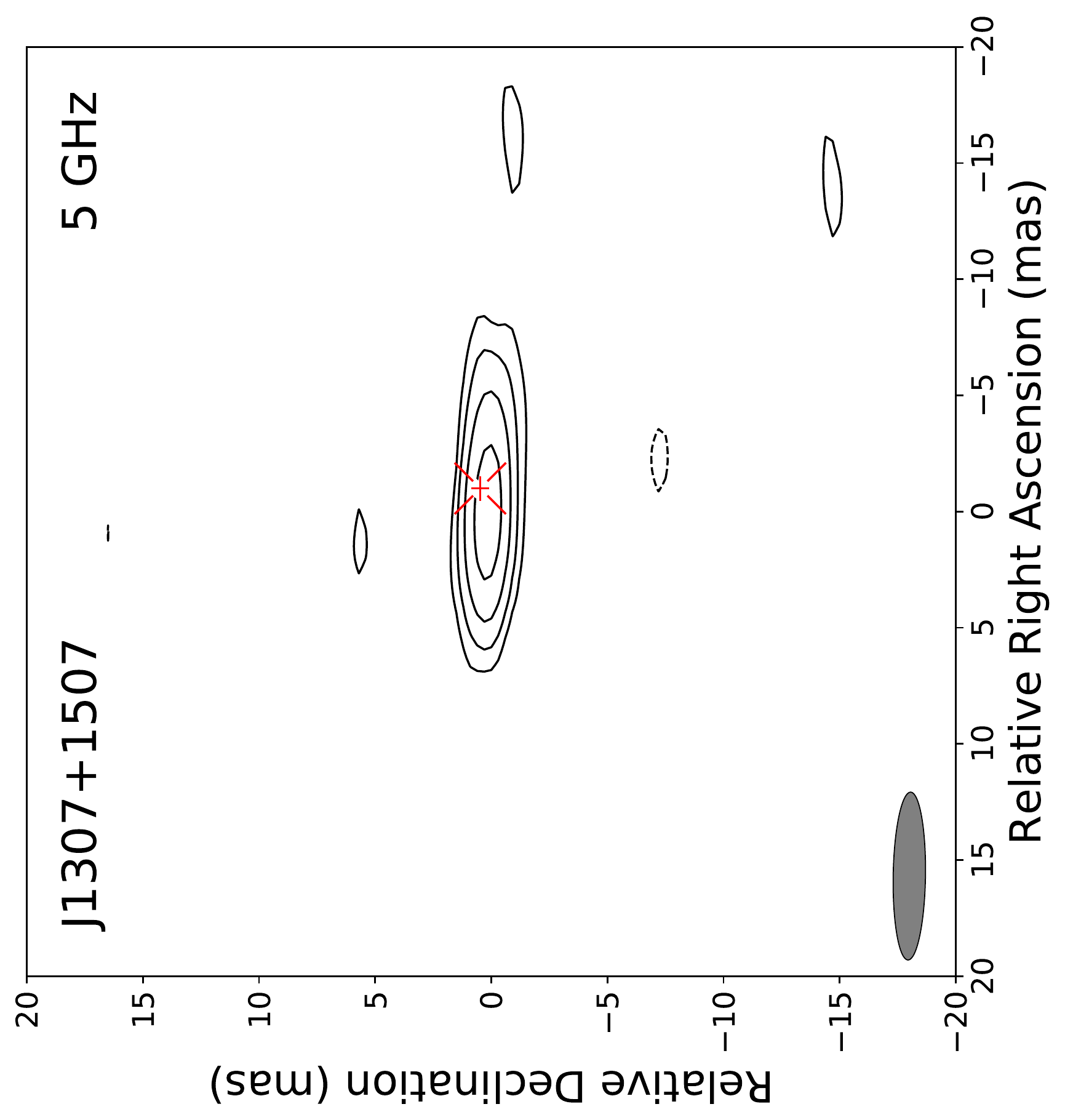}
            }
    \caption{$Continued$}
\end{figure*}
\begin{figure*}
    \setcounter{figure}{0}
\gridline{  \includegraphics[width=0.37\textwidth,angle=-90]{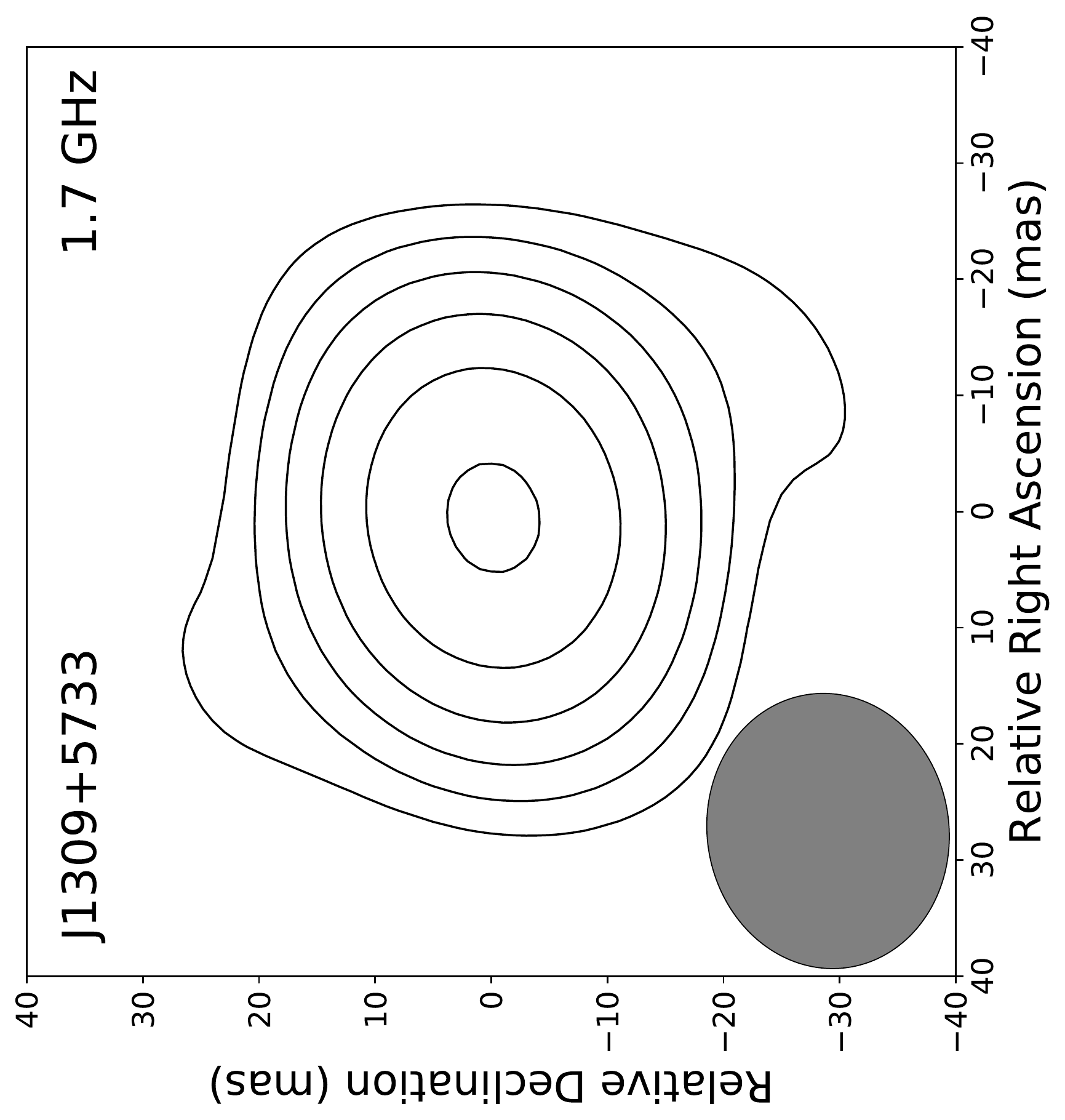}
            \includegraphics[width=0.37\textwidth,angle=-90]{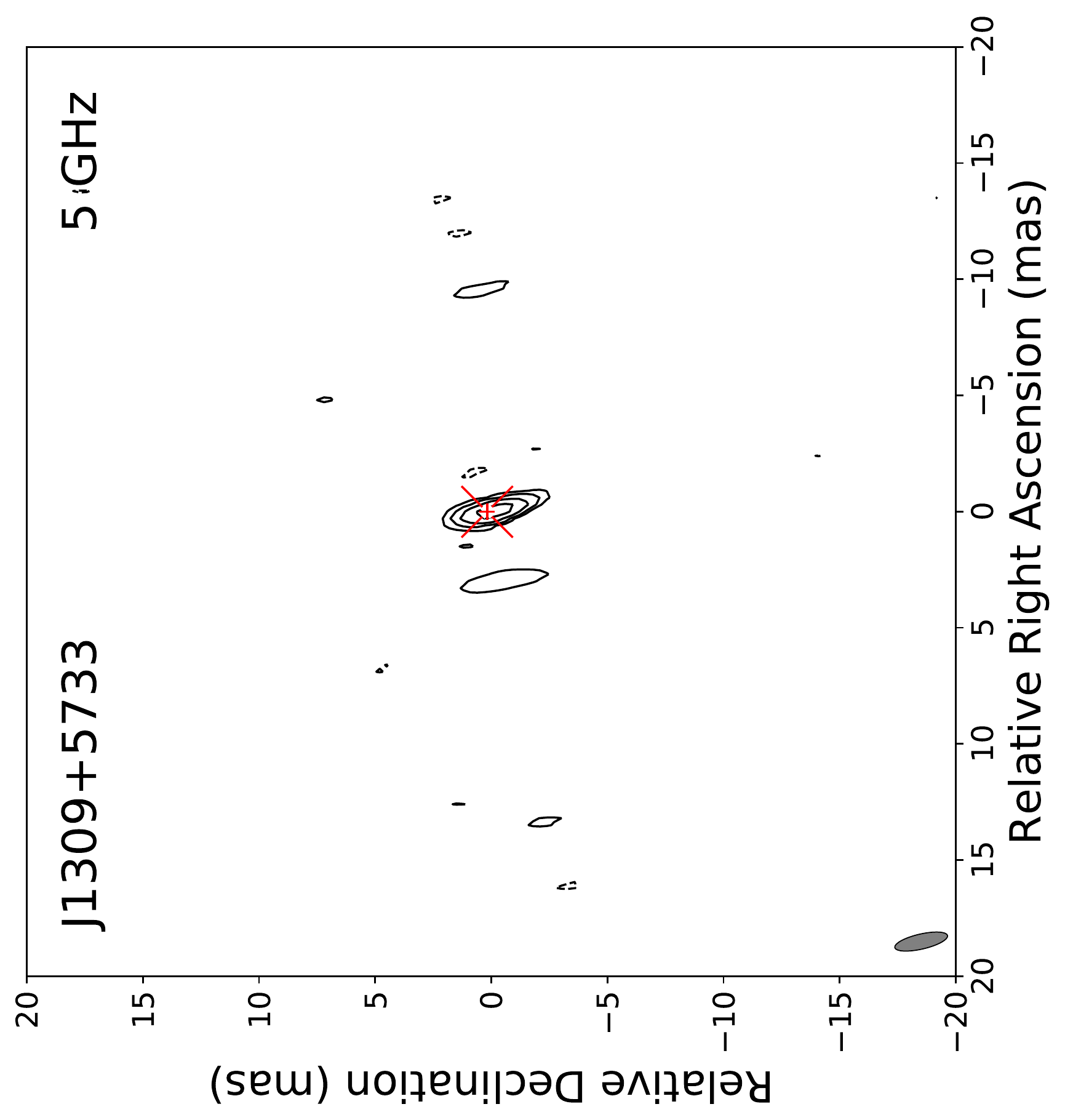}
            }
\gridline{  \includegraphics[width=0.37\textwidth,angle=-90]{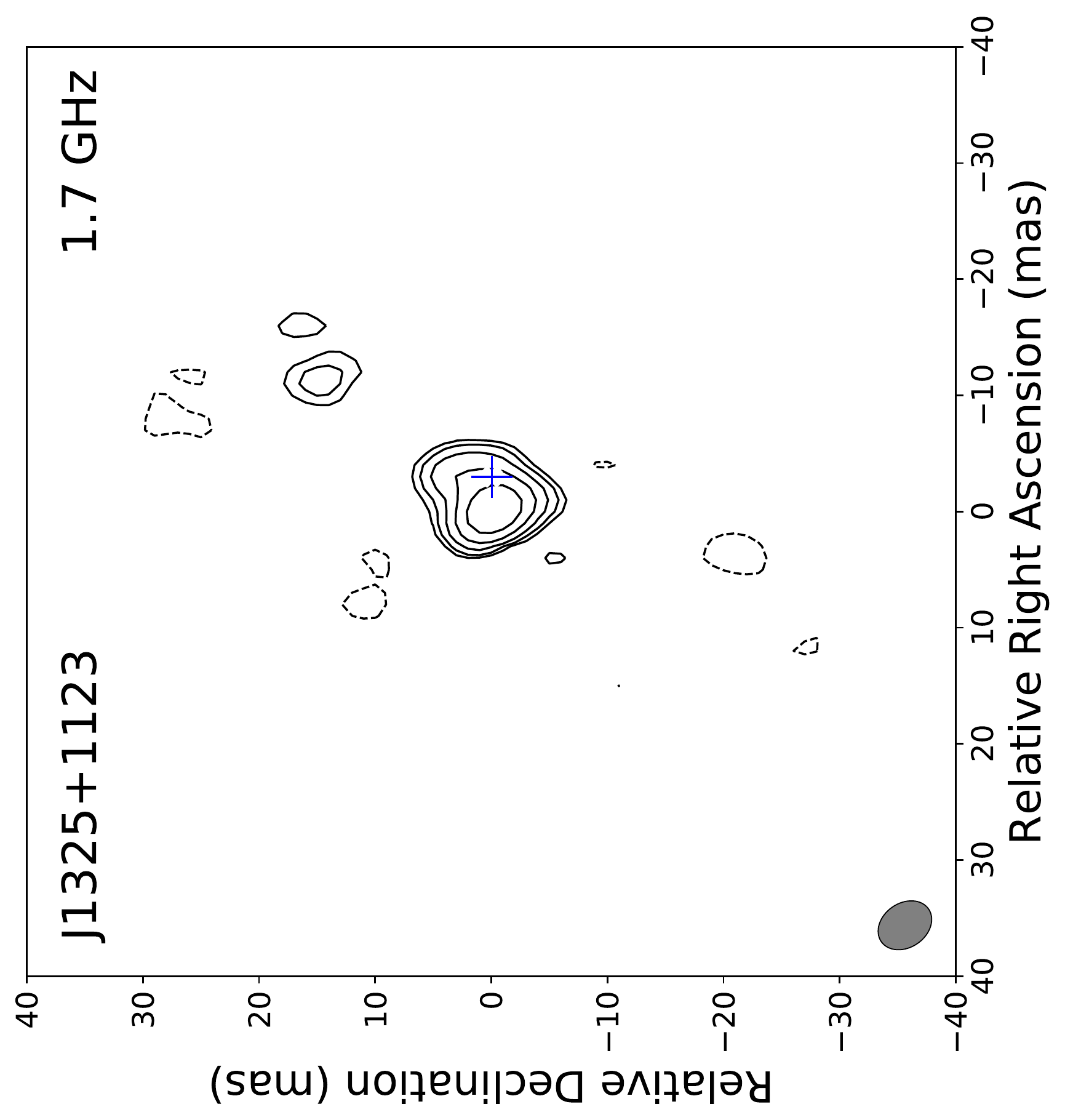}
            \includegraphics[width=0.37\textwidth,angle=-90]{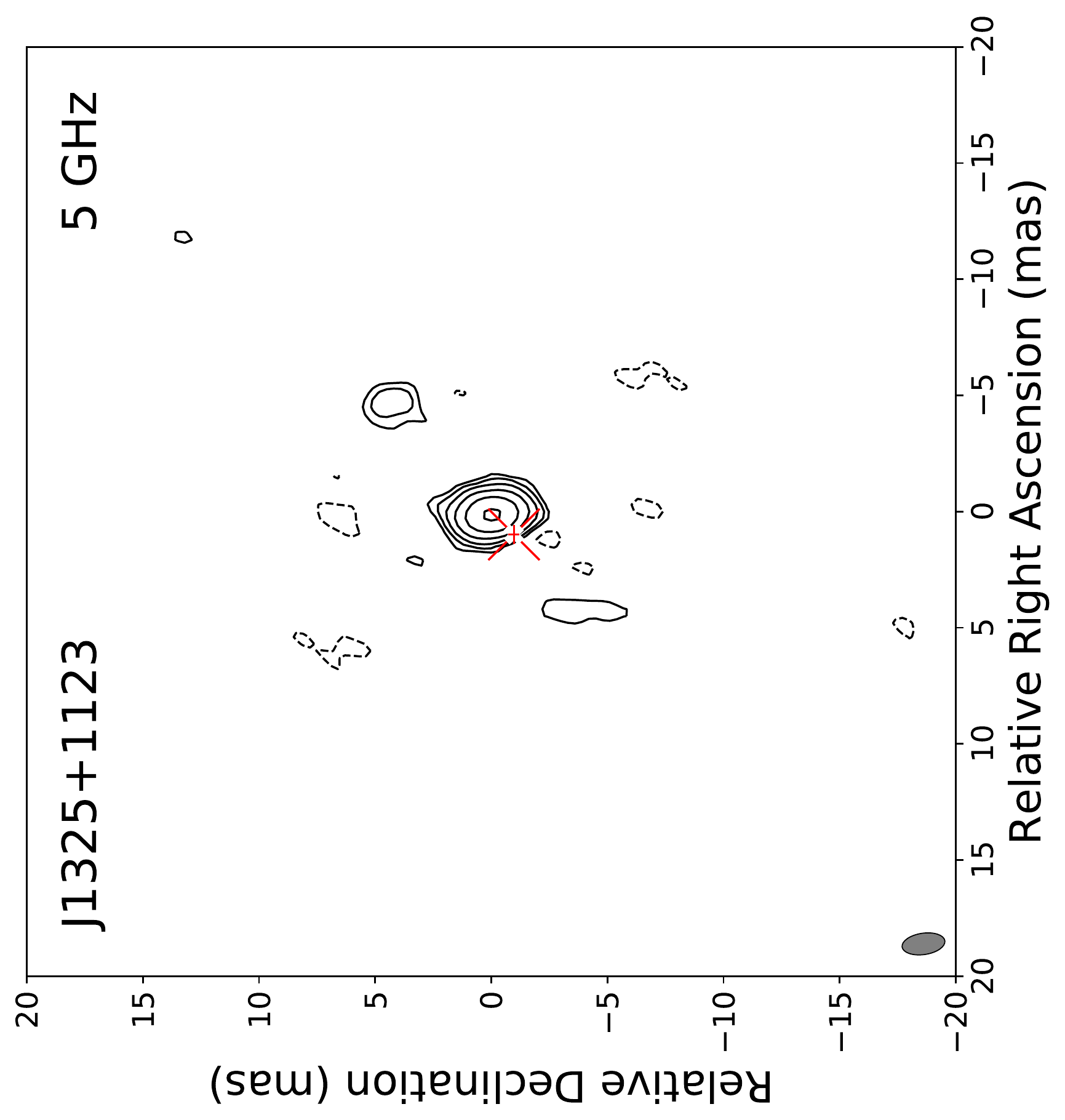}
            }
\gridline{  \includegraphics[width=0.37\textwidth,angle=-90]{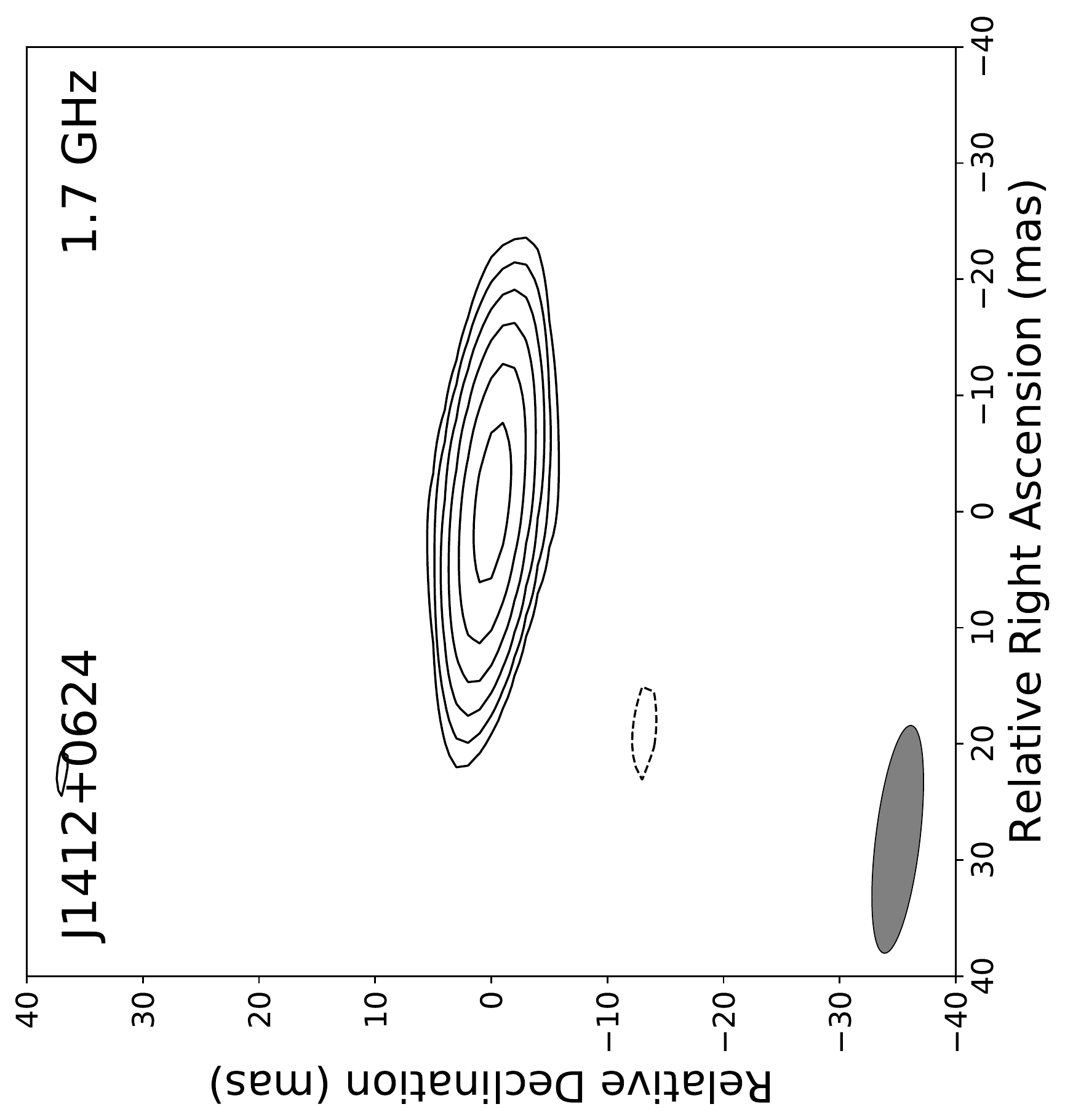}
            \includegraphics[width=0.37\textwidth,angle=-90]{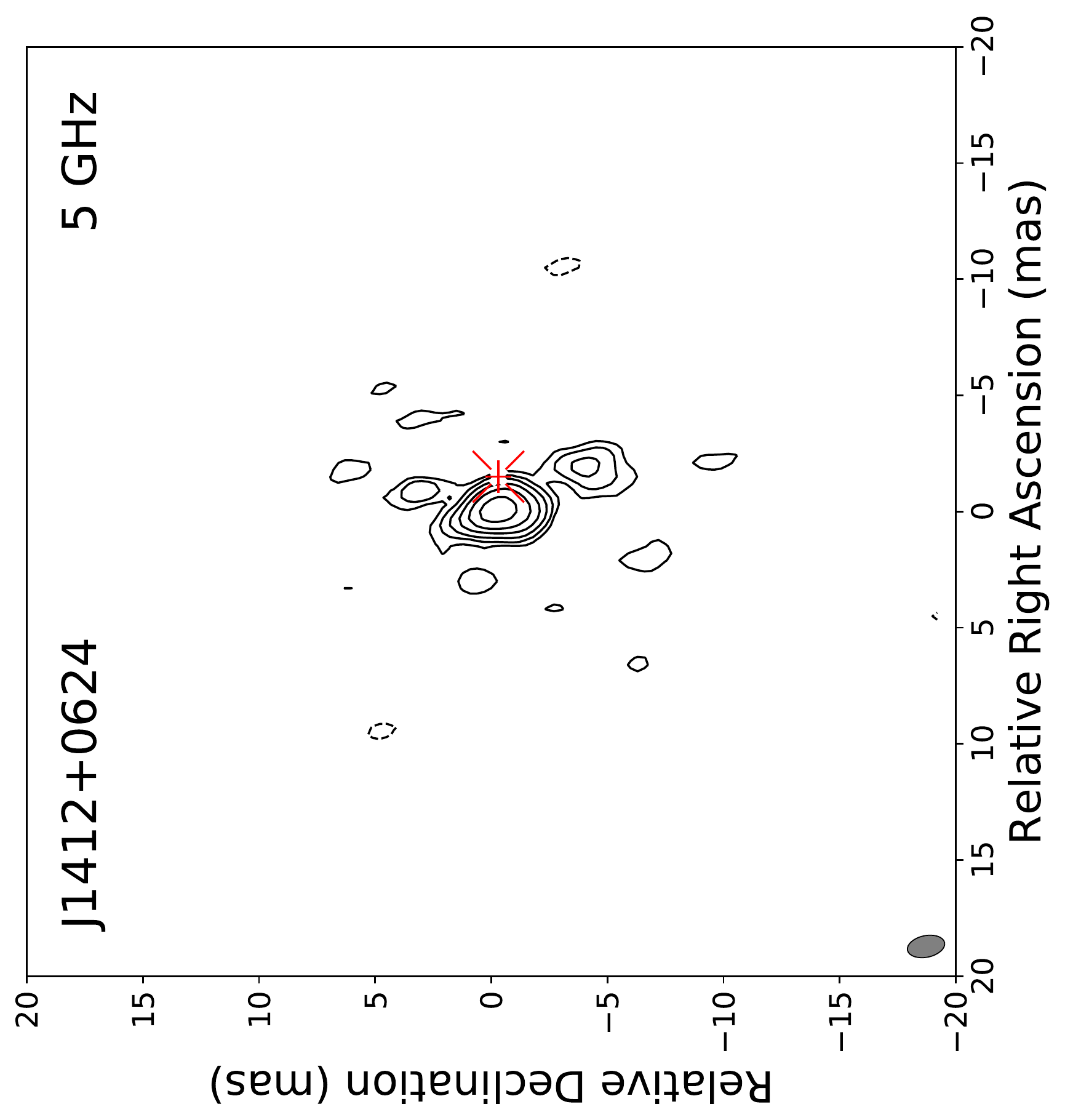}
            }
    \caption{$Continued$}
\end{figure*}
\begin{figure*}
    \setcounter{figure}{0}
\gridline{  \includegraphics[width=0.37\textwidth,angle=-90]{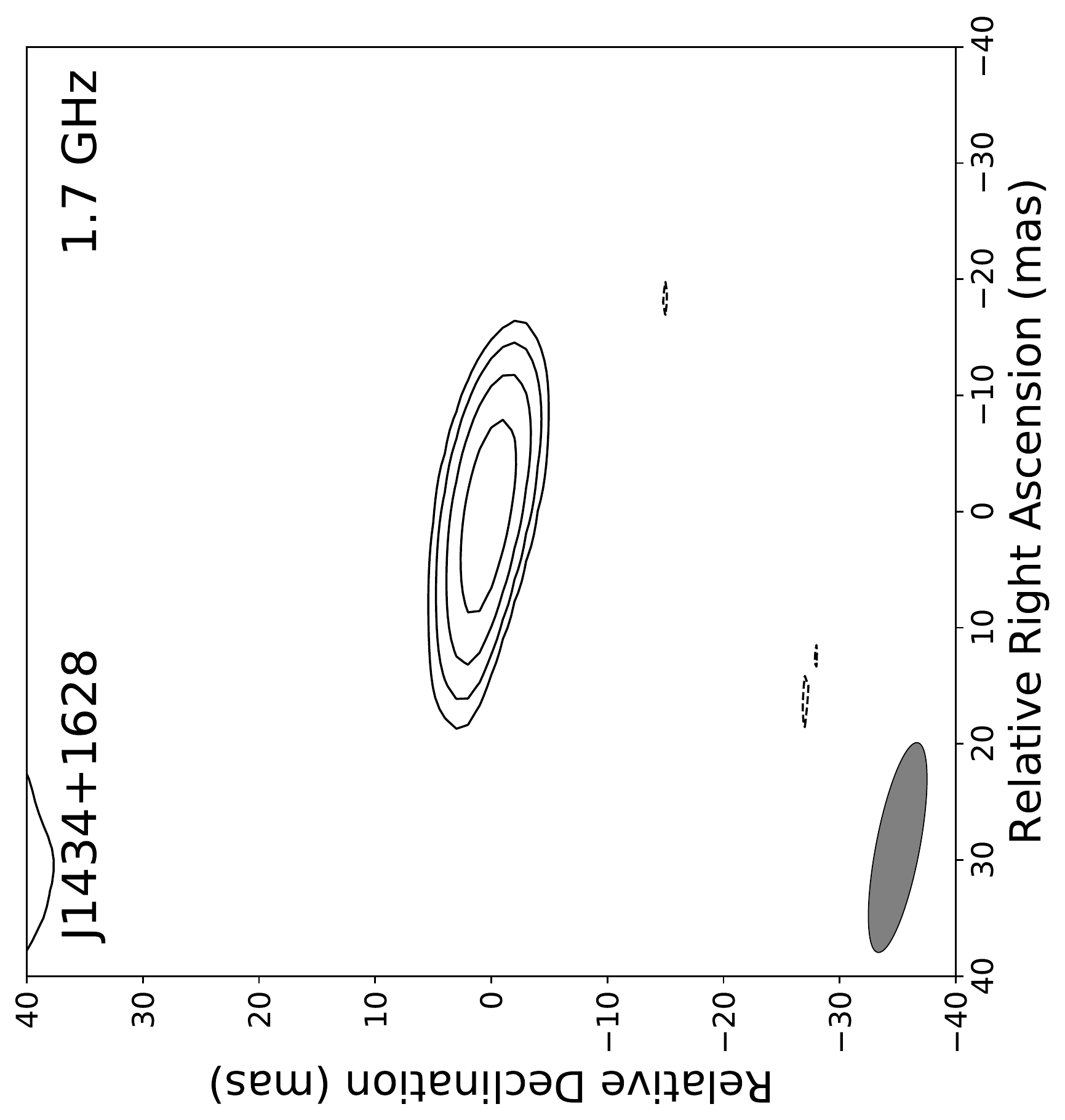}
            \includegraphics[width=0.37\textwidth,angle=-90]{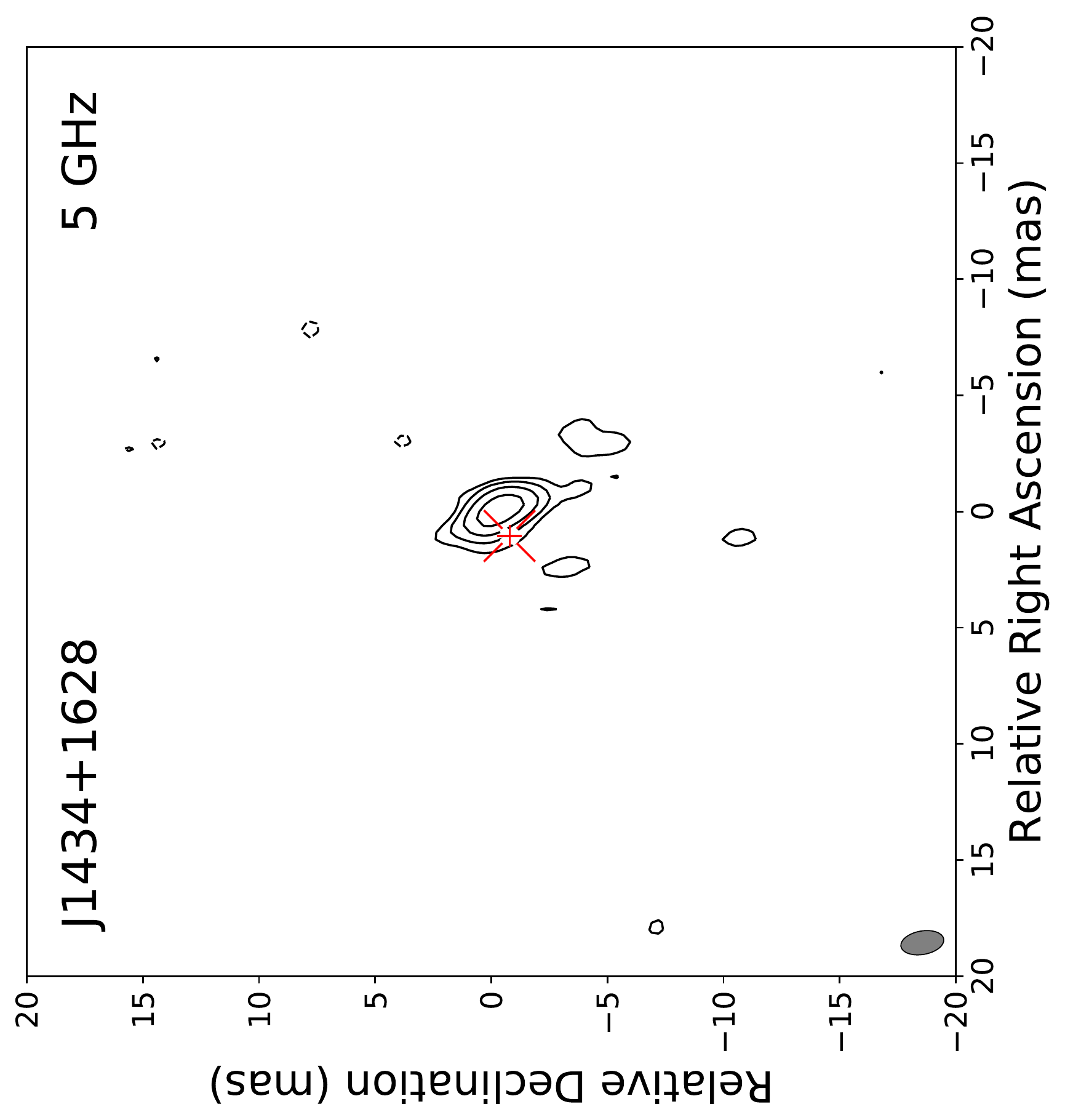}
            }
\gridline{  \includegraphics[width=0.37\textwidth,angle=-90]{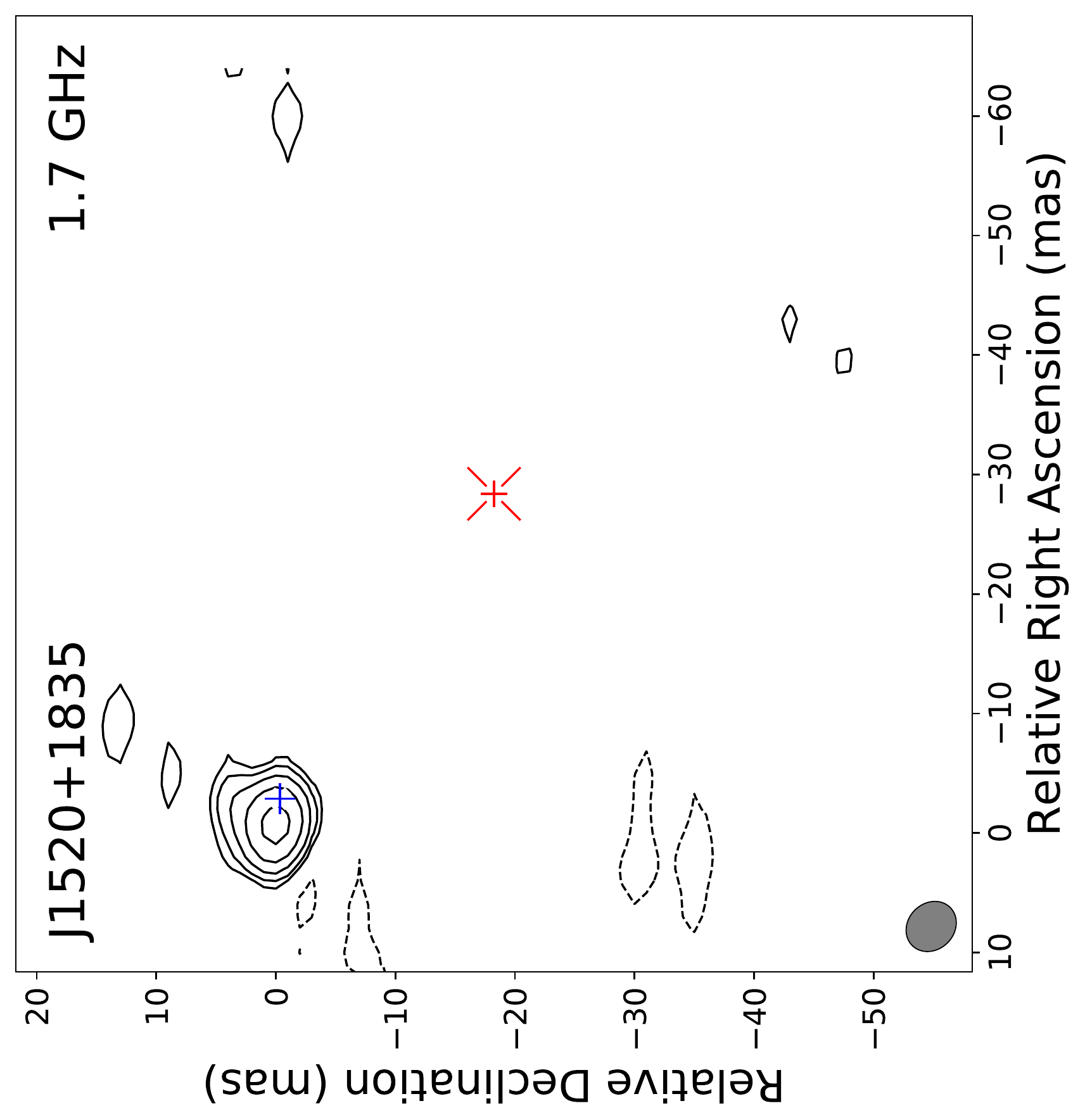}
            \includegraphics[width=0.37\textwidth,angle=-90]{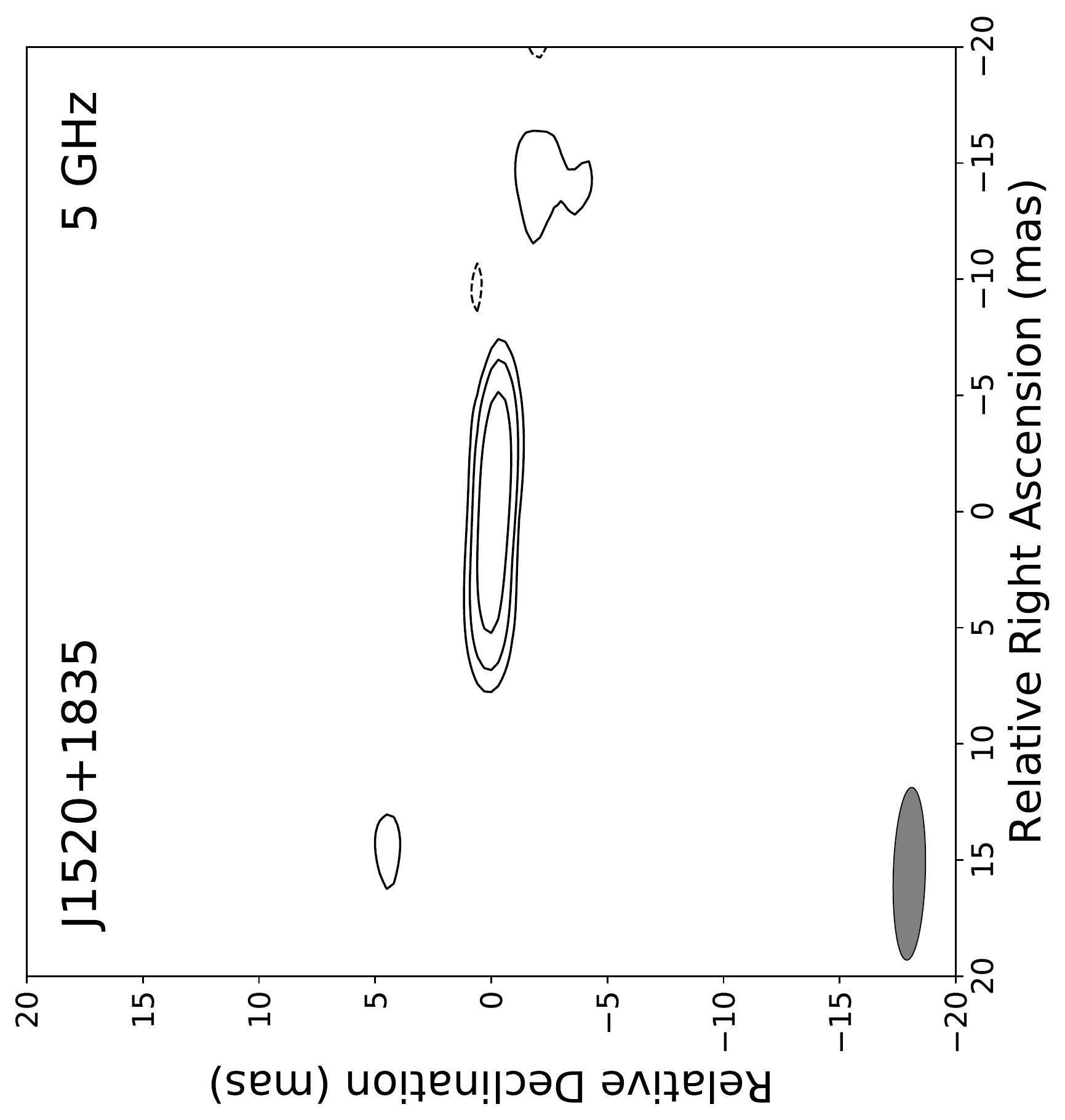}
            }
\gridline{  \includegraphics[width=0.37\textwidth,angle=-90]{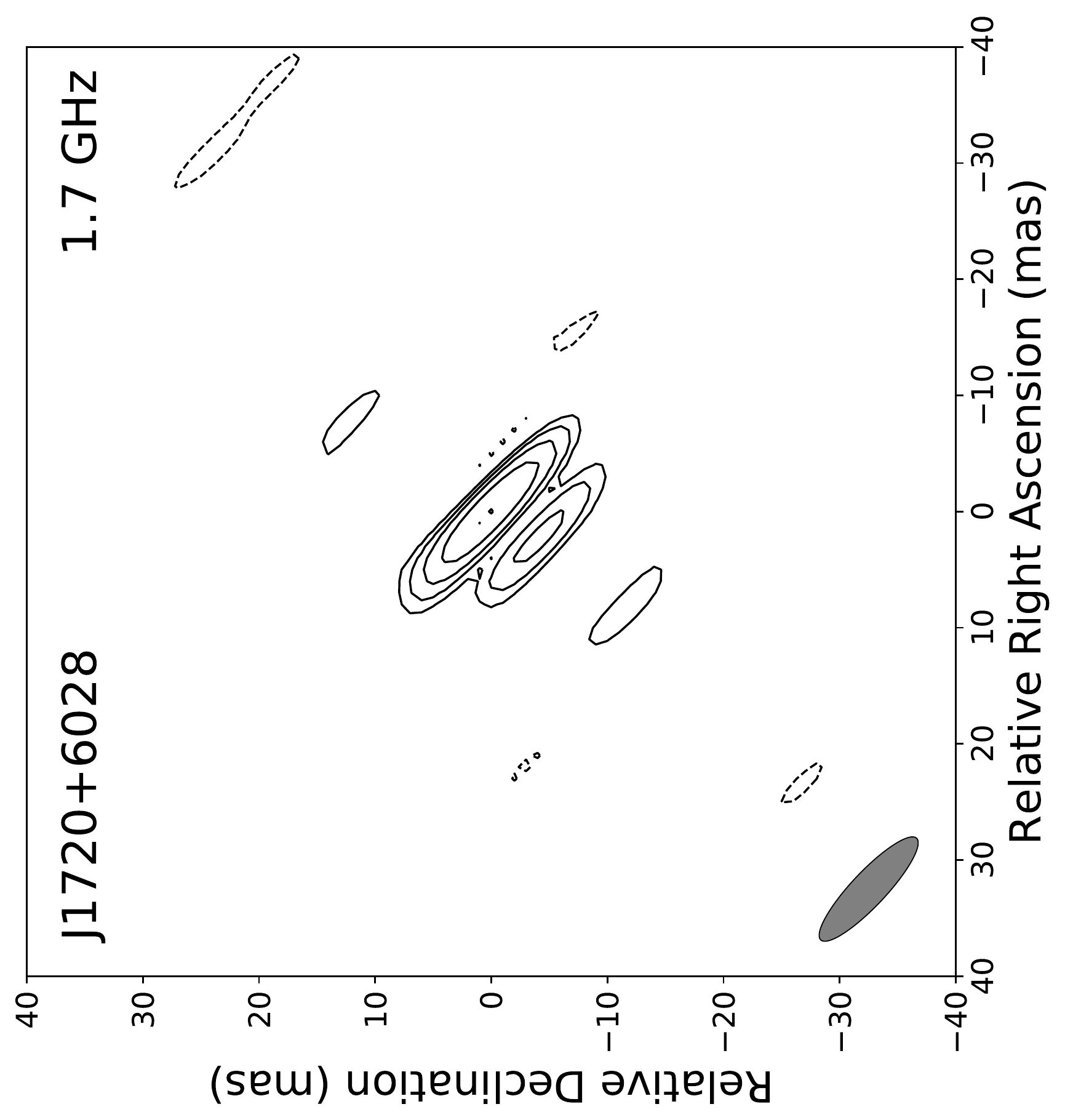}
            \includegraphics[width=0.37\textwidth,angle=-90]{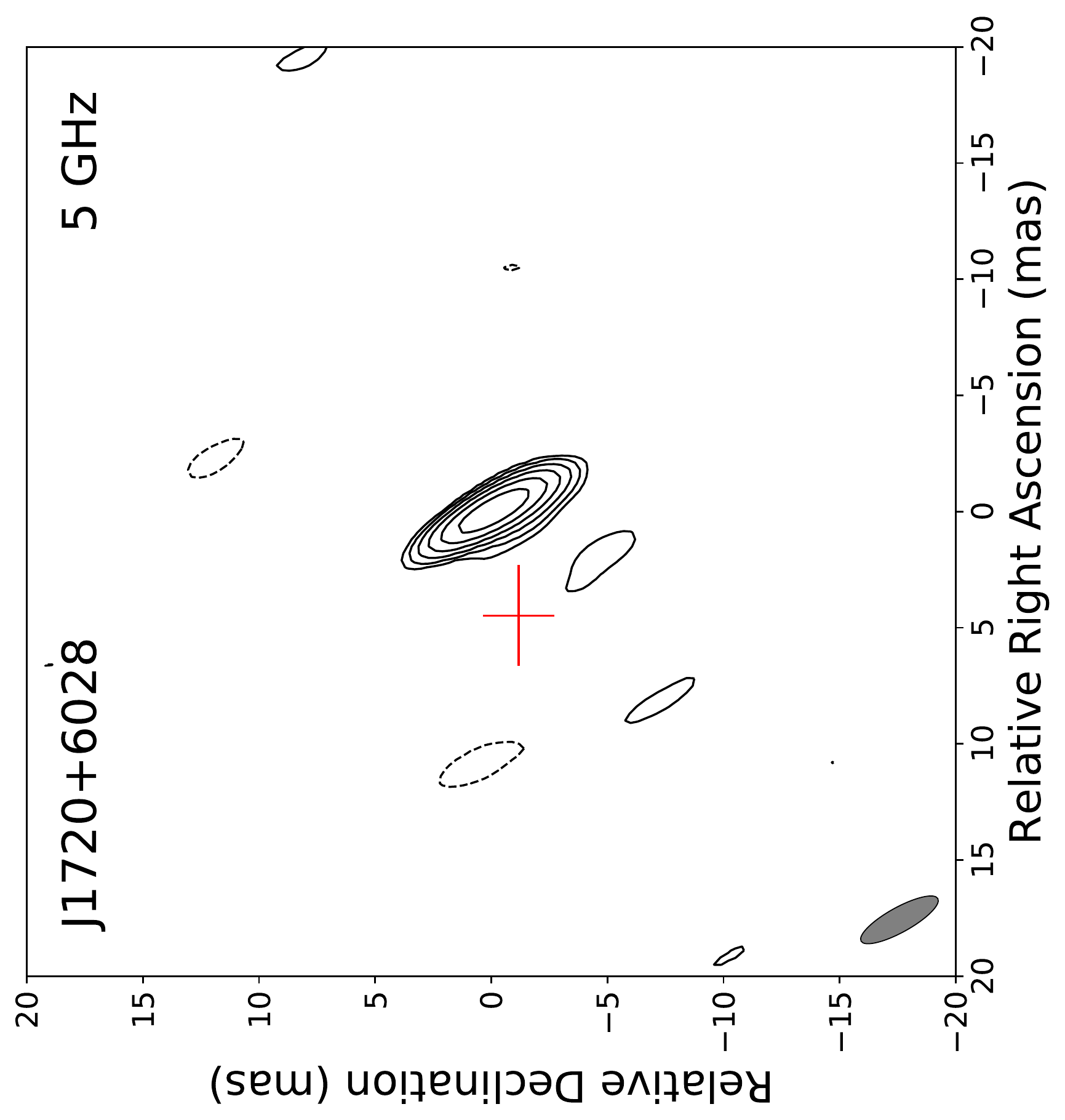}
            }
    \caption{$Continued$}
\label{evnimages}
\end{figure*}

\section{Results} 
\label{sec:results}

We present the naturally weighted \textsc{clean} images for each source in Figure~\ref{evnimages}, listing the image parameters in Table~\ref{imgparams}. The parameters of the fitted Gaussian model components are given in Table~\ref{phyparams}, with uncertainties calculated following \citet{1999ASPC..180..301F}. An additional $5\%$ error was added in quadrature to the flux densities, to account for the VLBI absolute amplitude calibration uncertainty \citep[e.g.][]{2012ApJS..198....5A,2015MNRAS.446.2921F}. Except for the two sources (J0304$+$0046 and J0918$+$0636) missing
reliable 5-GHz data, model parameters are given at both frequencies.  

The power-law spectral index $\alpha$ is defined as $S \propto \nu^{\alpha}$, where $S$ is the modeled flux density and $\nu$ the frequency. The two-point spectral indices were determined from flux density measurements at 1.7 and 5~GHz. VLBI spectral indices and their uncertainties are given in Col.~9 of Table~\ref{phyparams}, except for the two quasars without 5-GHz flux densities. We note that the measurements of the same target source at the two frequencies were conducted with a typical time difference of $\sim 1$~month, however, reaching up to $17$~months in some cases (Table~\ref{observations}). Hence, the calculated spectral indices should be treated with caution as they are based on non-simultaneous measurements, and some sources might be variable.

To check whether the sources are resolved by the interferometer, we calculated the minimum resolvable size for each project segment following equation (2) in \citet{2005AJ....130.2473K}. We found that out of the 13 targets, four sources (J0851$+$1423, J1037$+$1823, J1231$+$3816, and J1325$+$1123) are resolved and three (J0918$+$0636, J1309$+$5733, and J1720$+$6028) are unresolved at both frequency bands. The other six sources are unresolved by either of the 1.7- or 5-GHz EVN observations. Based on the derived model parameters and spectral indices, we calculated the redshift-corrected brightness temperatures \citep{1982ApJ...252..102C}
\begin{equation} \label{eq:tb}
    T_{\rm{b}} = 1.22 \times 10^{12} \, (1 + z) \frac{S_\nu}{\theta^2 \nu^2} \,\, [\mathrm{K}]
\end{equation}
and the monochromatic power \citep{2002astro.ph.10394H} 
\begin{equation} \label{eq:luminosity}
    P_{\rm \nu} = 1.20 \times 10^{20} \, D_{\rm{L}}^2 S_\nu (1+z)^{-1-\alpha} \,\, [\mathrm{W}\,\mathrm{Hz}^{-1}] 
\end{equation}
at both frequencies. Here $z$ is the redshift, $S_\nu$ the integrated flux density of the core in Jy, $\theta$ the fitted circular Gaussian diameter (full width at half-maximum, FWHM) in mas, $\nu$ the observing frequency in GHz, $\alpha=\alpha_{1.7}^5$ the  spectral index between the two measured frequencies, 1.7 and 5~GHz, and $D_{\text{L}}$ the luminosity distance in Mpc.  For the sources with unresolved core components, we can only give a lower limit to the brightness temperature, substituting the minimum resolvable angular size in Equation~(\ref{eq:tb}).

The phase-referenced VLBI positions of the sources were obtained from the 5-GHz images using the \textsc{aips} task \textsc{maxfit}. The right ascension (RA) and declination (Dec) values are presented in Cols.~2--3 of Table~\ref{imgparams}. The estimated astrometric errors at 5~GHz are comparable to those of the Gaia optical coordinates.

\begin{deluxetable*}{lll|cccc}
\tablenum{4}
\tablecaption{VLBI image parameters and source positions.}
\tablewidth{0pt}
\tablehead{
\colhead{ID} & \colhead{RA$_{\rm Gaia}$ ($^\mathrm{h~min~s}$)} & \colhead{Dec$_{\rm Gaia}$ ($^\circ$~$^\prime$~$^{\prime\prime}$)} & \colhead{$\nu$} & \multicolumn2c{Restoring beam} & \colhead{$1\sigma$ nosie} \\
\nocolhead{} & \colhead{RA$_{\rm VLBI,5GHz}$ ($^\mathrm{h~min~s}$)} &  \colhead{Dec$_{\rm VLBI,5GHz}$ ($^\circ$~$^\prime$~$^{\prime\prime}$)} & \colhead{(GHz)} & \colhead{(mas$\times$mas)} & \colhead{($^\circ$)} & \colhead{(mJy\,beam$^{-1}$)}  
}
\decimalcolnumbers
\startdata
J0304$+$0046 & $-$ &  $-$                                & 5 & 2.5 $\times$ 1.4 & 86.2 & $-$ \\
& 03 04 37.21595 (0.3) & 00 46 53.6159 (0.3)                      & 1.7 & 22.3  $\times$ 3.4 & 84.1 & 0.31 \\
J0851$+$1423 & 08 51 11.59953 (0.9) & 14 23 37.7017 (0.4)   & 5 & 1.8 $\times$ 0.9 & 7.6  & 0.06  \\
& 08 51 11.59954 (0.6) & 14 23 37.7013 (0.6)                            & 1.7 & 4.3 $\times$ 2.9 & 12.5 & 0.07 \\
J0918$+$0636 & 09 18 24.37998 (0.4) & 06 36 53.4111 (0.3)   & 5 & 1.6 $\times$ 0.9 & 9.3 & $-$ \\
& 09 18 24.37998 (0.4) & 06 36 53.4162 (0.6)                            & 1.7 & 18.1 $\times$ 2.9 & 89.5 & 0.39 \\
J1006$+$4627 & 10 06 45.59650 (0.7) & 46 27 17.2828 (0.8)   & 5 & 4.3 $\times$ 0.93 &  176.5  & 0.17 \\
& 10 06 45.59633 (0.5) & 46 27 17.2833 (0.5)                            & 1.7 & 14.1$\times$ 3.6 & 0.8 & 0.10 \\
J1037$+$1823 & 10 37 17.73037 (1.6) & 18 23 03.0931 (0.8)   & 5 & 5.2 $\times$ 0.8 & 4.0 & 0.10 \\
& 10 37 17.73040 (0.6) & 18 23 03.0918 (0.6)                            & 1.7 & 4.4 $\times$ 3.6 & 52.6 & 0.09 \\
J1231$+$3816 & 12 31 42.17213 (0.5) & 38 16 59.0408 (0.4)   & 5 & 5.2 $\times$ 0.9 & 0.2 & 0.08 \\
& 12 31 42.17206 (0.4) & 38 16 59.0429 (0.4)                            & 1.7 & 19.9 $\times$ 3.9 & 84.6 & 0.08 \\
J1307$+$1507 & 13 07 38.83415 (0.5) & 15 07 52.1226 (0.3)   & 5 & 7.2 $\times$ 1.4 & 89.1  & 0.03 \\
& 13 07 38.83423 (0.6) & 15 07 52.1223 (0.6)                           & 1.7 & 18.8 $\times$ 4.1 & 90 & 0.04 \\
J1309$+$5733 & 13 09 40.68855 (0.3) & 57 33 09.9371 (0.3)   & 5 & 2.4 $\times$ 0.6 & 12.9 & 0.21 \\
& 13 09 40.68854 (0.3) & 57 33 09.9368 (0.3)                            & 1.7 & 23.8  $\times$ 20.8  & 98.9 & 0.13 \\
J1325$+$1123 & 13 25 12.49327 (0.4) & 11 23 29.8380 (0.2)   & 5 & 1.9 $\times$ 0.9 & 8.3 & 0.20  \\
& 13 25 12.49322 (0.4) & 11 23 29.8389 (0.4)                            & 1.7 & 4.9 $\times$ 3.8 & 35.3 & 0.52 \\
J1412$+$0624 & 14 12 09.96955 (0.7) & 06 24 06.8665 (0.5)   & 5 & 1.6 $\times$ 0.9 & 11.7 & 0.06  \\
& 14 12 09.96965 (0.4) & 06 24 06.8665 (0.6)                            & 1.7 & 19.8 $\times$ 3.8 & 83.2 & 0.13 \\
J1434$+$1628 & 14 34 13.05469 (0.4) & 16 28 52.7346 (0.5)   & 5 & 1.9 $\times$ 1.0 & 10.0 & 0.02  \\
& 14 34 13.05461 (0.5) & 16 28 52.7350 (0.5)                            & 1.7 & 18.4 $\times$ 3.8 & 79.2 & 0.03 \\
J1520$+$1835 & 15 20 28.14216 (0.3) & 18 35 56.1587 (0.3)   & 5 & 7.4 $\times$ 1.4 & 88.3 & 0.03  \\
& 15 20 28.14386 (0.4) & 18 35 56.1767 (0.3)                            & 1.7 & 4.5 $\times$ 3.9 & 45.0 & 0.06 \\
J1720$+$6028 & 17 20 07.18399 (2.1) & 60 28 24.0118 (1.5)   & 5 & 3.9 $\times$ 1.1 & 29.0 & 0.02 \\
 & 17 20 07.18368 (0.3) & 60 28 24.0128 (0.2)                           & 1.7 & 12.0 $\times$ 2.9 & 46.7 & 0.08 \\
\enddata
\tablecomments{Col.~1 $-$ source designation; Col. 2--3 $-$ right ascension and declination coordinates of the sources, the first line per source lists the optical coordinates obtained by Gaia (where available), the second line per source lists the VLBI 5-GHz position; Col.~4 $-$ observed frequency; Col.~5 $-$ elliptical Gaussian restoring beam half-power width; Col.~6 $-$ major axis position angle of the restoring beam, measured from north through east; Col. 7 $-$ 1$\sigma$ image noise.}
\label{imgparams}
\end{deluxetable*}

\begin{longrotatetable}
\begin{deluxetable*}{lrrrrrrrrr}
\tablenum{5}
\tablecaption{Basic parameters and the derived physical properties of the sample of $z>4$ radio sources in the EG102 experiment. }
\tablewidth{0pt}
\tablehead{
\colhead{ID} &  \colhead{$S_\mathrm{VLBI,1.7~GHz}$} & \colhead{$S_\mathrm{VLBI,5~GHz}$}  & \colhead{$\theta_\mathrm{1.7~GHz}$} & \colhead{$\theta_\mathrm{5~GHz}$} & \colhead{$\alpha_{1.7}^5$} & \colhead{$T_\mathrm{b,1.7~GHz}$} & \colhead{$T_\mathrm{b,5~GHz}$} & \colhead{$P_\mathrm{1.7~GHz}$} & \colhead{$P_\mathrm{5~GHz}$}        \\
 & \colhead{(mJy)} & \colhead{(mJy)} & \colhead{(mas)} & \colhead{(mas)} & \nocolhead{}
 & \colhead{($10^9$ K)} & \colhead{($10^9$ K)} & \colhead{($10^{26}$ W~Hz$^{-1}$)} & \colhead{($10^{26}$ W~Hz$^{-1}$)}
}
\decimalcolnumbers
\startdata
J0304$+$0046 & 15.01* (2.32) & 	  $-$         & 	$<$ 4.78 & 	    $-$       &          $-$     & $>$ 1.47	&	$-$		    &	5.20 (0.98)$^{a}$	&	$-$	
  \\
J0851$+$1423 & 4.94* (0.82) & 	4.60* (0.67) & 	4.24 (0.64) & 	1.12 (0.15) &   $-0.12$ (0.19) &  0.62 (0.21)	&	0.94 (0.28)	&	2.10 (0.42)	&	1.95 (0.35)
  \\
J0918$+$0636 & 38.53 (3.86) & 	35.50$^{\dagger}$ (2.18) & $<$ 2.41 & $<$ 0.71$^{\dagger}$ & $-0.08$ (0.11)  & $>$ 14.52 & $>$ 17.85 &	14.62 (2.14) & 13.47 (1.66)
  \\
J1006$+$4627 & 8.78* (0.73) & 	6.69* (0.53) &  $<$ 1.75 & 	1.41 (0.08) &   $-0.25$ (0.11) & $>$ 6.58	&	0.89 (0.13)	&	4.90 (0.68)	&	3.73 (0.51)
  \\
J1037$+$1823 & 7.38* (0.77) & 	7.15* (1.05) & 	2.44 (0.20) & 	2.88 (0.38) &   $-0.05$ (0.29) & 2.64 (0.51)	&	0.21 (0.07)	&	2.54 (0.38)	&	2.46 (0.71)
  \\
J1231$+$3816 & 8.91* (0.82) & 	6.79* (0.72) & 	4.91 (0.34) & 	1.12 (0.11) &   $-0.25$ (0.13) & 0.80 (0.13)	&	1.36 (0.27)	&	4.41 (0.62)	&	3.36 (0.52)
  \\
J1307$+$1507 & 2.69* (0.19) & 	1.20* (0.14) & 	9.95 (0.15) & 	$<$ 1.14 &   $-0.76$ (0.14) & 0.06 (0.01)	&	$>$ 0.23	&	2.98 (0.38)	&	1.32 (0.21)
  \\
J1309$+$5733 & 22.30 (1.32) & 	12.78* (1.52) &  $<$ 1.90 &  $<$ 0.40 &   $-0.52$ (0.15) & $>$ 13.73 &	$>$ 21.03	&	18.10 (2.25)	&	10.40 (1.71)
  \\
J1325$+$1123 & 62.70 (3.91) & 	31.16 (1.86) & 	0.76 (0.02) & 	1.12 (0.03) &   $-0.61$ (0.09) & 248.00 (20.98)	& 6.56 (0.55)	&	63.40 (7.39)	&	31.50 (3.91)
  \\
J1412$+$0624 & 18.80 (1.89) & 	15.38 (0.85) &  $<$	2.66 & 	0.77 (0.02) & 	$-0.18$ (0.10) & $>$ 6.11	&	6.92 (0.47)	&	9.44 (1.37)	&	7.73 (0.95)
  \\
J1434$+$1628 & 1.99* (0.22) & 	1.23* (0.12) & 	$<$ 2.88 & 	0.79 (0.06) &   $-0.44$ (0.15) & $>$ 0.53 	&	0.49 (0.09)	&	1.38 (0.21)	&	0.85 (0.12)
  \\
J1520$+$1835 & 4.80* (0.61) & 	1.05* (0.17) & 	2.95 (0.31) & 	$<$ 1.44 &   $-1.42$ (0.21) & 1.19 (0.30)	&	$>$ 0.13 &	15.68 (2.64) &	3.43 (0.67)
  \\
J1720$+$6028 & 7.25* (0.50) &    4.20* (0.23) &  $<$ 1.27 & 	$<$ 0.29 &   $-0.51$ (0.08) & $>$ 10.32	&	$>$ 13.68	&	6.15 (0.79)	&	3.56 (0.44)
 \\
\enddata
\tablecomments{$^{\dagger}$ based on archival data from experiment ES034A. $^{a}$ assuming 0 spectral index. Col.~1 $-$ Source designation; Cols.~2-3 $-$ fitted 1.7- and 5-GHz EVN flux densities and uncertainties; flux density values marked with * are corrected for the $25\%$ coherence loss for deriving the physical parameters (see Section~\ref{datared}); Cols.~4-5 $-$ fitted circular Gaussian component diameter (FWHM) at 1.7 and 5~GHz, or upper limit equals to the minimum resolvable angular size \citep{2005AJ....130.2473K} for unresolved components; Col.~6 $-$ spectral index between 1.7 and 5~GHz; Cols.~7-8 $-$ brightness temperature and its uncertainty at 1.7 and 5~GHz, or lower limit for unresolved components; Cols.~9-10 $-$ monochromatic radio power and its uncertainty at 1.7 and 5~GHz.}
\label{phyparams}
\end{deluxetable*}
\end{longrotatetable}

\section{Discussion} 
\label{sec:discussion}

\subsection{The Origin of the Radio Emission and Doppler-boosting}
\label{disc1}

The brightness temperatures are $T_\mathrm{b} > 10^6$\,K for all sources, indicating non-thermal radio emission related to AGN activity \citep{2000ApJ...530..704K,2011A&A...526A..74M}. As non-thermal emission could also originate from supernova remnants or remnant complexes, sources have to meet a monochromatic power limit for the radio emission to be considered as driven by AGN activity \citep{2012MNRAS.423.1325A}. As \citet{2014MNRAS.442..682M} showed, radio emission of sources at $z > 2$ is powered by AGN if the 1.4-GHz monochromatic powers exceed $4 \times 10^{24}$ W~Hz$^{-1}$, while star formation is the dominant process for objects with lower powers. The rest-frame 1.4-GHz powers can be calculated according to Equation~(\ref{eq:luminosity}), where the $S_\mathrm{1.4\,GHz}$ flux density values of the brightest components are extrapolated from the 1.7- and 5-GHz values measured with VLBI, assuming power-law spectra with the determined $\alpha_{1.7}^5$ spectral index (Table~\ref{phyparams}). All 13 sources exceed this limit. For the one with missing 5-GHz flux density measurement (J0304+0046), we considered a broad range of possible spectral indices, $-1 \leq \alpha_{1.7}^5 \leq 0$.

The brightness temperature of the sources (Table~\ref{phyparams}) can be used to determine whether the emission is relativistically enhanced (Doppler-boosted). \citet{1994ApJ...426...51R} estimated the theoretical value of the intrinsic brightness temperature in case of the energy densities of the emitting plasma and the magnetic field are in equipartition, and found $T_\mathrm{b,eq} \approx 5 \times 10^{10}$\,K. By assuming equipartition condition in the sources, the intrinsic brightness temperature is $T_\mathrm{b,eq}$, and the Doppler factor can be calculated as
\begin{equation}
 \delta = \frac{T_\mathrm{b}}{T_\mathrm{b,eq}}.
\end{equation}
where $T_\mathrm{b}$ is the redshift-corrected (or source rest-frame) brightness temperature derived from VLBI measurements at a given frequency. If $\delta > 1$, the radio emission is considered Doppler-boosted. Among the sources examined, J1325$+$1123 is the only one with certainly Doppler-boosted radio emission at 1.7~GHz. On the other hand, there are three sources (J0851$+$1423, J1037$+$1823, and J1231$+$3816) where no indication of Doppler-boosting is seen. For the remaining, not clearly resolved objects whose brightness temperatures are lower limits at either of the frequency bands, the data are insufficient to decide whether the radio emission is Doppler-enhanced or not. 

Doppler-boosting in AGN jets can be observed under very special conditions only. Indeed, when there is no sign of Doppler-boosting (i.e. $\delta \leq 1$), it is either because of the jet inclination angle is large and thus relativistic beaming is not present, or the flux density is measured far away from the peak frequency of the spectrum. In the latter case, it is possible that the source would appear Doppler-boosted at other frequencies, closer to the radio spectral peak \citep[e.g.][]{1994ApJ...426...51R,2020ApJS..247...57C}.

\subsection{Continuum Radio Spectra}
\label{disc2}

\begin{figure*}
\gridline{  \includegraphics[width=0.328\textwidth]{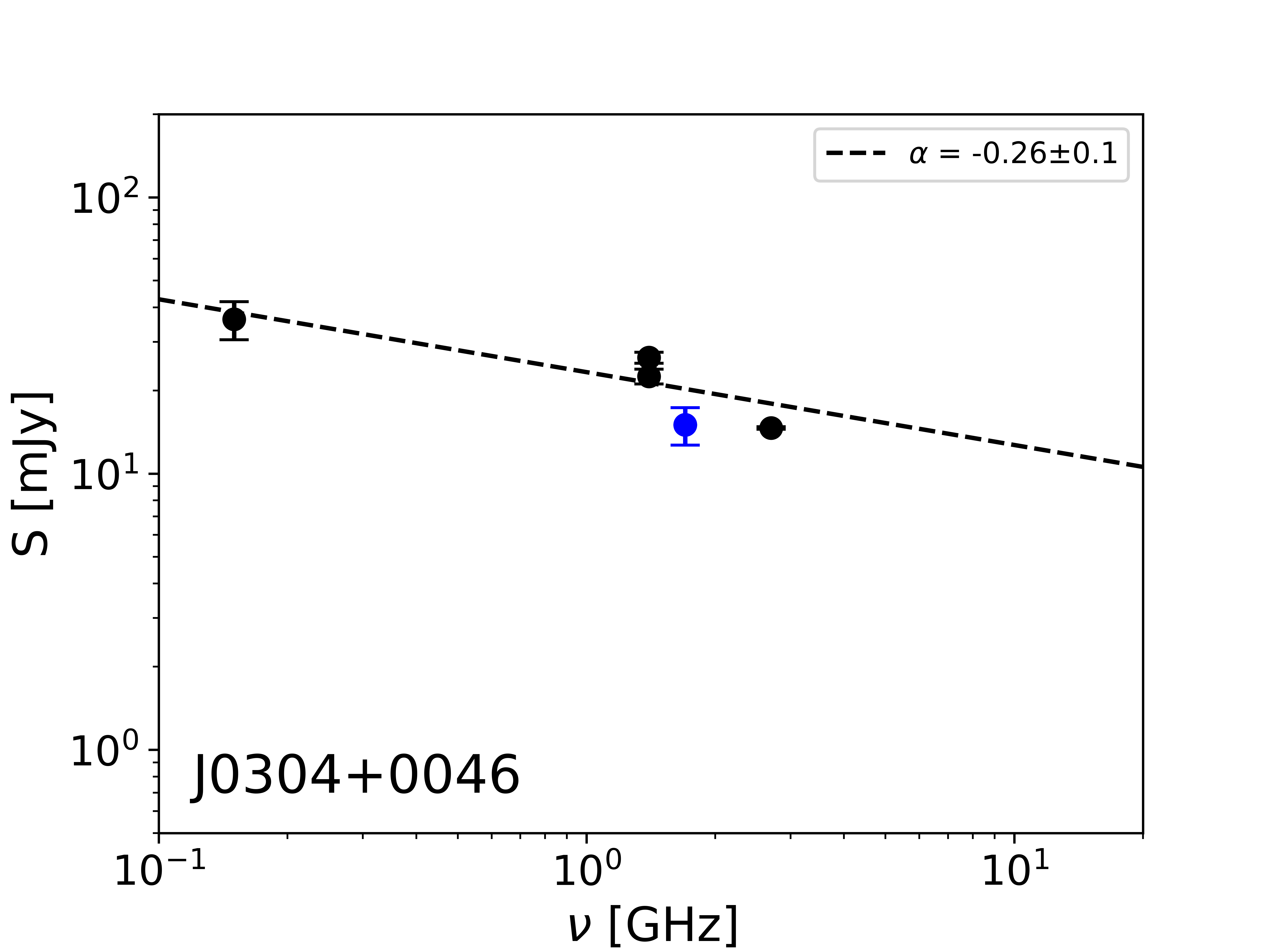}
            \includegraphics[width=0.328\textwidth]{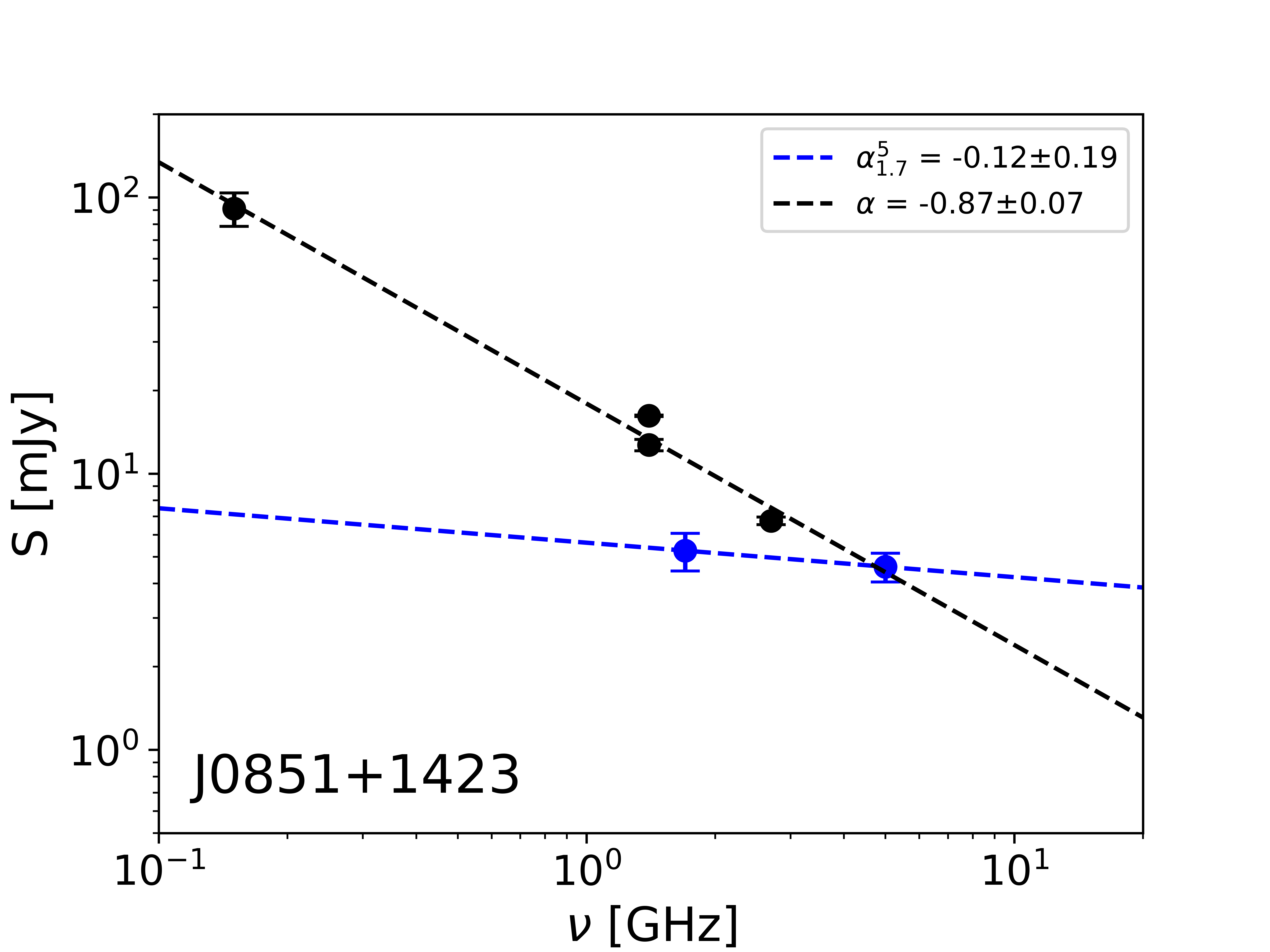}
            \includegraphics[width=0.328\textwidth]{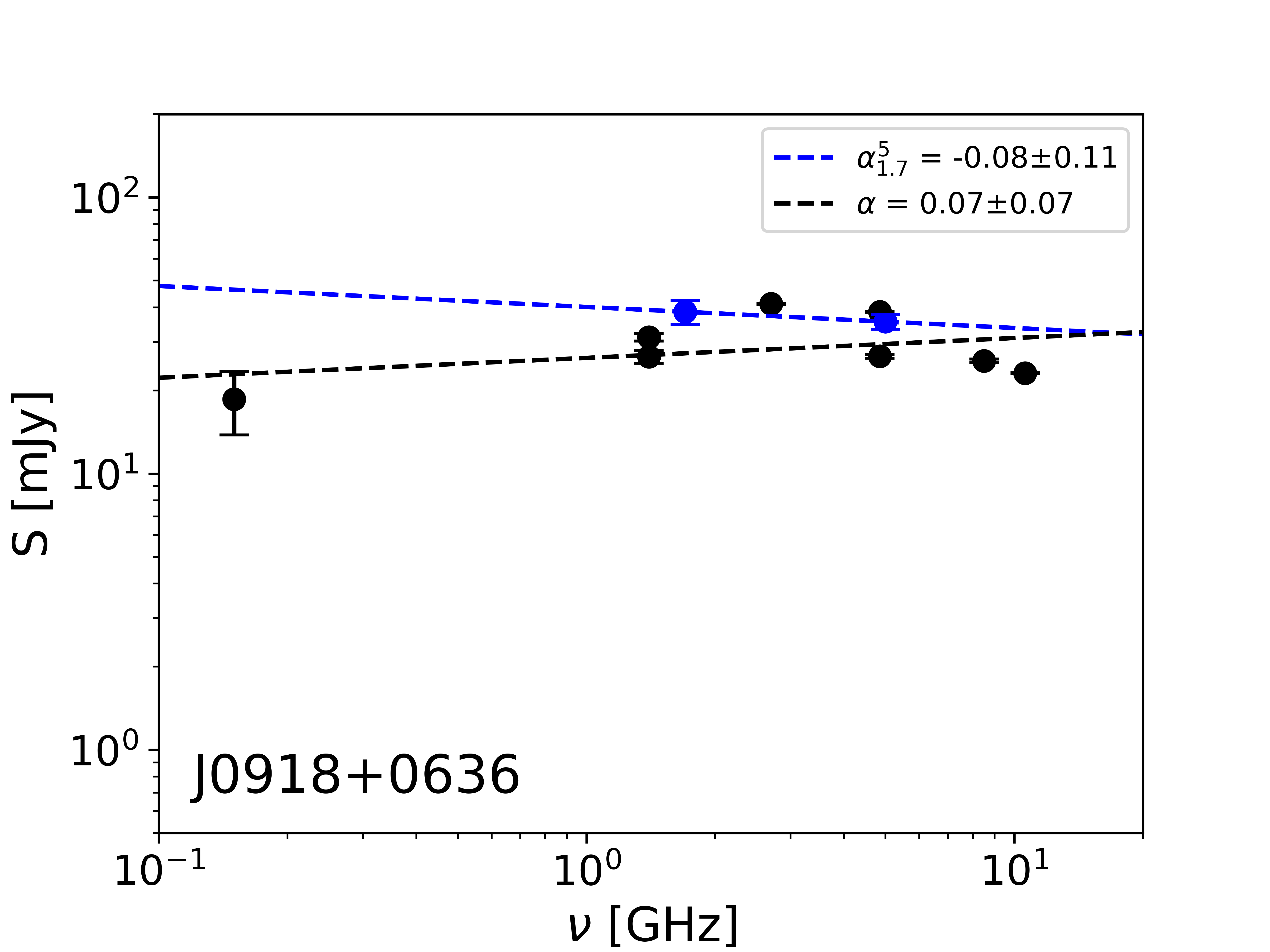}
            }
\gridline{  \includegraphics[width=0.328\textwidth]{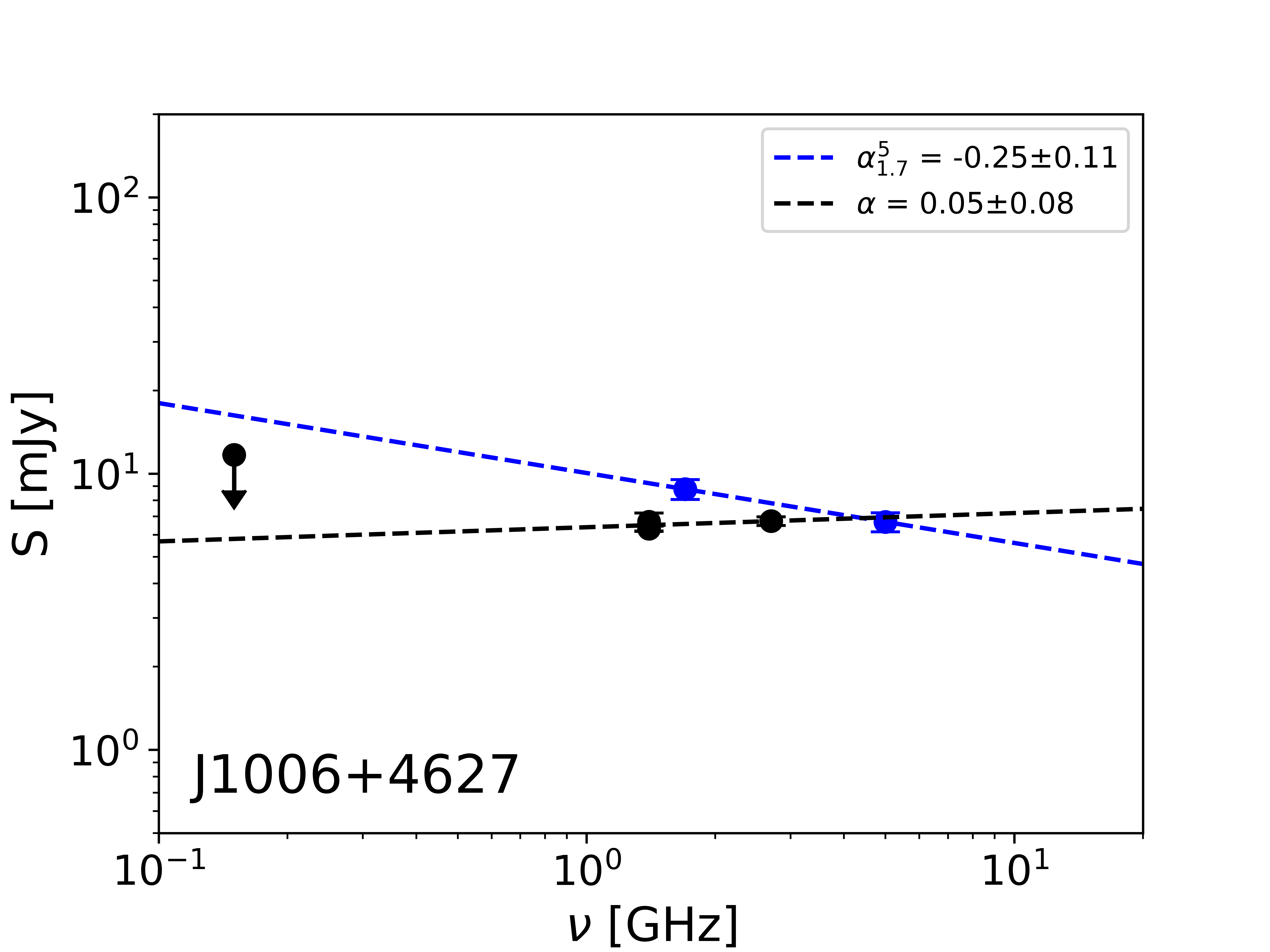}
            \includegraphics[width=0.328\textwidth]{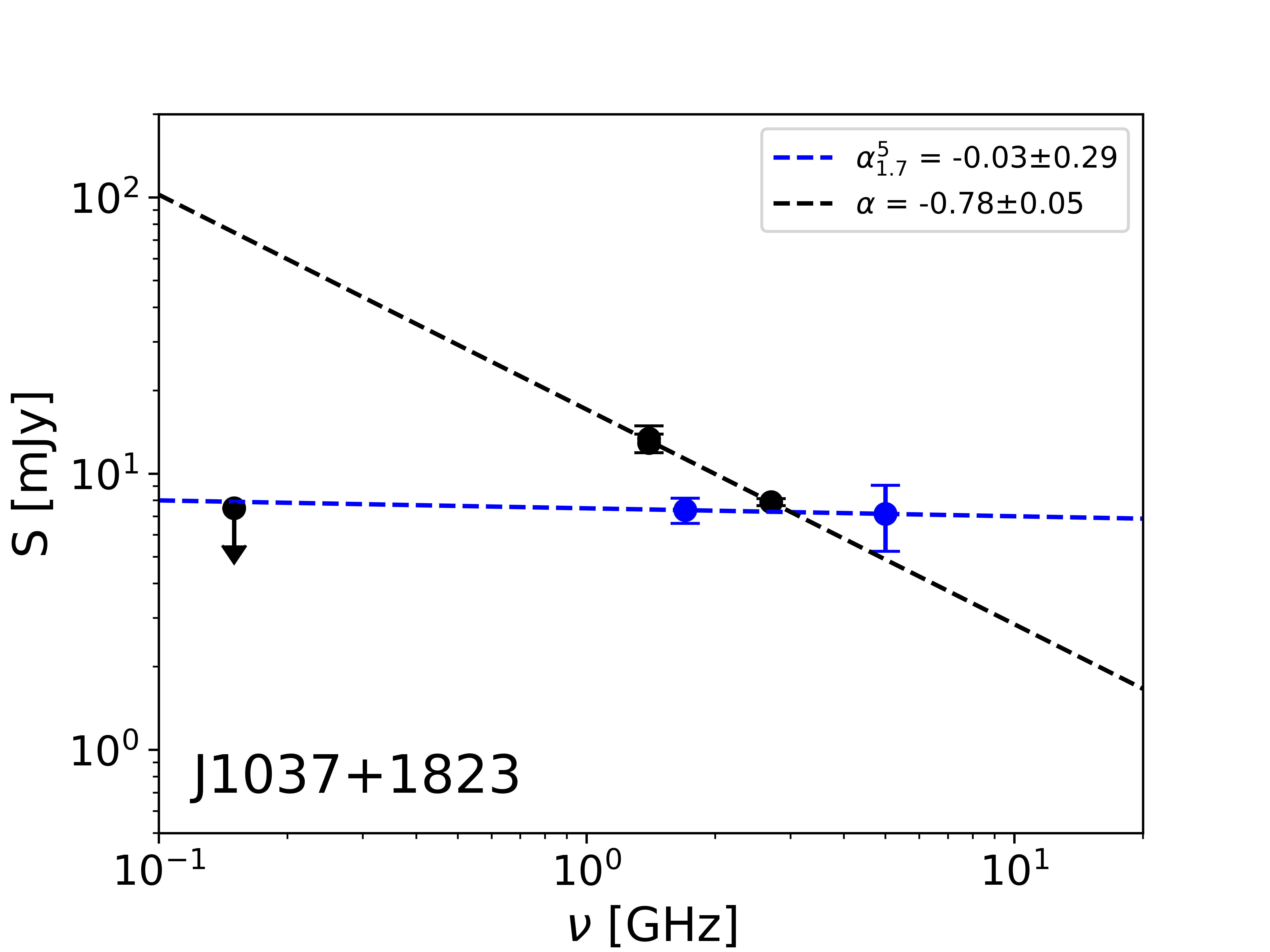}
            \includegraphics[width=0.328\textwidth]{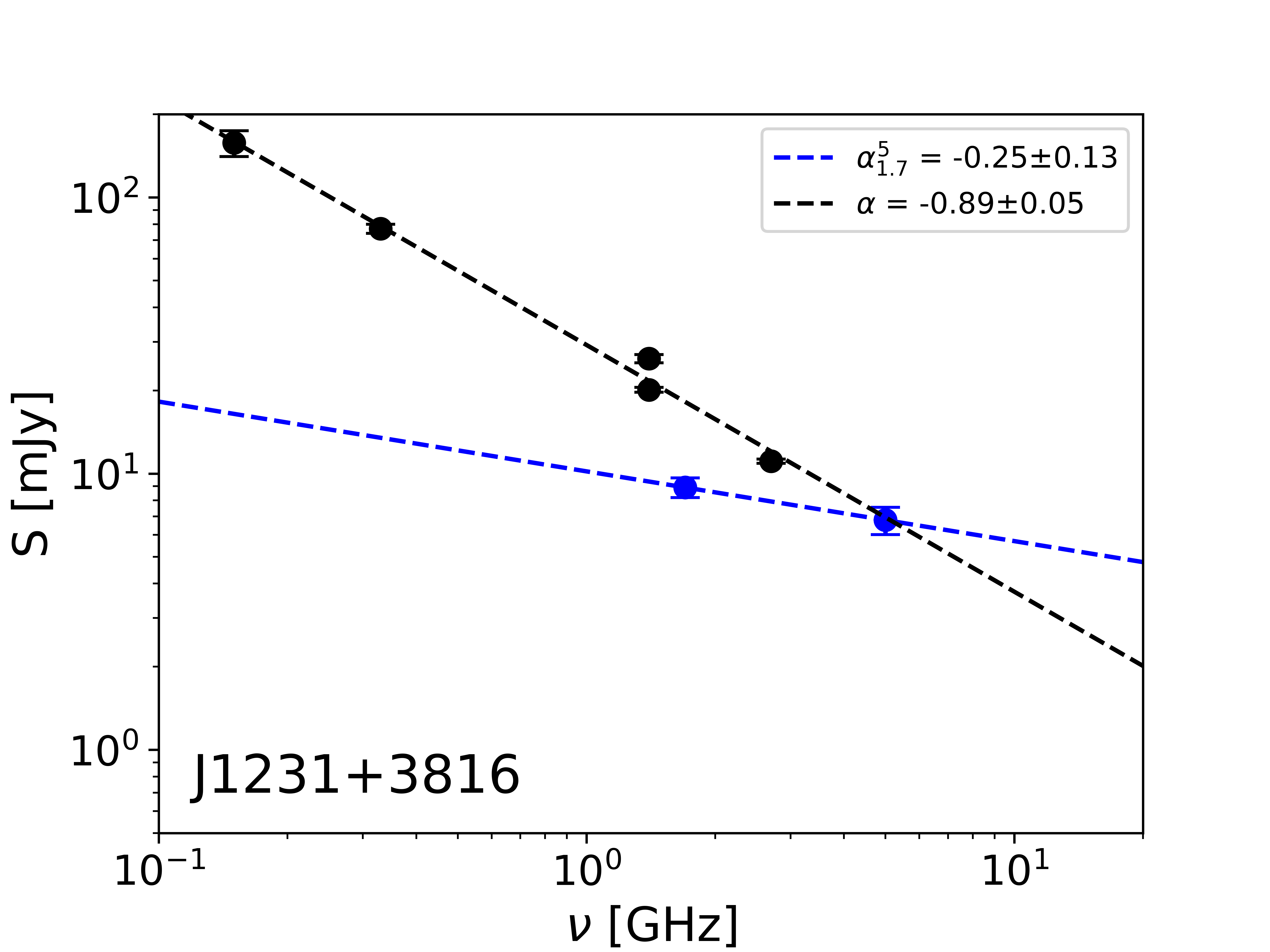}
            }
\gridline{  \includegraphics[width=0.328\textwidth]{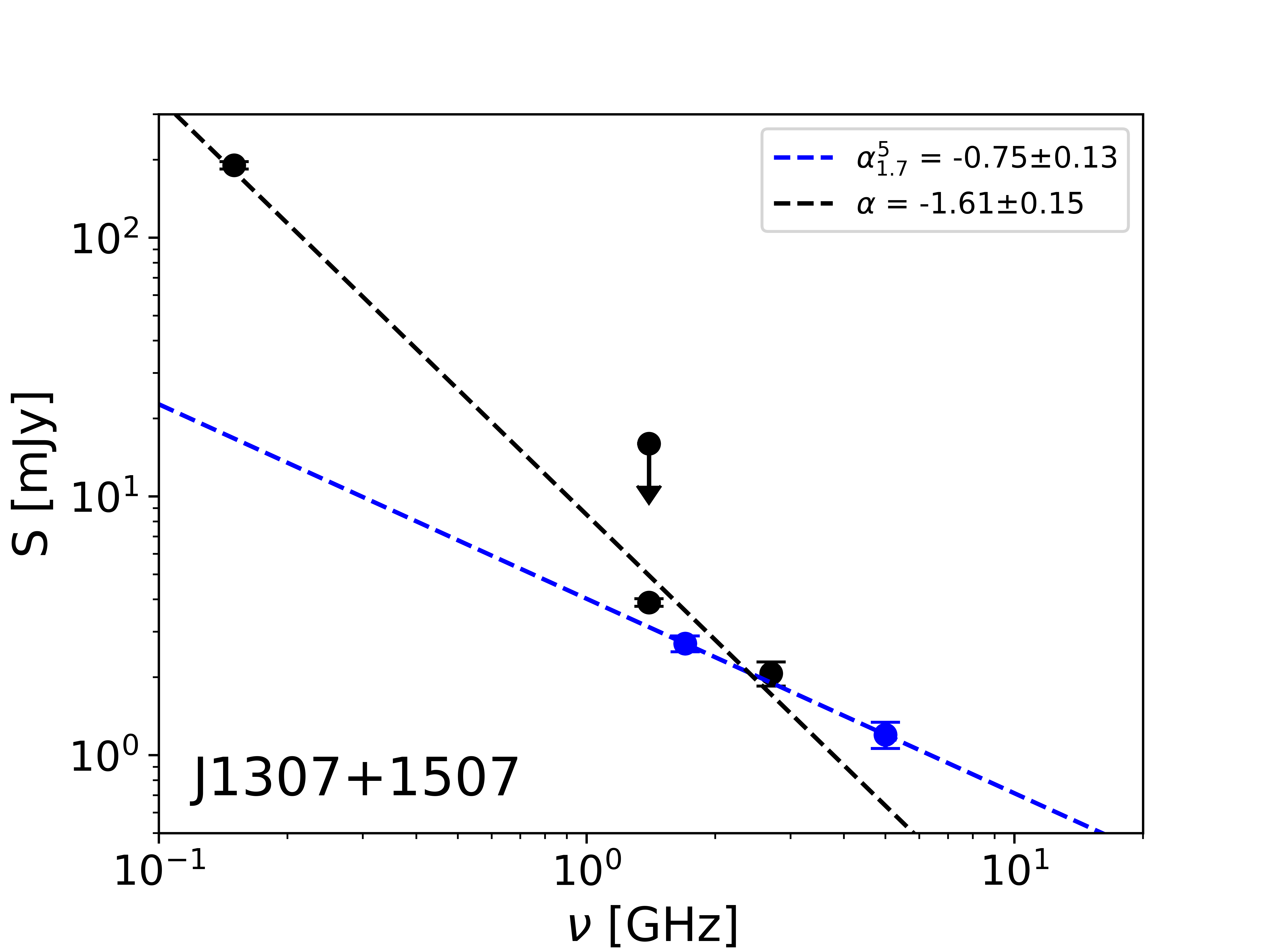}
            \includegraphics[width=0.328\textwidth]{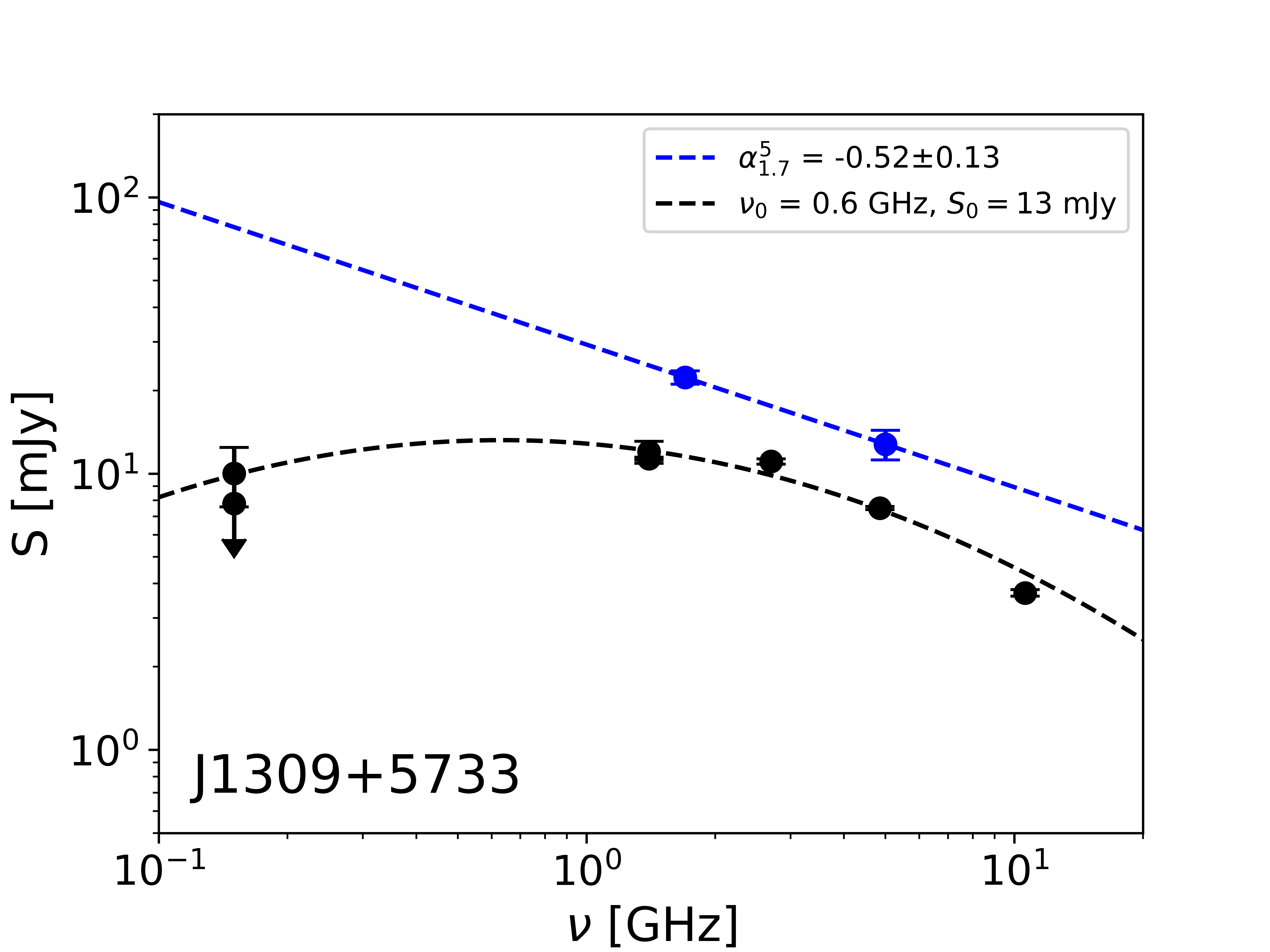}
            \includegraphics[width=0.328\textwidth]{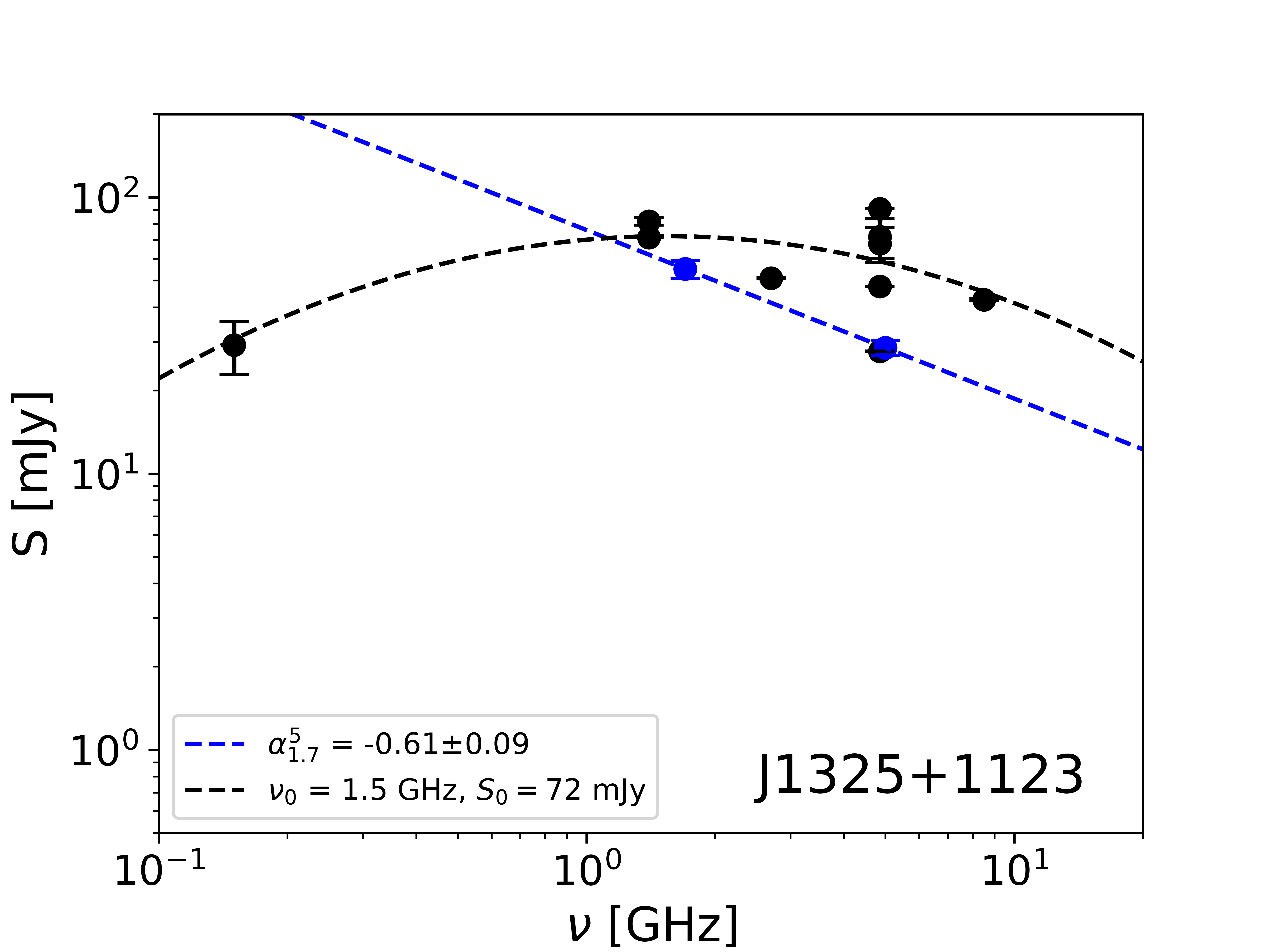}
            }
\gridline{  \includegraphics[width=0.328\textwidth]{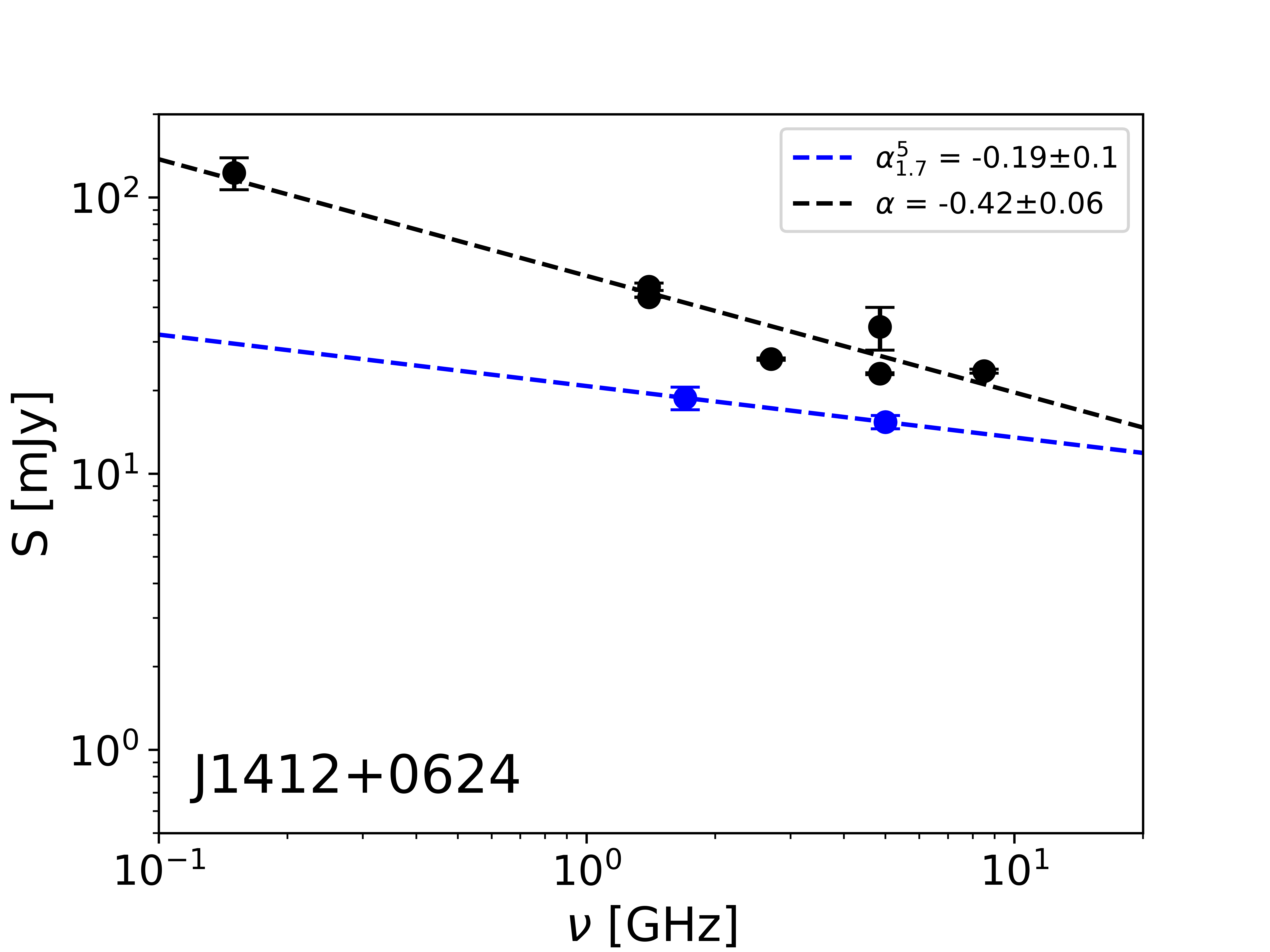}
            \includegraphics[width=0.328\textwidth]{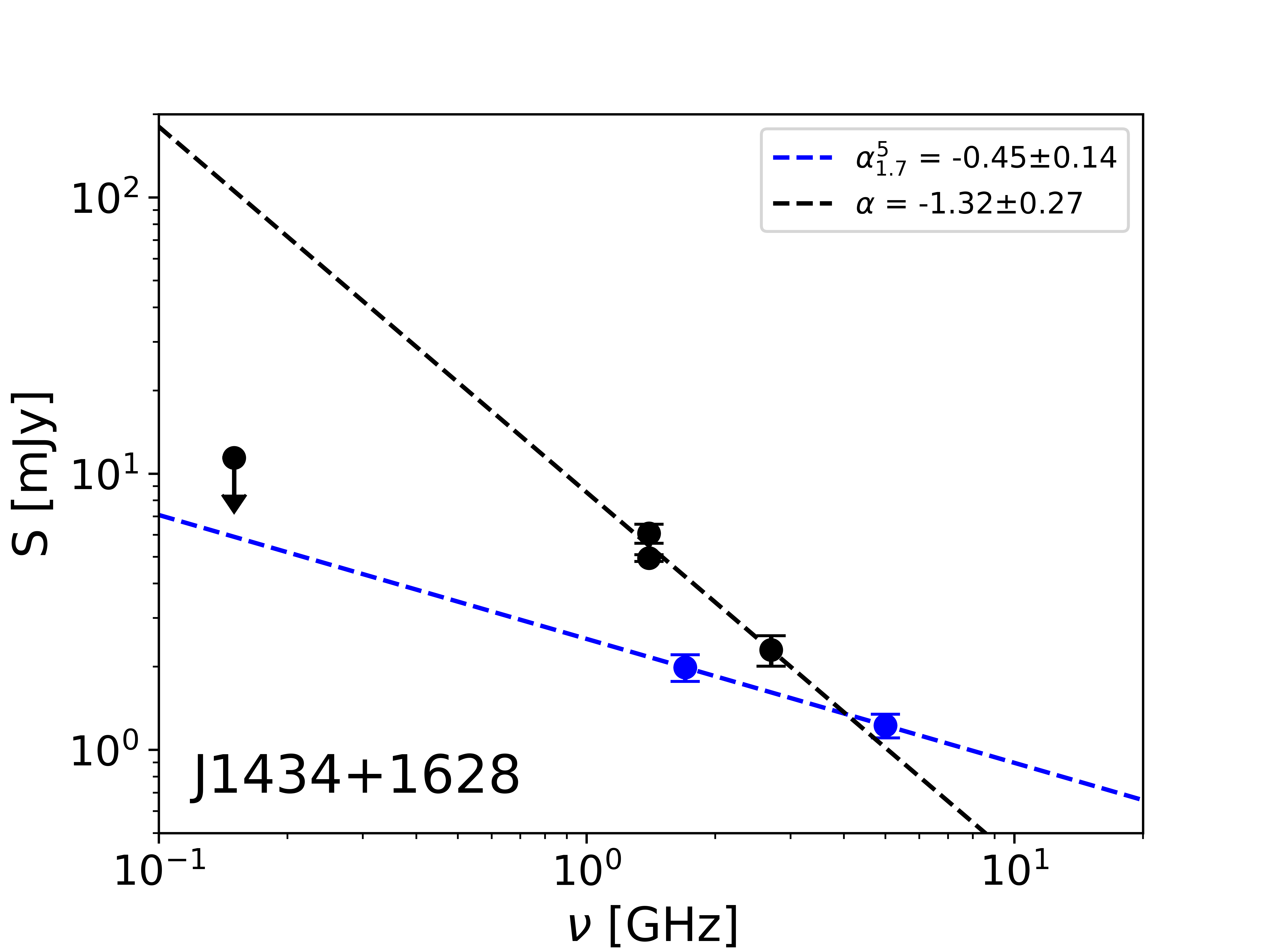}
            \includegraphics[width=0.328\textwidth]{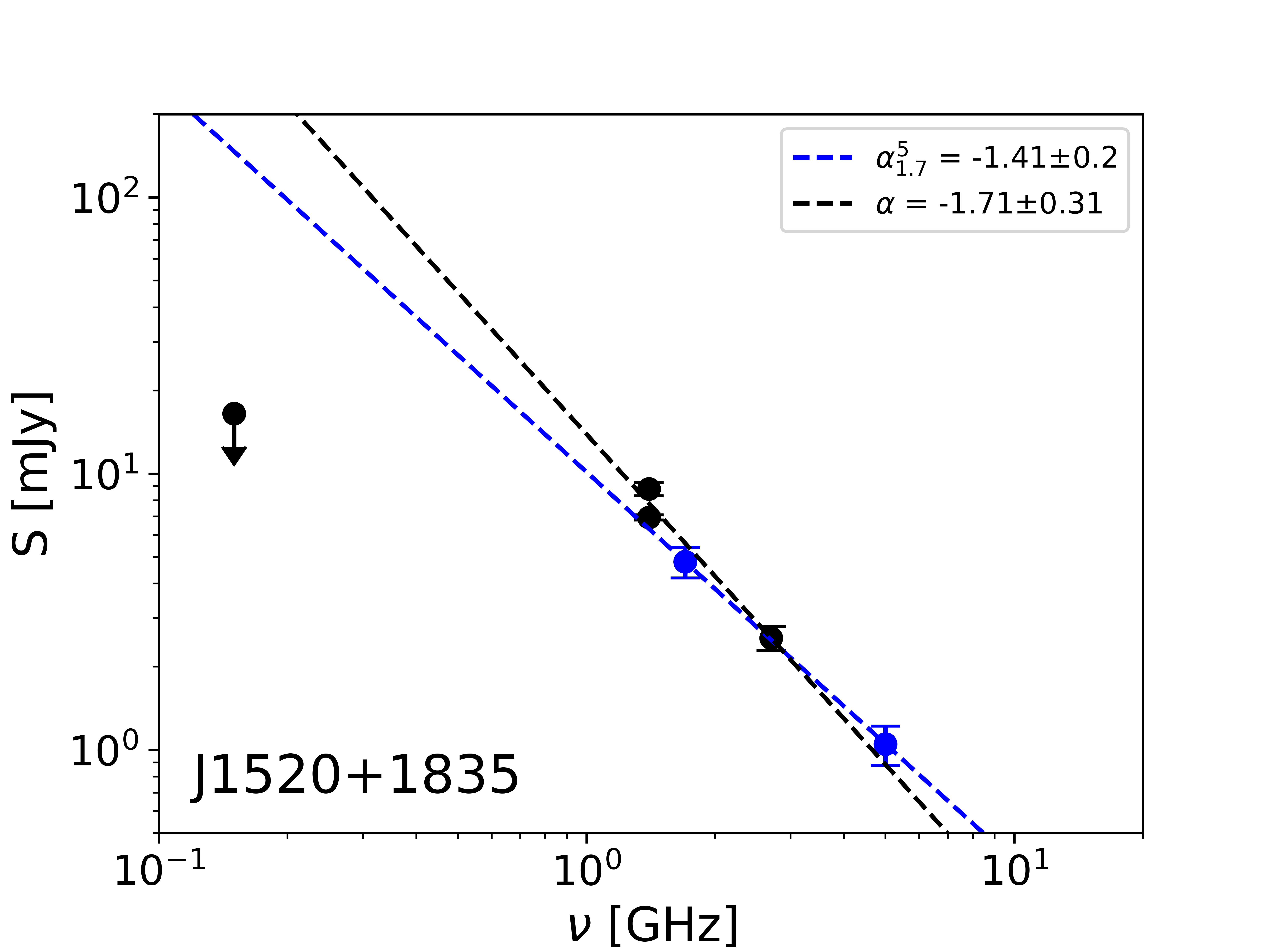}
            }
\gridline{  \includegraphics[width=0.328\textwidth]{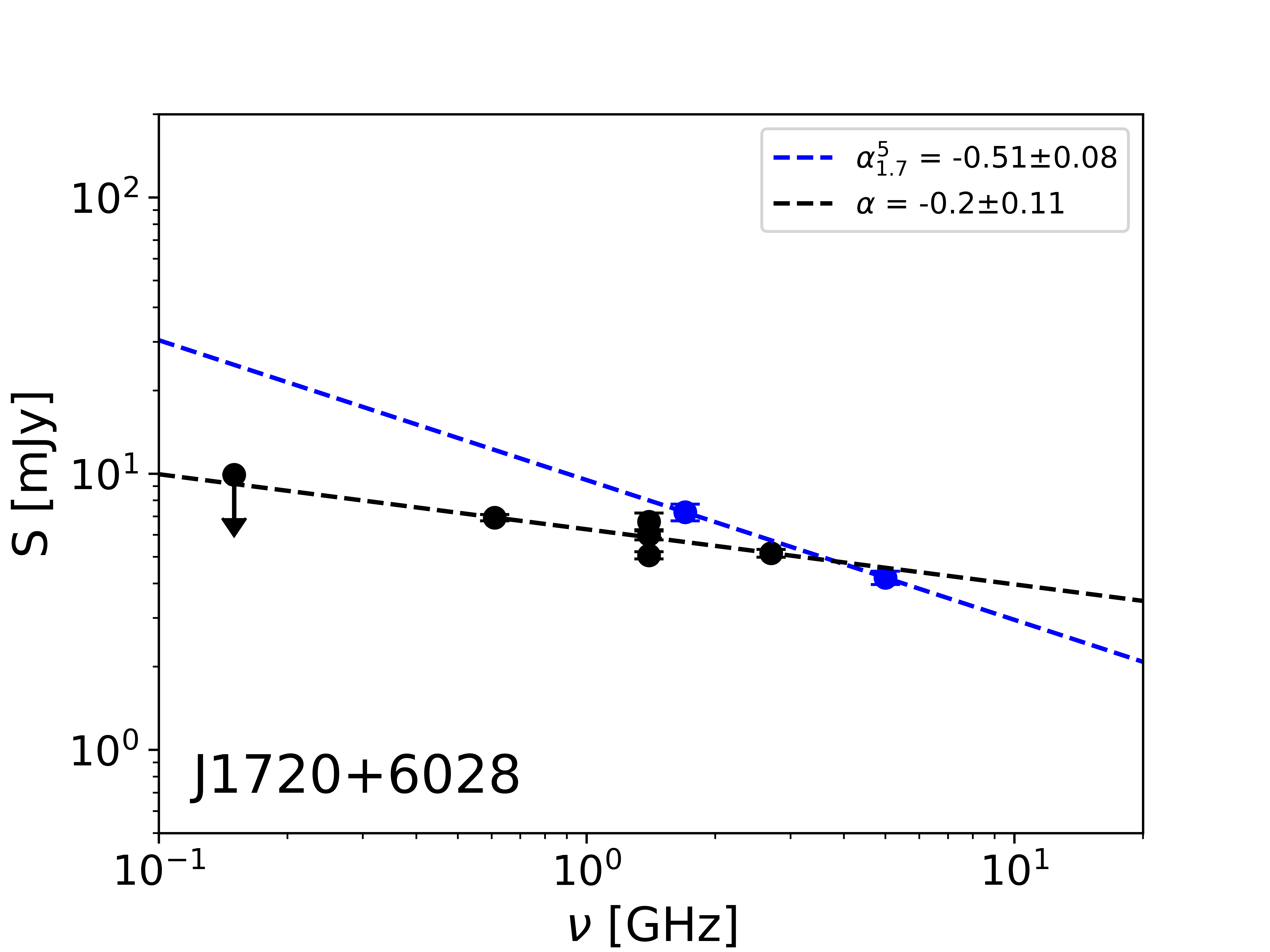}
            }
\caption{Radio continuum spectra of the 13 sources in our EVN sample. \textit{Black:} total flux density data collected from the literature. \textit{Blue:} compact component flux densities observed with high resolution in this project.}
\label{spectra}
\end{figure*}

Continuum radio spectra of the 13 sources are shown in Figure~\ref{spectra}, both using fitted compact component flux densities from our high-resolution EVN data, and total flux densities from low-resolution measurements from the literature (Table \ref{litspectral}). VLBI data are available at two frequencies (except for J0304+0046 with only 1.7-GHz flux density), therefore we determined the two-point spectral index $\alpha_{1.7}^5$ (Table~\ref{phyparams}) by assuming a power-law radio spectrum. We note that although most of the EVN measurements of the same target source at the two frequencies were conducted with a typical time difference of $\sim 1$\,month, in some cases up to $17$ months elapsed between the two observations (Table~\ref{observations}), and in the case of J0918$+$0636, the time difference is $\sim 20$\,yr. Hence the calculated spectral indices for those sources (especially for J0918$+$0636) should be treated with caution because of possible long-term flux density variability.    

Spectra with $\alpha < -0.5$ are called steep, with $-0.5 \leq \alpha \leq 0$ flat, and with $0 < \alpha$ inverted.
From the 12 quasars in our sample with spectral indices characteristic to the compact VLBI-imaged structure, $5$ sources ($42\%$) have steep and $7$ ($58\%$) have flat spectra. Taking into account the uncertainties of the spectral indices, there are 3 sources that could have either flat or steep spectrum within the errors. Similarly, there are 3 sources with flat spectra which could be inverted within the uncertainties. Similar flat-to-steep-spectrum ratio was found by \citet{2016MNRAS.463.3260C} in their study of  $z > 4.5$ quasars. The two smaller samples of $z > 4$ quasars of \citet{2010AA...524A..83F} and \citet{2017MNRAS.467..950C} also have similar ratios, with $50-50\%$ of objects with flat and steep spectra.                

For the total flux density spectra also plotted in Figure~\ref{spectra}, we collected available single-dish and low-resolution radio interferometric observations from the literature for all the sources in our sample. Since these quasars are relatively faint and typically below the survey flux density thresholds, only a handful of measurements were found for the majority of them. Fortunately, the 1.4-GHz FIRST \citep{1997ApJ...475..479W} and NVSS \citep{1998AJ....115.1693C}, and the recent 2.7-GHz  VLASS \citep{2020PASP..132c5001L,2020RNAAS...4..175G} surveys all have observations for all of our sources. In NVSS, J1307$+$1507 is blended with a nearby bright radio source, therefore, the corresponding spectral point is marked as an upper limit in Figure~\ref{spectra}. In addition to the surveys above, Table~\ref{litspectral} contains the references where flux densities were collected from for our sources.

\begin{deluxetable*}{lrrrrrrrrr}[h!]
\tablenum{6}
\tablecaption{Total flux density data used to make Fig. \ref{spectra} taken from the literature.}
\tiny
\tablewidth{0pt}
\tablehead{
\colhead{ID} & \colhead{$\alpha$} &  \colhead{$S_\mathrm{150~MHz}$} &  \colhead{$S_\mathrm{330~MHz}$} &  \colhead{$S_\mathrm{610~MHz}$}  &  \colhead{$S_\mathrm{1.4~GHz}$} &  \colhead{$S_\mathrm{2.7~GHz}$}  &  \colhead{$S_\mathrm{5~GHz}$} &  \colhead{$S_\mathrm{8.5~GHz}$}  &  \colhead{$S_\mathrm{10.6~GHz}$} \\
\colhead{} & \colhead{} &  \colhead{(mJy)} &  \colhead{(mJy)} &  \colhead{(mJy)}  &  \colhead{(mJy)} &  \colhead{(mJy)}  &  \colhead{(mJy)} &  \colhead{(mJy)}  &  \colhead{(mJy)} 
}
\decimalcolnumbers
\startdata  
J0304$+$0046 & $-$0.26 (0.10) & 36.2 (5.7)$^{10}$ &    &     & 22.5 (0.1)$^3$ & 14.63 (0.16)$^{12}$ &  &     &   \\
             &  &             &    &     & 24.6 (0.8)$^4$ &              &  &     &   \\
J0851$+$1423 & $-$0.87 (0.07) & 92.1 (12.6)$^{10}$ &    &     & 16.2 (0.1)$^3$ & 6.75 (0.21)$^{12}$ &  &     &   \\
             &  &  &    &     & 12.3 (0.5)$^4$ &   &  &     &   \\
J0918$+$0636 & 0.07 (0.07) & 18.6 (4.8)  &    &     & 26.5 (0.1)$^3$ & 41.18 (0.24)$^{12}$ & 38.5$^7$ & 25.6 (0.4)$^6$ & 23.1$^7$ \\
             &                        &  &    &     & 30.9 (1.0)$^4$ &              & 26.6 (0.4)$^8$ &     &   \\
J1006$+$4627 & $0.05$ (0.06) & $<$ 11.7*        &    &     & 6.3 (0.1)$^3$ & 6.73 (0.24)$^{12}$   &  &     &  \\
             &  &             &    &     & 6.4 (0.4)$^4$ &               &  &     &  \\
J1037$+$1823 & $-0.78$ (0.05) & $<$ 7.5*        &    &     & 13.4 (0.2)$^3$ & 7.89 (0.23)$^{12}$  &  &     &  \\
             &  &             &    &     & 11.4 (0.5)$^4$ &              &  &     &   \\
J1231$+$3816 & $-0.89$ (0.05) & 157.5 (16.9)$^{10}$ & 77 (2.9)$^5$ & & 24.04 (0.13)$^3$ & 11.09 (0.20)$^{11}$ &   &     &  \\
             &  &             &    &     & 25.7 (0.9)$^4$ &              &  &     &   \\
J1307$+$1507 & $-1.61$ (0.14) & 190.6 (6.3)* &    &     & 3.89 (0.13)$^3$ & 2.07 (0.22)$^{12}$ &  &     &  \\
             &  &             &    &     & $<$ 16.2 (0.6)$^4$ &               &  &     &   \\
J1309$+$5733 & $S_0 =$ 13 (1) & 10.01 (2.43)$^{11}$ &   &     & 11.33 (0.14)$^3$ & 11.07(0.24)$^{11}$ & 7.5$^7$ &  & 3.7$^7$  \\
             & $\nu_0 =$ 0.6 (0.1) & $<$ 7.8*           &    &     & 11.2 (0.9)$^4$ &               &  &     & \\
J1325$+$1123 & $S_0 =$ 72 (17) & 29.2 (6.3) &    &     & 71.5 (0.1)$^3$ & 51.03 (0.26)$^{12}$  & 72 (12)$^1$ & 42.6 (0.4)$^6$ &  \\
             & $\nu_0 =$ 1.5 (0.2) &             &    &     & 81.4 (2.5)$^4$ &               & 47.6$^7$    &     &   \\
             &  &             &    &     &            &               & 68 (10)$^2$ &     &   \\
J1412$+$0624 & $-$0.42 (0.06) & 122.8 (16.2)$^{10}$ &   &     & 43.5 (0.1)$^3$ & 25.98 (0.25)$^{12}$  & 23.0 (0.2)$^8$  & 23.5 (0.4)$^6$ & \\
             &  &             &    &     & 47.2 (1.5)$^4$ &               & 34 (6)$^7$ &     &   \\
J1434$+$1628 & $-$1.12 (0.27) &  $<$ 11.4*      &    &     & 4.21 (0.14)$^3$ & 2.30 (0.29)$^{12}$  &   &     &  \\
             &  &             &    &     & 5.0 (0.5)$^4$ &                &  &     &   \\
J1520$+$1835 & $-$1.71 (0.31) & $<$ 16.5*            &    &     & 6.94 (0.15)$^3$ & 2.54 (0.25)$^{12}$  &   &     &  \\
             &  &             &    &     & 8.8 (0.5)$^4$ &                &  &     &   \\
J1720$+$6028 & $-$0.20 (0.11) & $<$ 9.9*      & & 6.93 (0.18)$^8$ & 5.06 (0.15)$^3$ & 5.15 (0.17)$^{12}$  &   &     &  \\
             &  &             &    &     & 6.6 (0.4)$^4$ &                &  &     &   \\
             &  &             &    &     & 6.01 (0.26)$^6$ &              &  &     &   \\
\enddata
\tablecomments{
Col.~1 $-$ source designation; Col.~2 $-$ fitted spectral indices; where log-parabolic function was fitted, we give the fitted values $S_0$ in mJy and $\nu_0$ in GHz.; Col.~3--10 $-$ flux densities and their uncertainties (where available) collected from the literature; numbers in the upper indices are references for the flux density values. The symbol * marks the TGSS $3\sigma$ upper limits. References: 1: \citet{1991ApJS...75.1011G}; 2: \citet{1991ApJS...75....1B}; 3: \citet{1995ApJ...450..559B}; 4: \citet{1998AJ....115.1693C}; 5; \citet{2000yCat.8062....0D}; 6: \citet{2003MNRAS.341....1M}; 7: \citet{2004MNRAS.348..857H}; 8: \citet{2007MNRAS.376..371J}; 9: \citet{2007MNRAS.376.1251G}; 10: \citet{2017AA...598A..78I}; 11: \citet{2017AA...598A.104S}; 12: \citet{2020RNAAS...4..175G}
}
\label{litspectral}
\end{deluxetable*}
In most cases, we fitted the spectral data points with a power-law function, similarly to the high-resolution case. However, for J1309$+$5733 and J1325$+$1123, we found that the choice of a log-parabolic function
\begin{equation}
\log S = a (\log \nu - \log~\nu_0)^2 + b 
\label{logparabola}
\end{equation}
resulted in a better fit (i.e. lower $\chi^2$ values) than using the power-law. In Equation~(\ref{logparabola}), $\nu_0$ is the frequency corresponding to the peak flux density $S_0$, while $a$ and $b$ are numerical constants without any physical meaning \citep[e.g.][]{2017MNRAS.467.2039C}.
The fitted spectral index $\alpha$ (for the power-law cases), or the $\nu_0$ and $S_0$ values (for the log-parabolic fits) are given in Col.~2 of Table~\ref{spectra} and as insets in the spectrum plots (Figure~\ref{spectra}).

To obtain low-frequency flux density points for our spectra, we analysed TGSS Alternative Data Release 1 \citep[Tata Institute of Fundamental Research Giant Metrewave Radio Telescope Sky Survey,][]{2017AA...598A..78I} images, since all of our sources are in its sky coverage. Four out of the 13 sources are listed in the TGSS source catalogue. \citet{2019MNRAS.484..204C} found two other sources of our sample (J0918$+$0636 and J1325$+$1123) in the TGSS images detected with lower significance (at least $2\sigma$). For these two, and for another source J1307$+$1507 that we found in the image but not in the TGSS source catalogue, we fitted Gaussian brightness distribution models at the source positions with the \textsc{aips} task \textsc{jmfit} using the corresponding TGSS tiles. For the remaining six undetected objects, we consider the $3\sigma$ image noise (measured in the close vicinity of the objects using the task \textsc{imean}) to derive an upper limit to the flux density of an unresolved source (Table~\ref{litspectral}). This way we get important constrains on the low-frequency end of the spectra, even if there is no TGSS detection available.

Some of the sources whose spectrum is fitted with a power-law function might in reality have a peaked spectrum as well, that is better described with a log-parabolic function. However, because of the lack of low-frequency measurements, parameters of such a fit would be poorly constrained. Indeed, when low-frequency synchrotron emission is properly sampled, high-redshift ($z>5$) radio-loud quasars tend to show evidence for spectral turnover at rest-frame frequencies $\sim 1-50$\,GHz \citep{2022A&A...659A.159S}. Peaked continuum radio spectra are also common for bright $z>3$ radio quasars \citep{2021MNRAS.508.2798S}.

\subsection{Flux Density Variability}
\label{disc3}

Our possibility to investigate flux density variability is limited because there are generally no measurements available at the same frequencies and angular resolutions, but at different observing epochs. However, all of our sources have been detected in three VLA surveys (FIRST, NVSS, and VLASS), albeit the NVSS flux density value for J1307$+$1507 can only be regarded as an upper limit. We calculate flux density ratios from non-simultaneous measurements (Table~\ref{fluxratios}) and interpret them in the context of variability.  

As discussed by \citet{2016MNRAS.463.3260C}, the ratio between the FIRST and NVSS flux densities ($S_\mathrm{FIRST}/S_\mathrm{NVSS}$) can be a variability indicator. Both surveys were conducted at 1.4~GHz but at different epochs, and in different configurations of the VLA. The B configuration used by FIRST provided $\sim 5\arcsec$ angular resolution, while the most compact D configuration used by NVSS led to $\sim 45\arcsec$ resolution. Following \citet{2016MNRAS.463.3260C}, we consider a source definitely variable if the FIRST flux density exceeds the NVSS value by more than $10\%$. In our sample, there are two such sources, J0851$+$1423 and J1037$+$1823 (Table~\ref{fluxratios}). If $S_\mathrm{FIRST}/S_\mathrm{NVSS} \ll 1$, it may be caused either by variability or the presence of an emission region extended to $\sim 10\arcsec$ which remains unresolved in NVSS but becomes resolved in FIRST. In general, any other source may as well be variable, even with $S_\mathrm{FIRST}/S_\mathrm{NVSS} \approx 1$, just being unresolved and having, accidentally, nearly equal flux densities at the two epochs when the FIRST and NVSS measurements were made.  

In a similar way, we can compare VLBI and VLA flux densities at the same frequencies that were measured with substantially different angular resolutions. Table~\ref{fluxratios} gives 1.4- and 2.7-GHz flux density ratios based on FIRST and VLASS data, respectively. The VLBI values were calculated by extrapolating (to 1.4~GHz) and interpolating (to 2.7~GHz) our EVN measurements made at 1.7 and 5~GHz, assuming a power-law radio spectrum in the GHz frequency range, using spectral indices $\alpha_{1.7}^5$ calculated in Section~\ref{disc1} (Table~\ref{phyparams}). For J0304+0046 and J0918$+$0636, i.e. the sources with single-frequency EVN data available, we assumed here a zero spectral index. We repeat the cautionary note that the EVN spectral index values may be affected by variability since the measurements at the two frequencies were not simultaneous. Also, if any of our objects is a GPS or megahertz peaked-spectrum (MPS) source, their spectra are not necessarily well described by a power-law function. However, the range where extrapolation and interpolation are done is quite narrow in frequency. If $S_\mathrm{VLBI,1.4\,GHz}/S_\mathrm{FIRST} > 1.1$ or $S_\mathrm{VLBI,2.7\,GHz}/S_\mathrm{VLASS} > 1.1$, we consider the source variable. If these ratios are below unity, the most likely explanation is that a significant fraction of the total radio emission originates from an extended region completely resolved with VLBI. However, variability cannot be ruled out entirely in any of the sources. 

In summary, we consider a source variable when at least one of the three flux density ratios described above exceeds unity by more than 10\%. By this criterion, we find the following quasars definitely variable: J0851$+$1423, J0918$+$0636, J1006$+$4627, J1037+1823, J1309$+$5733, and J1720$+$6028 (Table~\ref{fluxratios}).    \\

\begin{deluxetable*}{lccc}
\tablenum{7}
\tablecaption{Flux density ratios of the target sources.}
\tablewidth{0pt}
\tablehead{
\colhead{ID} & \colhead{$S_\mathrm{FIRST}/S_\mathrm{NVSS}$} & \colhead{$S_\mathrm{VLBI,1.4\,GHz}/S_\mathrm{FIRST}$}  & \colhead{$S_\mathrm{VLBI,2.7\,GHz}/S_\mathrm{VLASS}$} 
}
\decimalcolnumbers
\startdata
J0304$+$0046 & 0.85 (0.03)  & 0.71$^*$    & 1.03$^*$    \\
J0851$+$1423 & 1.32 (0.05)	& 0.33 (0.03) &	0.74 (0.14) \\
J0918$+$0636 & 0.86 (0.03)	& 1.45$^*$    &	0.94$^*$    \\
J1006$+$4627 & 0.98 (0.06)	& 1.47 (0.10) &	1.16 (0.07) \\
J1037$+$1823 & 1.20 (0.06)	& 0.54 (0.06) &	0.92 (0.21) \\
J1231$+$3816 & 0.93 (0.03)	& 0.39 (0.01) &	0.72 (0.04) \\
J1307$+$1507 & --       	& 0.80 (0.03) &	0.92 (0.12) \\
J1309$+$5733 & 1.01 (0.08)	& 2.28 (0.02) &	1.62 (0.04) \\
J1325$+$1123 & 0.87 (0.03)	& 1.09 (0.01) &	1.02 (0.01) \\
J1412$+$0624 & 0.92 (0.03)	& 0.56 (0.01) &	0.83 (0.02) \\
J1434$+$1628 & 0.83 (0.09)	& 0.52 (0.04) &	0.70 (0.17) \\
J1520$+$1835 & 0.78 (0.05)	& 0.91 (0.03) &	0.98 (0.10) \\
J1720$+$6028 & 0.77 (0.03)	& 1.57 (0.06) &	1.11 (0.05) \\
\enddata
\tablecomments{$^*$ assuming 0 spectral index. Col.~1 $-$ Source designation; Col.~2 $-$ Ratio of 1.4-GHz FIRST and NVSS flux densities and its uncertainty; Col.~3 $-$ Ratio of the extrapolated 1.4-GHz VLBI and FIRST flux densities and its uncertainty; Col.~4 $-$ Ratio of the interpolated 2.7-GHz VLBI and VLASS flux densities and its uncertainty}
\label{fluxratios}
\end{deluxetable*}

\subsection{Gaia Optical Positions}
\label{disc4}

The coordinates from the recent Gaia \citep{2016A&A...595A...1G} EDR3 catalog are the most accurate optical astrometric positions available \citep{2021A&A...649A...1G}. The high sensitivity of Gaia enables the detection of faint extragalactic sources as well, like the high-redshift AGN in our sample. We checked the reliability of Gaia positions by comparing the newest EDR3 solutions with those in the earlier data release \citep[DR2,][]{2018A&A...616A...1G}, and found that they are consistent within the positional accuracy of the survey (from a few $\mu$as to a few mas, depending on the actual object). Comparing the Gaia EDR3 optical coordinates with radio positions of quasars obtained from our VLBI observations can help to reveal additional information on the nature of these sources. 

The VLBI radio position corresponds to the brightest compact region of the jet, which may be either a synchrotron self-absorbed core or a compact hotspot associated with a shock front between the jet and the surrounding medium. On the other hand, the optical position is mostly determined by the location of the accretion disk, which is in the closest vicinity of the central SMBH. However, the preferred direction of occasional offsets between the radio and optical AGN positions, up to a few mas, is found to statistically coincide with the VLBI jet direction, suggesting the presence of strong pc-scale optical jet emission in certain objects \citep{2017A&A...598L...1K,2019ApJ...871..143P}. In misaligned jetted sources like CSOs, the brightest features marking the radio position are hotspots in the radio lobes, which appear further away from the central object. Therefore, if a significant offset exceeding the usual values found for beamed AGN is detected between the radio (VLBI) and optical (Gaia) coordinates, it strongly suggests the misaligned nature of the quasar. For faint radio sources, like the typical high-redshift quasars, revealing a Gaia--VLBI positional mismatch could help constraining the class.

All but one of our sources (J0304$+$0046) have been observed with Gaia and have EDR3 optical position available. The positional uncertainties in EDR3 are below $1$~mas in most cases, except for J1037$+$1823 and J1720$+$6028, for which the errors are somewhat higher, but still within $\sim 2$~mas. We indicate Gaia positions in our VLBI images with red crosses, while 5-GHz VLBI peak positions are also shown in the 1.7-GHz images (with blue crosses) when the offset is non-negligible (Figure~\ref{evnimages}). Radio and optical positions are listed in Table~\ref{imgparams}, along with their uncertainties. We consider a positional offset significant if the optical position differs by more than $3\sigma_\mathrm{pos}$ from the 5-GHz position. Here $\sigma_\mathrm{pos}$ is the uncertainty of the positional difference, $\sigma_\mathrm{pos} = \sqrt{\sigma_\mathrm{5\,GHz}^2+\sigma_\mathrm{Gaia}^2}$, where $\sigma_\mathrm{5\,GHz}$ is the uncertainty of the 5-GHz radio coordinate and $\sigma_\mathrm{Gaia}$ refer to the uncertainty of the Gaia optical coordinate. We tag a source with `Slight' in Table~\ref{classification} when the radio--optical offset is between $(1-3) \times \sigma_\mathrm{pos}$. There are a few sources where the 1.7- and 5-GHz radio positions do not agree. These are indicated in Figure~\ref{evnimages} and the notable individual cases are discussed in Section~\ref{disc5}.

\subsection{Classification and Notes on Individual Sources}
\label{disc5}

Here we attempt to classify the radio quasars in our sample as blazar or non-blazar sources. We take various aspects into account, namely the radio structure revealed by our EVN images (Section~\ref{sec:results}, Figure~\ref{evnimages}) and the fitted brightness distribution model parameters (Table~\ref{phyparams}), the continuum radio spectra (Section~\ref{disc2}), the possible flux density variability (Section~\ref{disc3}), and the Gaia EDR3 optical coordinates (Section~\ref{disc4}). We also checked in the literature whether there are any available archival X-ray data for them. The criteria we considered for the classification and the results are summarized in Table~\ref{classification}. Two main classes are defined, FSRQs and steep-spectrum sources (SS). Based on the measured brightness temperatures as well as the 1.4-GHz monochromatic powers, the radio emission of all 13 sources originates from AGN activity (Section~\ref{disc1}). We also briefly comment on individual sources. 

\begin{deluxetable*}{lccccccc}
\tablenum{8}
\tablecaption{Summary of the classification of the target sources.}
\tablewidth{0pt}
\tablehead{
\colhead{ID} & \colhead{X-ray} & \colhead{Doppler-boosted} & \colhead{Total flux density} & \colhead{VLBI} & \colhead{\quad\quad Variable\quad\quad} & \colhead{Radio--optical} & \colhead{Classification} \\ 
\nocolhead{} & \nocolhead{} & \nocolhead{} & \colhead{spectrum} & \colhead{spectrum} & \nocolhead{} & \colhead{offset} & \nocolhead{}
}
\decimalcolnumbers
\startdata
J0304$+$0046 & Yes & Possibly  & Flat   & --    & Possibly  & --            & FSRQ? \\
J0851$+$1423 & No  & No        & Steep  & Flat  & Yes       & No            & SS \\
J0918$+$0636 & Yes & Possibly  & Inverted & Flat & Yes      & Yes           & FSRQ \\
J1006$+$4627 & No  & Possibly  & Inverted & Flat & Yes      & Slight        & FSRQ \\
J1037$+$1823 & No  & No        & Steep  & Flat  & Yes       & No            & SS \\
J1231$+$3816 & Yes & No        & Steep  & Flat  & Possibly  & No            & SS \\
J1307$+$1507 & No  & Possibly  & Steep  & Steep & Possibly  & No            & SS \\
J1309$+$5733 & Yes & Possibly  & Peaked & Steep & Yes       & No            & FSRQ \\
J1325$+$1123 & Yes & Yes       & Peaked & Steep & Possibly  & No            & FSRQ \\
J1412$+$0624 & Yes & Possibly  & Flat   & Flat  & Possibly  & Slight        & FSRQ \\
J1434$+$1628 & No  & Possibly  & Steep  & Flat  & Possibly  & No            & SS \\
J1520$+$1835 & No  & Possibly  & Steep  & Steep & Possibly  & Yes           & SS (CSO) \\
J1720$+$6028 & No  & Possibly  & Flat   & Steep & Yes       & Slight        & FSRQ \\
\enddata
\tablecomments{Col.~1 $-$ Source designation; Col.~2 $-$ X-ray detection; Col.~3$-$ Doppler boosting (`Possibly' means $T_{\rm b}$ is lower limit at either frequency); Col.~4 $-$ Spectral classification based on the collected total flux density data; Col.~5 $-$ Spectral classification based on the dual-frequency EVN data; Col.~6 $-$ Source variability; Col.~7 $-$ Radio--optical positional offset; Col.~8 $-$ Proposed classification; SS stands for steep spectrum, the question mark indicates if contradictory properties are found and discussed in Section~\ref{disc5}}
\label{classification}
\end{deluxetable*}

\textit{J0304$+$0046:} This source has reliable EVN data only at 1.7~GHz. A single compact component without any extended emission is present which could indicate a blazar core, and the flat total flux density spectrum suggests an FSRQ. However, in the absence of a Gaia optical position, a VLBI spectral index, and a clear evidence for Doppler-boosted emission, we cannot classify this object with complete certainty. Chandra X-ray measurement is presented in \citet{2019MNRAS.482.2016Z} with a $1.8$ lower limit to the photon index in the interval of 0.5$-$8 keV.

\textit{J0851$+$1423:} At 5~GHz, we see an indication of an extended structure in the northwest--southeast (NW--SE) direction. The faint NE feature might as well be a noise peak at $\sim 5 \sigma$. There is a $\sim 5$\,mas offset (within $3\sigma$) between the two radio positions. Based on the total flux density spectrum, we classify this as a steep-spectrum source. However, as it seems to be variable and its VLBI spectrum is flat, simultaneous sensitive multi-frequency follow-up VLBI observations would be ideal to securely confirm the nature of this source.

\textit{J0918$+$0636:} This source, together with a few others from our sample, can be found in a multi-wavelength study of \citet{2019MNRAS.489.2732I}. They gave an X-ray photon index of $1.3 \pm 0.4$ (between 0.5$-$10 keV) and derived a flat radio spectrum. Their classification is uncertain. Note that both the total flux density spectrum and the two-point VLBI spectrum (which is based on measurements about two decades apart) are likely affected by variability. The radio and optical coordinates have a significant offset of 5~mas in declination. The calibrator J0915$+$0745 shows compact emission thus it cannot explain the positional difference. The X-ray detection \citep{2004AJ....128..523B} strengthens the FSRQ classification.

\textit{J1006$+$4627:} There is a slight radio--optical offset, about 3~mas in the jet direction. The flux density upper limit at 150~MHz is consistent with the inverted/flat continuum spectrum. Only an upper limit to the X-ray flux from ROSAT data is available \citep{2005AJ....129.2519V}.

\textit{J1037$+$1823:} The total flux density spectum is steep based on the data points in the GHz range, but considering the upper limit at 150~MHz, it seems that the spectrum is in fact peaked. The VLBI position is different at the two observed frequencies, with an offset of $\sim 6$\,mas in the NW direction. However, this value is still within $3\sigma$. A likely explanation is that the phase-reference calibrator source (J1045$+$1735) has an extended jet structure roughly in this direction, which may not be fully accounted for. The agreement between the 5-GHz VLBI position and the Gaia position suggests that the 5-GHz component coincides with the AGN core.

\textit{J1231$+$3816:} This object is detected with Chandra \citep{2019MNRAS.482.2016Z} with derived photon index of $1.4 \pm 0.3$ between 0.5$-$8 keV. We classify it as a steep-spectrum source.

\textit{J1307$+$1507:} The 1.7- and 5-GHz VLBI positions have $\sim 5$\,mas ($2\sigma$) offset in the N--S direction. The Gaia optical position agrees well with the 5-GHz VLBI position.

\textit{J1309$+$5733:} The source is detected with the EVN at both frequencies but only short baselines between five European antennas were available at 1.7~GHz. Therefore, the angular resolution achieved is poorer than for other sources in the sample (Figure~\ref{evnimages}). To characterize the total flux density spectrum, we applied a log-parabolic function as, with a 150-MHz Low-Frequency Array (LOFAR) measurement \citep{2017AA...598A.104S} included, it gave a better fit than a simple power-law. The spectral peak is $(13 \pm 1)$~mJy at $(600 \pm 100)$~MHz. However, the VLBI flux densities are well above the total flux density spectrum (Figure~\ref{spectra}), suggesting high-amplitude variability. Based on this, the peaked shape of the total flux density spectrum constructed from non-simultaneous measurements may be caused by flux density variability, thus we classify the source as FSRQ. This is also supported by the Chandra X-ray measurement \citep{2004AJ....128..523B} with photon index with $1.8 \pm 0.5$ (between 0.5$-$8 keV) and the detection in the XMM-Newton Slew Survey \citep{2008AA...480..611S}.   

\textit{J1325$+$1123:} This is the second source where we fitted log-parabolic function to the total flux density data. A spectrum has a peak of $(72 \pm 17)$~mJy at $(1.5 \pm 0.2)$~GHz. The scatter of the flux density values suggests possible variability. The two radio positions have $\sim 4$\,mas offset ($< 3\sigma$), similarly to J1037$+$1823 and J1307$+$1507. The location of the 5-GHz peak falls in an extended part of the 1.7-GHz structure (Figure~\ref{evnimages}). In this case, the calibrator (J1327$+$1223) does not have an extended jet. The source is detected in X-rays with Chandra \citep{2004AJ....128..523B}. The photon index between 0.5$-$10 keV is $1.5 \pm 0.5$ as given by  \citet{2019MNRAS.489.2732I}, who found that the radio spectrum is peaked, similarly to our finding. Their classification of this source is uncertain.

\textit{J1412$+$0624:} This source is tentatively classified as FSRQ. However, there is no clear evidence for Doppler-boosted emission. There is a slight radio--optical positional offset, still within $< 3\sigma_\mathrm{pos}$, almost perpendicular to the structure seen in the 5-GHz image (Figure~\ref{evnimages}) which shows somewhat spurious extended features around the central core. It has Chandra X-ray detection \citep{2013ApJ...763..109W}. The source may be a quasar with jet inclination angle larger than that typical for blazars. \citet{2019MNRAS.489.2732I} gave an X-ray photon index of $1.6 \pm 0.5$ between 0.5$-$10 keV, and also found the radio spectrum flat. However, they were uncertain whether this source was a blazar.

\textit{J1434$+$1628:} It is the faintest source with the lowest radio powers in our sample (Table~\ref{phyparams}). Similarly to J1037$+$1823, the 150-MHz flux density upper limit suggests a peaked spectrum instead of the steep spectrum fitted to the GHz data points. We found no strong evidence for the blazar nature so we classify it as a misaligned radio source.

\textit{J1520$+$1835:} Arguably, it is the most puzzling source in our sample, and one of the faintest. Both the total flux density and VLBI spectra are ultra-steep ($\alpha < -1$). As in most other cases, however, low-frequency total flux density measurements would better constrain the spectral fit. The TGSS upper limit indeed suggests a peaked overall spectrum, in contrast to the steep spectrum fitted to the high-frequency data points in the GHz range. The most remarkable feature is that the Gaia optical position has a highly significant offset with respect to the VLBI position which is coincident at both frequencies. The angular separation between the optical and radio positions is $\sim 30$\,mas, which at the source redshift corresponds a projected linear distance of $250$\,pc. This alone suggests the presence of a CSO-like source whose brighter (NE) hotspot is detected with the EVN but the radio emission of the other lobe on the opposite (SW) side of the core is completely resolved out. In the 1.7-GHz image of J1520$+$1835, while the radio brightness peak is located at $(0,0)$ as for all other sources, the field is centered on the Gaia position marked by a red cross (Figure~\ref{evnimages}). The core, also invisible in the radio, is marked by the Gaia position. Another observed property seems to support this scenario: it is not uncommon that a CSO component has extremely steep VLBI spectrum in the optically thin regime \citep[e.g.][]{2013A&A...550A.113W}. 

An alternative explanation is a positional mismatch. Certainly, we cannot exclude the possibility that the radio and optical AGN are two different background/foreground sources seen in projection. However, given the very small angular separation, this possibility is extremely low. Therefore, we suggest that J1520$+$1835 is a CSO candidate. The number of high-redshift CSOs (and candidates) observed so far is extremely small \citep{2008AA...484L..39F, 2022MNRAS.tmp..166A}. Although limited observational data are available on this source, we note that qualitatively, the suspected morphology on the angular scale of tens of mas described above  makes it similar to the low-redshift CSO quasar PKS~1117$+$146 \citep{1998MNRAS.297..559B}. Sensitive follow-up e-MERLIN imaging observations could be able to confirm the presence of the suspected emission in the SW.

\textit{J1720$+$6028:} The flux density upper limit at 150~MHz does not contradict with the flat total flux density spectrum. The radio--optical positional offset is $\sim 5$\,mas in right ascension. However, the Gaia value has the highest uncertainty among our targets, $\sim 2$~mas in both coordinates. The radio--optical offset, taking the positional uncertainties into account, is close to the jet direction. The optical position can be associated with the fainter jet component seen in the 1.7-GHz image (Figure~\ref{evnimages}). All other parameters of this source suggest that it is an FSRQ.


\section{Summary} 
\label{sec:conclusion}

In this paper, we investigated 13 high-redshift ($4 < z < 4.5$) blazar candidates, to reveal the nature of these radio sources. To this end, we performed a series of dual-frequency (1.7 and 5~GHz) VLBI observations using the EVN and e-MERLIN arrays. Based on these high-resolution observations and additional low-resolution radio data obtained from the literature, we derived physical properties such as the origin of the radio emission, flux density variability, spectral indices, and the potential presence of Doppler-boosting in the jet. The Gaia EDR3 optical coordinates were also collected for 12 of the 13 sources in the sample. Moreover, we found X-ray detections in the literature for 6 sources. Based on the available information, we made an attempt to decide whether our target sources belong to blazars or AGN with misaligned radio jets. This sample increases the number of known VLBI-observed $z > 4$ radio AGN by about $25\%$.  

All sources could be classified based on the observed and derived properties. Out of the 13 radio sources studied, six objects (46\%) turned out to be blazars, with core--jet or compact core morphology. Six other sources (46\%) fall in the category of misaligned objects like steep-spectrum (GPS/MPS) sources. Finally, one source (8\%) is tentatively classified as blazar with some uncertainty. Although it shows more characteristics of an FSRQ, additional data are needed to confirm this classification. Previous studies of high-redshift radio AGN \citep{2016MNRAS.463.3260C,2017MNRAS.467..950C} reached similar conclusions for the blazar/non-blazar ratio in their sample. In this new $z>4$ radio quasar sample, J1520$+$1835 appears as a potential CSO candidate, based on the large offset of its VLBI radio and Gaia optical positions. To confirm this, high-sensitivity follow-up VLBI and e-MERLIN observations would be needed. Other sources are also good targets for follow-up observations. However, none of them seems particularly promising for long-term jet proper motion monitoring, as prominent jet structures are not detected, in accordance with the expectation for highly redshifted steep-spectrum radio features. 

High-resolution VLBI observations are essential for studies of jetted AGN in the early Universe. Combined with other data, e.g. precise Gaia optical astrometry or X-ray observations of jet emission, VLBI data give us a useful tool to classify $z > 4$ objects. An important conclusion of this study, in accordance with the earlier findings of \citet{2017MNRAS.467..950C}, is that a pre-selection of blazar candidates based on their infrared, optical or X-ray properties combined with the presence of strong radio emission detected with low-resolution observations does not guarantee their true blazar nature, as some of them turn out to be misaligned jetted sources.

\begin{acknowledgments}
The EVN is a joint facility of independent European, African, Asian and North American radio astronomy institutes. Scientific results from data presented in this publication are derived from the following EVN project code: EG102. The e-MERLIN is a National Facility operated by the University of Manchester at Jodrell Bank Observatory on behalf of STFC. The research leading to these results has received funding from the European Commission Horizon 2020 Research and Innovation Programme under grant agreement No. 730562 (RadioNet). We thank the Hungarian National Research, Development and Innovation Office (OTKA K134213 and 2018-2.1.14-T\'ET-CN-2018-00001) for support. H. M. C. acknowledges support by the National Natural Science  Foundation of China (Grants No. U2031116 and U1731103). 
This work presents results from the European Space Agency (ESA) space mission Gaia. Gaia data are being processed by the Gaia Data Processing and Analysis Consortium (DPAC). Funding for the DPAC is provided by national institutions, in particular the institutions participating in the Gaia MultiLateral Agreement (MLA). The Gaia mission website is \url{https://www.cosmos.esa.int/gaia}. The Gaia archive website is \url{https://archives.esac.esa.int/gaia}.
This research has made use of the NASA/IPAC Extragalactic Database (NED) which is operated by the Jet Propulsion Laboratory, California Institute of Technology, under contract with the National Aeronautics and Space Administration.
This research has made use of the VizieR catalog access tool, CDS, Strasbourg, France (DOI: 10.26093/cds/vizier). The original description  of the VizieR service was published in \citet{2000A&AS..143...23O}. 
\end{acknowledgments}

%

\vspace{5mm}
\facilities{EVN, e-MERLIN}


\software{
{\sc aips} \citep{2003ASSL..285..109G},
{\sc Difmap} \citep{1997ASPC..125...77S},
{\sc Astropy} \citep{2013A&A...558A..33A,2018AJ....156..123A},
{\sc Matplotlib} \citep{2007CSE.....9...90H}
}





\bibliography{EG102}{}

\begin{thebibliography}{}
\expandafter\ifx\csname natexlab\endcsname\relax\def\natexlab#1{#1}\fi
\providecommand{\url}[1]{\href{#1}{#1}}
\providecommand{\dodoi}[1]{doi:~\href{http://doi.org/#1}{\nolinkurl{#1}}}
\providecommand{\doeprint}[1]{\href{http://ascl.net/#1}{\nolinkurl{http://ascl.net/#1}}}
\providecommand{\doarXiv}[1]{\href{https://arxiv.org/abs/#1}{\nolinkurl{https://arxiv.org/abs/#1}}}

\bibitem[{{Ajello} {et~al.}(2009){Ajello}, {Costamante}, {Sambruna}, {Gehrels},
  {Chiang}, {Rau}, {Escala}, {Greiner}, {Tueller}, {Wall}, \&
  {Mushotzky}}]{2009ApJ...699..603A}
{Ajello}, M., {Costamante}, L., {Sambruna}, R.~M., {et~al.} 2009, \apj, 699,
  603, \dodoi{10.1088/0004-637X/699/1/603}

\bibitem[{{Alef} \& {Porcas}(1986)}]{1986A&A...168..365A}
{Alef}, W., \& {Porcas}, R.~W. 1986, \aap, 168, 365

\bibitem[{{Alexandroff} {et~al.}(2012){Alexandroff}, {Overzier}, {Paragi},
  {Basu-Zych}, {Heckman}, {Kauffmann}, {Bourke}, {Lobanov}, {Ptak}, \&
  {Schiminovich}}]{2012MNRAS.423.1325A}
{Alexandroff}, R., {Overzier}, R.~A., {Paragi}, Z., {et~al.} 2012, \mnras, 423,
  1325, \dodoi{10.1111/j.1365-2966.2012.20959.x}

\bibitem[{{An} \& {Baan}(2012)}]{2012ApJ...760...77A}
{An}, T., \& {Baan}, W.~A. 2012, \apj, 760, 77,
  \dodoi{10.1088/0004-637X/760/1/77}

\bibitem[{{An} {et~al.}(2022){An}, {Wang}, {Zhang}, {Aditya}, {Hong}, \&
  {Cui}}]{2022MNRAS.tmp..166A}
{An}, T., {Wang}, A., {Zhang}, Y., {et~al.} 2022, \mnras,
  \dodoi{10.1093/mnras/stac205}

\bibitem[{{An} {et~al.}(2012){An}, {Wu}, {Yang}, {Taylor}, {Hong}, {Baan},
  {Liu}, {Wang}, {Zhang}, {Wang}, {Chen}, {Cui}, {Hao}, \&
  {Zhu}}]{2012ApJS..198....5A}
{An}, T., {Wu}, F., {Yang}, J., {et~al.} 2012, \apjs, 198, 5,
  \dodoi{10.1088/0067-0049/198/1/5}

\bibitem[{{An} {et~al.}(2020){An}, {Mohan}, {Zhang}, {Frey}, {Yang},
  {Gab{\'a}nyi}, {Gurvits}, {Paragi}, {Perger}, \&
  {Zheng}}]{2020NatCo..11..143A}
{An}, T., {Mohan}, P., {Zhang}, Y., {et~al.} 2020, Nature Communications, 11,
  143, \dodoi{10.1038/s41467-019-14093-2}

\bibitem[{{Astropy Collaboration} {et~al.}(2013){Astropy Collaboration},
  {Robitaille}, {Tollerud}, {Greenfield}, {Droettboom}, {Bray}, {Aldcroft},
  {Davis}, {Ginsburg}, {Price-Whelan}, {Kerzendorf}, {Conley}, {Crighton},
  {Barbary}, {Muna}, {Ferguson}, {Grollier}, {Parikh}, {Nair}, {Unther},
  {Deil}, {Woillez}, {Conseil}, {Kramer}, {Turner}, {Singer}, {Fox}, {Weaver},
  {Zabalza}, {Edwards}, {Azalee Bostroem}, {Burke}, {Casey}, {Crawford},
  {Dencheva}, {Ely}, {Jenness}, {Labrie}, {Lim}, {Pierfederici}, {Pontzen},
  {Ptak}, {Refsdal}, {Servillat}, \& {Streicher}}]{2013A&A...558A..33A}
{Astropy Collaboration}, {Robitaille}, T.~P., {Tollerud}, E.~J., {et~al.} 2013,
  \aap, 558, A33, \dodoi{10.1051/0004-6361/201322068}

\bibitem[{{Astropy Collaboration} {et~al.}(2018){Astropy Collaboration},
  {Price-Whelan}, {Sip{\H{o}}cz}, {G{\"u}nther}, {Lim}, {Crawford}, {Conseil},
  {Shupe}, {Craig}, {Dencheva}, {Ginsburg}, {VanderPlas}, {Bradley},
  {P{\'e}rez-Su{\'a}rez}, {de Val-Borro}, {Aldcroft}, {Cruz}, {Robitaille},
  {Tollerud}, {Ardelean}, {Babej}, {Bach}, {Bachetti}, {Bakanov}, {Bamford},
  {Barentsen}, {Barmby}, {Baumbach}, {Berry}, {Biscani}, {Boquien}, {Bostroem},
  {Bouma}, {Brammer}, {Bray}, {Breytenbach}, {Buddelmeijer}, {Burke},
  {Calderone}, {Cano Rodr{\'\i}guez}, {Cara}, {Cardoso}, {Cheedella}, {Copin},
  {Corrales}, {Crichton}, {D'Avella}, {Deil}, {Depagne}, {Dietrich}, {Donath},
  {Droettboom}, {Earl}, {Erben}, {Fabbro}, {Ferreira}, {Finethy}, {Fox},
  {Garrison}, {Gibbons}, {Goldstein}, {Gommers}, {Greco}, {Greenfield},
  {Groener}, {Grollier}, {Hagen}, {Hirst}, {Homeier}, {Horton}, {Hosseinzadeh},
  {Hu}, {Hunkeler}, {Ivezi{\'c}}, {Jain}, {Jenness}, {Kanarek}, {Kendrew},
  {Kern}, {Kerzendorf}, {Khvalko}, {King}, {Kirkby}, {Kulkarni}, {Kumar},
  {Lee}, {Lenz}, {Littlefair}, {Ma}, {Macleod}, {Mastropietro}, {McCully},
  {Montagnac}, {Morris}, {Mueller}, {Mumford}, {Muna}, {Murphy}, {Nelson},
  {Nguyen}, {Ninan}, {N{\"o}the}, {Ogaz}, {Oh}, {Parejko}, {Parley}, {Pascual},
  {Patil}, {Patil}, {Plunkett}, {Prochaska}, {Rastogi}, {Reddy Janga},
  {Sabater}, {Sakurikar}, {Seifert}, {Sherbert}, {Sherwood-Taylor}, {Shih},
  {Sick}, {Silbiger}, {Singanamalla}, {Singer}, {Sladen}, {Sooley},
  {Sornarajah}, {Streicher}, {Teuben}, {Thomas}, {Tremblay}, {Turner},
  {Terr{\'o}n}, {van Kerkwijk}, {de la Vega}, {Watkins}, {Weaver}, {Whitmore},
  {Woillez}, {Zabalza}, \& {Astropy Contributors}}]{2018AJ....156..123A}
{Astropy Collaboration}, {Price-Whelan}, A.~M., {Sip{\H{o}}cz}, B.~M., {et~al.}
  2018, \aj, 156, 123, \dodoi{10.3847/1538-3881/aabc4f}

\bibitem[{{Bassett} {et~al.}(2004){Bassett}, {Brandt}, {Schneider}, {Vignali},
  {Chartas}, \& {Garmire}}]{2004AJ....128..523B}
{Bassett}, L.~C., {Brandt}, W.~N., {Schneider}, D.~P., {et~al.} 2004, \aj, 128,
  523, \dodoi{10.1086/422019}

\bibitem[{{Beasley} \& {Conway}(1995)}]{1995ASPC...82..327B}
{Beasley}, A.~J., \& {Conway}, J.~E. 1995, in Astronomical Society of the
  Pacific Conference Series, Vol.~82, Very Long Baseline Interferometry and the
  VLBA, ed. J.~A. {Zensus}, P.~J. {Diamond}, \& P.~J. {Napier}, 327

\bibitem[{{Beasley} {et~al.}(2002){Beasley}, {Gordon}, {Peck}, {Petrov},
  {MacMillan}, {Fomalont}, \& {Ma}}]{2002ApJS..141...13B}
{Beasley}, A.~J., {Gordon}, D., {Peck}, A.~B., {et~al.} 2002, \apjs, 141, 13,
  \dodoi{10.1086/339806}

\bibitem[{{Becker} {et~al.}(1991){Becker}, {White}, \&
  {Edwards}}]{1991ApJS...75....1B}
{Becker}, R.~H., {White}, R.~L., \& {Edwards}, A.~L. 1991, \apjs, 75, 1,
  \dodoi{10.1086/191529}

\bibitem[{{Becker} {et~al.}(1995){Becker}, {White}, \&
  {Helfand}}]{1995ApJ...450..559B}
{Becker}, R.~H., {White}, R.~L., \& {Helfand}, D.~J. 1995, \apj, 450, 559,
  \dodoi{10.1086/176166}

\bibitem[{{Best} {et~al.}(2005){Best}, {Kauffmann}, {Heckman}, {Brinchmann},
  {Charlot}, {Ivezi{\'c}}, \& {White}}]{2005MNRAS.362...25B}
{Best}, P.~N., {Kauffmann}, G., {Heckman}, T.~M., {et~al.} 2005, \mnras, 362,
  25, \dodoi{10.1111/j.1365-2966.2005.09192.x}

\bibitem[{{Bondi} {et~al.}(1998){Bondi}, {Garrett}, \&
  {Gurvits}}]{1998MNRAS.297..559B}
{Bondi}, M., {Garrett}, M.~A., \& {Gurvits}, L.~I. 1998, \mnras, 297, 559,
  \dodoi{10.1046/j.1365-8711.1998.01518.x}

\bibitem[{{Caccianiga} {et~al.}(2019){Caccianiga}, {Moretti}, {Belladitta},
  {Della Ceca}, {Ant{\'o}n}, {Ballo}, {Cicone}, {Dallacasa}, {Gargiulo},
  {Ighina}, {March{\~a}}, \& {Severgnini}}]{2019MNRAS.484..204C}
{Caccianiga}, A., {Moretti}, A., {Belladitta}, S., {et~al.} 2019, \mnras, 484,
  204, \dodoi{10.1093/mnras/sty3526}

\bibitem[{{Cao} {et~al.}(2017){Cao}, {Frey}, {Gab{\'a}nyi}, {Paragi}, {Yang},
  {Cseh}, {Hong}, \& {An}}]{2017MNRAS.467..950C}
{Cao}, H.~M., {Frey}, S., {Gab{\'a}nyi}, K.~{\'E}., {et~al.} 2017, \mnras, 467,
  950, \dodoi{10.1093/mnras/stx160}

\bibitem[{{Cao} {et~al.}(2014){Cao}, {Frey}, {Gurvits}, {Yang}, {Hong},
  {Paragi}, {Deller}, \& {Ivezi{\'c}}}]{2014AA...563A.111C}
{Cao}, H.~M., {Frey}, S., {Gurvits}, L.~I., {et~al.} 2014, \aap, 563, A111,
  \dodoi{10.1051/0004-6361/201323328}

\bibitem[{{Cheng} {et~al.}(2020){Cheng}, {An}, {Frey}, {Hong}, {He},
  {Kellermann}, {Lister}, {Lao}, {Li}, {Mohan}, {Yang}, {Wu}, {Zhang}, {Zhang},
  \& {Zhao}}]{2020ApJS..247...57C}
{Cheng}, X.~P., {An}, T., {Frey}, S., {et~al.} 2020, \apjs, 247, 57,
  \dodoi{10.3847/1538-4365/ab791f}

\bibitem[{{Cheung} {et~al.}(2012){Cheung}, {Stawarz}, {Siemiginowska},
  {Gobeille}, {Wardle}, {Harris}, \& {Schwartz}}]{2012ApJ...756L..20C}
{Cheung}, C.~C., {Stawarz}, {\L}., {Siemiginowska}, A., {et~al.} 2012, \apjl,
  756, L20, \dodoi{10.1088/2041-8205/756/1/L20}

\bibitem[{{Condon} {et~al.}(1982){Condon}, {Condon}, {Gisler}, \&
  {Puschell}}]{1982ApJ...252..102C}
{Condon}, J.~J., {Condon}, M.~A., {Gisler}, G., \& {Puschell}, J.~J. 1982,
  \apj, 252, 102, \dodoi{10.1086/159538}

\bibitem[{{Condon} {et~al.}(1998){Condon}, {Cotton}, {Greisen}, {Yin},
  {Perley}, {Taylor}, \& {Broderick}}]{1998AJ....115.1693C}
{Condon}, J.~J., {Cotton}, W.~D., {Greisen}, E.~W., {et~al.} 1998, \aj, 115,
  1693, \dodoi{10.1086/300337}

\bibitem[{{Coppejans} {et~al.}(2016){Coppejans}, {Frey}, {Cseh}, {M{\"u}ller},
  {Paragi}, {Falcke}, {Gab{\'a}nyi}, {Gurvits}, {An}, \&
  {Titov}}]{2016MNRAS.463.3260C}
{Coppejans}, R., {Frey}, S., {Cseh}, D., {et~al.} 2016, \mnras, 463, 3260,
  \dodoi{10.1093/mnras/stw2236}

\bibitem[{{Coppejans} {et~al.}(2017){Coppejans}, {van Velzen}, {Intema},
  {M{\"u}ller}, {Frey}, {Coppejans}, {Cseh}, {Williams}, {Falcke},
  {K{\"o}rding}, {Orr{\'u}}, {Paragi}, \& {Gab{\'a}nyi}}]{2017MNRAS.467.2039C}
{Coppejans}, R., {van Velzen}, S., {Intema}, H.~T., {et~al.} 2017, \mnras, 467,
  2039, \dodoi{10.1093/mnras/stx215}

\bibitem[{{de Bruyn} {et~al.}(2000){de Bruyn}, {Miley}, {Rengelink}, {Tang},
  {Bremer}, {Rottgering}, {Raimond}, {Bremer}, \&
  {Fullagar}}]{2000yCat.8062....0D}
{de Bruyn}, G., {Miley}, G., {Rengelink}, R., {et~al.} 2000, VizieR Online Data
  Catalog, VIII/62

\bibitem[{{Fabian}(2012)}]{2012ARA&A..50..455F}
{Fabian}, A.~C. 2012, \araa, 50, 455,
  \dodoi{10.1146/annurev-astro-081811-125521}

\bibitem[{{Fan} {et~al.}(2020){Fan}, {Chen}, {An}, {Xie}, {Han}, {Knudsen}, \&
  {Yang}}]{2020ApJ...905L..32F}
{Fan}, L., {Chen}, W., {An}, T., {et~al.} 2020, \apjl, 905, L32,
  \dodoi{10.3847/2041-8213/abcebf}

\bibitem[{{Fey} {et~al.}(2004){Fey}, {Ma}, {Arias}, {Charlot},
  {Feissel-Vernier}, {Gontier}, {Jacobs}, {Li}, \&
  {MacMillan}}]{2004AJ....127.3587F}
{Fey}, A.~L., {Ma}, C., {Arias}, E.~F., {et~al.} 2004, \aj, 127, 3587,
  \dodoi{10.1086/420998}

\bibitem[{{Fomalont}(1999)}]{1999ASPC..180..301F}
{Fomalont}, E.~B. 1999, in Astronomical Society of the Pacific Conference
  Series, Vol. 180, Synthesis Imaging in Radio Astronomy II, ed. G.~B.
  {Taylor}, C.~L. {Carilli}, \& R.~A. {Perley}, 301

\bibitem[{{Fomalont} {et~al.}(2003){Fomalont}, {Petrov}, {MacMillan}, {Gordon},
  \& {Ma}}]{2003AJ....126.2562F}
{Fomalont}, E.~B., {Petrov}, L., {MacMillan}, D.~S., {Gordon}, D., \& {Ma}, C.
  2003, \aj, 126, 2562, \dodoi{10.1086/378712}

\bibitem[{{Fossati} {et~al.}(1999){Fossati}, {Celotti}, {Ghisellini}, \&
  {Maraschi}}]{1999ASPC..159..351F}
{Fossati}, G., {Celotti}, A., {Ghisellini}, G., \& {Maraschi}, L. 1999, in
  Astronomical Society of the Pacific Conference Series, Vol. 159, BL Lac
  Phenomenon, ed. L.~O. {Takalo} \& A.~{Sillanp{\"a}{\"a}}, 351.
\newblock \doarXiv{astro-ph/9812158}

\bibitem[{{Frey} {et~al.}(2013){Frey}, {Fogasy}, {Paragi}, \&
  {Gurvits}}]{2013MNRAS.431.1314F}
{Frey}, S., {Fogasy}, J.~O., {Paragi}, Z., \& {Gurvits}, L.~I. 2013, \mnras,
  431, 1314, \dodoi{10.1093/mnras/stt249}

\bibitem[{{Frey} {et~al.}(1997){Frey}, {Gurvits}, {Kellermann}, {Schilizzi}, \&
  {Pauliny-Toth}}]{1997AA...325..511F}
{Frey}, S., {Gurvits}, L.~I., {Kellermann}, K.~I., {Schilizzi}, R.~T., \&
  {Pauliny-Toth}, I.~I.~K. 1997, \aap, 325, 511

\bibitem[{{Frey} {et~al.}(2008){Frey}, {Gurvits}, {Paragi}, \& {{\'E}.
  Gab{\'a}nyi}}]{2008AA...484L..39F}
{Frey}, S., {Gurvits}, L.~I., {Paragi}, Z., \& {{\'E}. Gab{\'a}nyi}, K. 2008,
  \aap, 484, L39, \dodoi{10.1051/0004-6361:200810040}

\bibitem[{{Frey} {et~al.}(2003){Frey}, {Mosoni}, {Paragi}, \&
  {Gurvits}}]{2003MNRAS.343L..20F}
{Frey}, S., {Mosoni}, L., {Paragi}, Z., \& {Gurvits}, L.~I. 2003, \mnras, 343,
  L20, \dodoi{10.1046/j.1365-8711.2003.06869.x}

\bibitem[{{Frey} {et~al.}(2015){Frey}, {Paragi}, {Fogasy}, \&
  {Gurvits}}]{2015MNRAS.446.2921F}
{Frey}, S., {Paragi}, Z., {Fogasy}, J.~O., \& {Gurvits}, L.~I. 2015, \mnras,
  446, 2921, \dodoi{10.1093/mnras/stu2294}

\bibitem[{{Frey} {et~al.}(2010){Frey}, {Paragi}, {Gurvits}, {Cseh}, \&
  {Gab{\'a}nyi}}]{2010AA...524A..83F}
{Frey}, S., {Paragi}, Z., {Gurvits}, L.~I., {Cseh}, D., \& {Gab{\'a}nyi},
  K.~{\'E}. 2010, \aap, 524, A83, \dodoi{10.1051/0004-6361/201015554}

\bibitem[{{Frey} {et~al.}(2011){Frey}, {Paragi}, {Gurvits}, {Gab{\'a}nyi}, \&
  {Cseh}}]{2011AA...531L...5F}
{Frey}, S., {Paragi}, Z., {Gurvits}, L.~I., {Gab{\'a}nyi}, K.~{\'E}., \&
  {Cseh}, D. 2011, \aap, 531, L5, \dodoi{10.1051/0004-6361/201117341}

\bibitem[{{Frey} {et~al.}(2005){Frey}, {Paragi}, {Mosoni}, \&
  {Gurvits}}]{2005AA...436L..13F}
{Frey}, S., {Paragi}, Z., {Mosoni}, L., \& {Gurvits}, L.~I. 2005, \aap, 436,
  L13, \dodoi{10.1051/0004-6361:200500112}

\bibitem[{{Frey} {et~al.}(2018){Frey}, {Titov}, {Melnikov}, {de Vicente}, \&
  {Shu}}]{2018AA...618A..68F}
{Frey}, S., {Titov}, O., {Melnikov}, A.~E., {de Vicente}, P., \& {Shu}, F.
  2018, \aap, 618, A68, \dodoi{10.1051/0004-6361/201832771}

\bibitem[{{Gab{\'a}nyi} {et~al.}(2015){Gab{\'a}nyi}, {Cseh}, {Frey}, {Paragi},
  {Gurvits}, {An}, \& {Zhang}}]{2015MNRAS.450L..57G}
{Gab{\'a}nyi}, K.~{\'E}., {Cseh}, D., {Frey}, S., {et~al.} 2015, \mnras, 450,
  L57, \dodoi{10.1093/mnrasl/slv046}

\bibitem[{{Gab{\'a}nyi} {et~al.}(2018){Gab{\'a}nyi}, {Frey}, {Gurvits},
  {Paragi}, \& {Perger}}]{2018RNAAS...2..200G}
{Gab{\'a}nyi}, K.~{\'E}., {Frey}, S., {Gurvits}, L.~I., {Paragi}, Z., \&
  {Perger}, K. 2018, Research Notes of the American Astronomical Society, 2,
  200, \dodoi{10.3847/2515-5172/aaec82}

\bibitem[{{Gab{\'a}nyi} {et~al.}(2019){Gab{\'a}nyi}, {Frey}, {Satyapal},
  {Constantin}, \& {Pfeifle}}]{2019A&A...630L...5G}
{Gab{\'a}nyi}, K.~{\'E}., {Frey}, S., {Satyapal}, S., {Constantin}, A., \&
  {Pfeifle}, R.~W. 2019, \aap, 630, L5, \dodoi{10.1051/0004-6361/201936519}

\bibitem[{{Gab{\'a}nyi} {et~al.}(2021){Gab{\'a}nyi}, {Frey}, {An}, {Cao},
  {Paragi}, {Gurvits}, {Zhang}, {Sbarrato}, {Krezinger}, {Perger}, \&
  {Mez{\"o}}}]{2021AN....342.1092G}
{Gab{\'a}nyi}, K.~{\'E}., {Frey}, S., {An}, T., {et~al.} 2021, Astronomische
  Nachrichten, 342, 1092, \dodoi{10.1002/asna.20210057}

\bibitem[{{Gaia Collaboration} {et~al.}(2016){Gaia Collaboration}, {Prusti},
  {de Bruijne}, {Brown}, {Vallenari}, {Babusiaux}, {Bailer-Jones}, {Bastian},
  {Biermann}, {Evans}, {Eyer}, {Jansen}, {Jordi}, {Klioner}, {Lammers},
  {Lindegren}, {Luri}, {Mignard}, {Milligan}, {Panem}, {Poinsignon},
  {Pourbaix}, {Randich}, {Sarri}, {Sartoretti}, {Siddiqui}, {Soubiran},
  {Valette}, {van Leeuwen}, {Walton}, {Aerts}, {Arenou}, {Cropper}, {Drimmel},
  {H{\o}g}, {Katz}, {Lattanzi}, {O'Mullane}, {Grebel}, {Holland}, {Huc},
  {Passot}, {Bramante}, {Cacciari}, {Casta{\~n}eda}, {Chaoul}, {Cheek}, {De
  Angeli}, {Fabricius}, {Guerra}, {Hern{\'a}ndez}, {Jean-Antoine-Piccolo},
  {Masana}, {Messineo}, {Mowlavi}, {Nienartowicz}, {Ord{\'o}{\~n}ez-Blanco},
  {Panuzzo}, {Portell}, {Richards}, {Riello}, {Seabroke}, {Tanga},
  {Th{\'e}venin}, {Torra}, {Els}, {Gracia-Abril}, {Comoretto},
  {Garcia-Reinaldos}, {Lock}, {Mercier}, {Altmann}, {Andrae}, {Astraatmadja},
  {Bellas-Velidis}, {Benson}, {Berthier}, {Blomme}, {Busso}, {Carry},
  {Cellino}, {Clementini}, {Cowell}, {Creevey}, {Cuypers}, {Davidson}, {De
  Ridder}, {de Torres}, {Delchambre}, {Dell'Oro}, {Ducourant}, {Fr{\'e}mat},
  {Garc{\'\i}a-Torres}, {Gosset}, {Halbwachs}, {Hambly}, {Harrison}, {Hauser},
  {Hestroffer}, {Hodgkin}, {Huckle}, {Hutton}, {Jasniewicz}, {Jordan},
  {Kontizas}, {Korn}, {Lanzafame}, {Manteiga}, {Moitinho}, {Muinonen},
  {Osinde}, {Pancino}, {Pauwels}, {Petit}, {Recio-Blanco}, {Robin}, {Sarro},
  {Siopis}, {Smith}, {Smith}, {Sozzetti}, {Thuillot}, {van Reeven}, {Viala},
  {Abbas}, {Abreu Aramburu}, {Accart}, {Aguado}, {Allan}, {Allasia},
  {Altavilla}, {{\'A}lvarez}, {Alves}, {Anderson}, {Andrei}, {Anglada Varela},
  {Antiche}, {Antoja}, {Ant{\'o}n}, {Arcay}, {Atzei}, {Ayache}, {Bach},
  {Baker}, {Balaguer-N{\'u}{\~n}ez}, {Barache}, {Barata}, {Barbier}, {Barblan},
  {Baroni}, {Barrado y Navascu{\'e}s}, {Barros}, {Barstow}, {Becciani},
  {Bellazzini}, {Bellei}, {Bello Garc{\'\i}a}, {Belokurov}, {Bendjoya},
  {Berihuete}, {Bianchi}, {Bienaym{\'e}}, {Billebaud}, {Blagorodnova},
  {Blanco-Cuaresma}, {Boch}, {Bombrun}, {Borrachero}, {Bouquillon}, {Bourda},
  {Bouy}, {Bragaglia}, {Breddels}, {Brouillet}, {Br{\"u}semeister},
  {Bucciarelli}, {Budnik}, {Burgess}, {Burgon}, {Burlacu}, {Busonero}, {Buzzi},
  {Caffau}, {Cambras}, {Campbell}, {Cancelliere}, {Cantat-Gaudin}, {Carlucci},
  {Carrasco}, {Castellani}, {Charlot}, {Charnas}, {Charvet}, {Chassat},
  {Chiavassa}, {Clotet}, {Cocozza}, {Collins}, {Collins}, {Costigan}, {Crifo},
  {Cross}, {Crosta}, {Crowley}, {Dafonte}, {Damerdji}, {Dapergolas}, {David},
  {David}, {De Cat}, {de Felice}, {de Laverny}, {De Luise}, {De March}, {de
  Martino}, {de Souza}, {Debosscher}, {del Pozo}, {Delbo}, {Delgado},
  {Delgado}, {di Marco}, {Di Matteo}, {Diakite}, {Distefano}, {Dolding}, {Dos
  Anjos}, {Drazinos}, {Dur{\'a}n}, {Dzigan}, {Ecale}, {Edvardsson}, {Enke},
  {Erdmann}, {Escolar}, {Espina}, {Evans}, {Eynard Bontemps}, {Fabre},
  {Fabrizio}, {Faigler}, {Falc{\~a}o}, {Farr{\`a}s Casas}, {Faye}, {Federici},
  {Fedorets}, {Fern{\'a}ndez-Hern{\'a}ndez}, {Fernique}, {Fienga}, {Figueras},
  {Filippi}, {Findeisen}, {Fonti}, {Fouesneau}, {Fraile}, {Fraser}, {Fuchs},
  {Furnell}, {Gai}, {Galleti}, {Galluccio}, {Garabato}, {Garc{\'\i}a-Sedano},
  {Gar{\'e}}, {Garofalo}, {Garralda}, {Gavras}, {Gerssen}, {Geyer}, {Gilmore},
  {Girona}, {Giuffrida}, {Gomes}, {Gonz{\'a}lez-Marcos},
  {Gonz{\'a}lez-N{\'u}{\~n}ez}, {Gonz{\'a}lez-Vidal}, {Granvik}, {Guerrier},
  {Guillout}, {Guiraud}, {G{\'u}rpide}, {Guti{\'e}rrez-S{\'a}nchez}, {Guy},
  {Haigron}, {Hatzidimitriou}, {Haywood}, {Heiter}, {Helmi}, {Hobbs},
  {Hofmann}, {Holl}, {Holland}, {Hunt}, {Hypki}, {Icardi}, {Irwin}, {Jevardat
  de Fombelle}, {Jofr{\'e}}, {Jonker}, {Jorissen}, {Julbe}, {Karampelas},
  {Kochoska}, {Kohley}, {Kolenberg}, {Kontizas}, {Koposov}, {Kordopatis},
  {Koubsky}, {Kowalczyk}, {Krone-Martins}, {Kudryashova}, {Kull}, {Bachchan},
  {Lacoste-Seris}, {Lanza}, {Lavigne}, {Le Poncin-Lafitte}, {Lebreton},
  {Lebzelter}, {Leccia}, {Leclerc}, {Lecoeur-Taibi}, {Lemaitre}, {Lenhardt},
  {Leroux}, {Liao}, {Licata}, {Lindstr{\o}m}, {Lister}, {Livanou}, {Lobel},
  {L{\"o}ffler}, {L{\'o}pez}, {Lopez-Lozano}, {Lorenz}, {Loureiro},
  {MacDonald}, {Magalh{\~a}es Fernandes}, {Managau}, {Mann}, {Mantelet},
  {Marchal}, {Marchant}, {Marconi}, {Marie}, {Marinoni}, {Marrese},
  {Marschalk{\'o}}, {Marshall}, {Mart{\'\i}n-Fleitas}, {Martino}, {Mary},
  {Matijevi{\v{c}}}, {Mazeh}, {McMillan}, {Messina}, {Mestre}, {Michalik},
  {Millar}, {Miranda}, {Molina}, {Molinaro}, {Molinaro}, {Moln{\'a}r},
  {Moniez}, {Montegriffo}, {Monteiro}, {Mor}, {Mora}, {Morbidelli}, {Morel},
  {Morgenthaler}, {Morley}, {Morris}, {Mulone}, {Muraveva}, {Musella},
  {Narbonne}, {Nelemans}, {Nicastro}, {Noval}, {Ord{\'e}novic},
  {Ordieres-Mer{\'e}}, {Osborne}, {Pagani}, {Pagano}, {Pailler}, {Palacin},
  {Palaversa}, {Parsons}, {Paulsen}, {Pecoraro}, {Pedrosa}, {Pentik{\"a}inen},
  {Pereira}, {Pichon}, {Piersimoni}, {Pineau}, {Plachy}, {Plum}, {Poujoulet},
  {Pr{\v{s}}a}, {Pulone}, {Ragaini}, {Rago}, {Rambaux}, {Ramos-Lerate},
  {Ranalli}, {Rauw}, {Read}, {Regibo}, {Renk}, {Reyl{\'e}}, {Ribeiro},
  {Rimoldini}, {Ripepi}, {Riva}, {Rixon}, {Roelens}, {Romero-G{\'o}mez},
  {Rowell}, {Royer}, {Rudolph}, {Ruiz-Dern}, {Sadowski}, {Sagrist{\`a}
  Sell{\'e}s}, {Sahlmann}, {Salgado}, {Salguero}, {Sarasso}, {Savietto},
  {Schnorhk}, {Schultheis}, {Sciacca}, {Segol}, {Segovia}, {Segransan},
  {Serpell}, {Shih}, {Smareglia}, {Smart}, {Smith}, {Solano}, {Solitro},
  {Sordo}, {Soria Nieto}, {Souchay}, {Spagna}, {Spoto}, {Stampa}, {Steele},
  {Steidelm{\"u}ller}, {Stephenson}, {Stoev}, {Suess}, {S{\"u}veges}, {Surdej},
  {Szabados}, {Szegedi-Elek}, {Tapiador}, {Taris}, {Tauran}, {Taylor},
  {Teixeira}, {Terrett}, {Tingley}, {Trager}, {Turon}, {Ulla}, {Utrilla},
  {Valentini}, {van Elteren}, {Van Hemelryck}, {van Leeuwen}, {Varadi},
  {Vecchiato}, {Veljanoski}, {Via}, {Vicente}, {Vogt}, {Voss}, {Votruba},
  {Voutsinas}, {Walmsley}, {Weiler}, {Weingrill}, {Werner}, {Wevers},
  {Whitehead}, {Wyrzykowski}, {Yoldas}, {{\v{Z}}erjal}, {Zucker}, {Zurbach},
  {Zwitter}, {Alecu}, {Allen}, {Allende Prieto}, {Amorim},
  {Anglada-Escud{\'e}}, {Arsenijevic}, {Azaz}, {Balm}, {Beck}, {Bernstein},
  {Bigot}, {Bijaoui}, {Blasco}, {Bonfigli}, {Bono}, {Boudreault}, {Bressan},
  {Brown}, {Brunet}, {Bunclark}, {Buonanno}, {Butkevich}, {Carret}, {Carrion},
  {Chemin}, {Ch{\'e}reau}, {Corcione}, {Darmigny}, {de Boer}, {de Teodoro}, {de
  Zeeuw}, {Delle Luche}, {Domingues}, {Dubath}, {Fodor}, {Fr{\'e}zouls},
  {Fries}, {Fustes}, {Fyfe}, {Gallardo}, {Gallegos}, {Gardiol}, {Gebran},
  {Gomboc}, {G{\'o}mez}, {Grux}, {Gueguen}, {Heyrovsky}, {Hoar}, {Iannicola},
  {Isasi Parache}, {Janotto}, {Joliet}, {Jonckheere}, {Keil}, {Kim},
  {Klagyivik}, {Klar}, {Knude}, {Kochukhov}, {Kolka}, {Kos}, {Kutka}, {Lainey},
  {LeBouquin}, {Liu}, {Loreggia}, {Makarov}, {Marseille}, {Martayan},
  {Martinez-Rubi}, {Massart}, {Meynadier}, {Mignot}, {Munari}, {Nguyen},
  {Nordlander}, {Ocvirk}, {O'Flaherty}, {Olias Sanz}, {Ortiz}, {Osorio},
  {Oszkiewicz}, {Ouzounis}, {Palmer}, {Park}, {Pasquato}, {Peltzer}, {Peralta},
  {P{\'e}turaud}, {Pieniluoma}, {Pigozzi}, {Poels}, {Prat}, {Prod'homme},
  {Raison}, {Rebordao}, {Risquez}, {Rocca-Volmerange}, {Rosen}, {Ruiz-Fuertes},
  {Russo}, {Sembay}, {Serraller Vizcaino}, {Short}, {Siebert}, {Silva},
  {Sinachopoulos}, {Slezak}, {Soffel}, {Sosnowska}, {Strai{\v{z}}ys}, {ter
  Linden}, {Terrell}, {Theil}, {Tiede}, {Troisi}, {Tsalmantza}, {Tur},
  {Vaccari}, {Vachier}, {Valles}, {Van Hamme}, {Veltz}, {Virtanen}, {Wallut},
  {Wichmann}, {Wilkinson}, {Ziaeepour}, \& {Zschocke}}]{2016A&A...595A...1G}
{Gaia Collaboration}, {Prusti}, T., {de Bruijne}, J.~H.~J., {et~al.} 2016,
  \aap, 595, A1, \dodoi{10.1051/0004-6361/201629272}

\bibitem[{{Gaia Collaboration} {et~al.}(2018){Gaia Collaboration}, {Brown},
  {Vallenari}, {Prusti}, {de Bruijne}, {Babusiaux}, {Bailer-Jones}, {Biermann},
  {Evans}, {Eyer}, \& et~al.}]{2018A&A...616A...1G}
{Gaia Collaboration}, {Brown}, A.~G.~A., {Vallenari}, A., {et~al.} 2018, \aap,
  616, A1, \dodoi{10.1051/0004-6361/201833051}

\bibitem[{{Gaia Collaboration} {et~al.}(2021){Gaia Collaboration}, {Brown},
  {Vallenari}, {Prusti}, {de Bruijne}, {Babusiaux}, {Biermann}, {Creevey},
  {Evans}, {Eyer}, {Hutton}, {Jansen}, {Jordi}, {Klioner}, {Lammers},
  {Lindegren}, {Luri}, {Mignard}, {Panem}, {Pourbaix}, {Randich}, {Sartoretti},
  {Soubiran}, {Walton}, {Arenou}, {Bailer-Jones}, {Bastian}, {Cropper},
  {Drimmel}, {Katz}, {Lattanzi}, {van Leeuwen}, {Bakker}, {Cacciari},
  {Casta{\~n}eda}, {De Angeli}, {Ducourant}, {Fabricius}, {Fouesneau},
  {Fr{\'e}mat}, {Guerra}, {Guerrier}, {Guiraud}, {Jean-Antoine Piccolo},
  {Masana}, {Messineo}, {Mowlavi}, {Nicolas}, {Nienartowicz}, {Pailler},
  {Panuzzo}, {Riclet}, {Roux}, {Seabroke}, {Sordo}, {Tanga}, {Th{\'e}venin},
  {Gracia-Abril}, {Portell}, {Teyssier}, {Altmann}, {Andrae}, {Bellas-Velidis},
  {Benson}, {Berthier}, {Blomme}, {Brugaletta}, {Burgess}, {Busso}, {Carry},
  {Cellino}, {Cheek}, {Clementini}, {Damerdji}, {Davidson}, {Delchambre},
  {Dell'Oro}, {Fern{\'a}ndez-Hern{\'a}ndez}, {Galluccio}, {Garc{\'\i}a-Lario},
  {Garcia-Reinaldos}, {Gonz{\'a}lez-N{\'u}{\~n}ez}, {Gosset}, {Haigron},
  {Halbwachs}, {Hambly}, {Harrison}, {Hatzidimitriou}, {Heiter},
  {Hern{\'a}ndez}, {Hestroffer}, {Hodgkin}, {Holl}, {Jan{\ss}en}, {Jevardat de
  Fombelle}, {Jordan}, {Krone-Martins}, {Lanzafame}, {L{\"o}ffler}, {Lorca},
  {Manteiga}, {Marchal}, {Marrese}, {Moitinho}, {Mora}, {Muinonen}, {Osborne},
  {Pancino}, {Pauwels}, {Petit}, {Recio-Blanco}, {Richards}, {Riello},
  {Rimoldini}, {Robin}, {Roegiers}, {Rybizki}, {Sarro}, {Siopis}, {Smith},
  {Sozzetti}, {Ulla}, {Utrilla}, {van Leeuwen}, {van Reeven}, {Abbas}, {Abreu
  Aramburu}, {Accart}, {Aerts}, {Aguado}, {Ajaj}, {Altavilla}, {{\'A}lvarez},
  {{\'A}lvarez Cid-Fuentes}, {Alves}, {Anderson}, {Anglada Varela}, {Antoja},
  {Audard}, {Baines}, {Baker}, {Balaguer-N{\'u}{\~n}ez}, {Balbinot}, {Balog},
  {Barache}, {Barbato}, {Barros}, {Barstow}, {Bartolom{\'e}}, {Bassilana},
  {Bauchet}, {Baudesson-Stella}, {Becciani}, {Bellazzini}, {Bernet}, {Bertone},
  {Bianchi}, {Blanco-Cuaresma}, {Boch}, {Bombrun}, {Bossini}, {Bouquillon},
  {Bragaglia}, {Bramante}, {Breedt}, {Bressan}, {Brouillet}, {Bucciarelli},
  {Burlacu}, {Busonero}, {Butkevich}, {Buzzi}, {Caffau}, {Cancelliere},
  {C{\'a}novas}, {Cantat-Gaudin}, {Carballo}, {Carlucci}, {Carnerero},
  {Carrasco}, {Casamiquela}, {Castellani}, {Castro-Ginard}, {Castro Sampol},
  {Chaoul}, {Charlot}, {Chemin}, {Chiavassa}, {Cioni}, {Comoretto}, {Cooper},
  {Cornez}, {Cowell}, {Crifo}, {Crosta}, {Crowley}, {Dafonte}, {Dapergolas},
  {David}, {David}, {de Laverny}, {De Luise}, {De March}, {De Ridder}, {de
  Souza}, {de Teodoro}, {de Torres}, {del Peloso}, {del Pozo}, {Delbo},
  {Delgado}, {Delgado}, {Delisle}, {Di Matteo}, {Diakite}, {Diener},
  {Distefano}, {Dolding}, {Eappachen}, {Edvardsson}, {Enke}, {Esquej}, {Fabre},
  {Fabrizio}, {Faigler}, {Fedorets}, {Fernique}, {Fienga}, {Figueras},
  {Fouron}, {Fragkoudi}, {Fraile}, {Franke}, {Gai}, {Garabato},
  {Garcia-Gutierrez}, {Garc{\'\i}a-Torres}, {Garofalo}, {Gavras}, {Gerlach},
  {Geyer}, {Giacobbe}, {Gilmore}, {Girona}, {Giuffrida}, {Gomel}, {Gomez},
  {Gonzalez-Santamaria}, {Gonz{\'a}lez-Vidal}, {Granvik},
  {Guti{\'e}rrez-S{\'a}nchez}, {Guy}, {Hauser}, {Haywood}, {Helmi}, {Hidalgo},
  {Hilger}, {H{\l}adczuk}, {Hobbs}, {Holland}, {Huckle}, {Jasniewicz},
  {Jonker}, {Juaristi Campillo}, {Julbe}, {Karbevska}, {Kervella}, {Khanna},
  {Kochoska}, {Kontizas}, {Kordopatis}, {Korn}, {Kostrzewa-Rutkowska},
  {Kruszy{\'n}ska}, {Lambert}, {Lanza}, {Lasne}, {Le Campion}, {Le Fustec},
  {Lebreton}, {Lebzelter}, {Leccia}, {Leclerc}, {Lecoeur-Taibi}, {Liao},
  {Licata}, {Lindstr{\o}m}, {Lister}, {Livanou}, {Lobel}, {Madrero Pardo},
  {Managau}, {Mann}, {Marchant}, {Marconi}, {Marcos Santos}, {Marinoni},
  {Marocco}, {Marshall}, {Martin Polo}, {Mart{\'\i}n-Fleitas}, {Masip},
  {Massari}, {Mastrobuono-Battisti}, {Mazeh}, {McMillan}, {Messina},
  {Michalik}, {Millar}, {Mints}, {Molina}, {Molinaro}, {Moln{\'a}r},
  {Montegriffo}, {Mor}, {Morbidelli}, {Morel}, {Morris}, {Mulone}, {Munoz},
  {Muraveva}, {Murphy}, {Musella}, {Noval}, {Ord{\'e}novic}, {Orr{\`u}},
  {Osinde}, {Pagani}, {Pagano}, {Palaversa}, {Palicio}, {Panahi}, {Pawlak},
  {Pe{\~n}alosa Esteller}, {Penttil{\"a}}, {Piersimoni}, {Pineau}, {Plachy},
  {Plum}, {Poggio}, {Poretti}, {Poujoulet}, {Pr{\v{s}}a}, {Pulone}, {Racero},
  {Ragaini}, {Rainer}, {Raiteri}, {Rambaux}, {Ramos}, {Ramos-Lerate}, {Re
  Fiorentin}, {Regibo}, {Reyl{\'e}}, {Ripepi}, {Riva}, {Rixon}, {Robichon},
  {Robin}, {Roelens}, {Rohrbasser}, {Romero-G{\'o}mez}, {Rowell}, {Royer},
  {Rybicki}, {Sadowski}, {Sagrist{\`a} Sell{\'e}s}, {Sahlmann}, {Salgado},
  {Salguero}, {Samaras}, {Sanchez Gimenez}, {Sanna}, {Santove{\~n}a},
  {Sarasso}, {Schultheis}, {Sciacca}, {Segol}, {Segovia}, {S{\'e}gransan},
  {Semeux}, {Shahaf}, {Siddiqui}, {Siebert}, {Siltala}, {Slezak}, {Smart},
  {Solano}, {Solitro}, {Souami}, {Souchay}, {Spagna}, {Spoto}, {Steele},
  {Steidelm{\"u}ller}, {Stephenson}, {S{\"u}veges}, {Szabados}, {Szegedi-Elek},
  {Taris}, {Tauran}, {Taylor}, {Teixeira}, {Thuillot}, {Tonello}, {Torra},
  {Torra}, {Turon}, {Unger}, {Vaillant}, {van Dillen}, {Vanel}, {Vecchiato},
  {Viala}, {Vicente}, {Voutsinas}, {Weiler}, {Wevers}, {Wyrzykowski}, {Yoldas},
  {Yvard}, {Zhao}, {Zorec}, {Zucker}, {Zurbach}, \&
  {Zwitter}}]{2021A&A...649A...1G}
---. 2021, \aap, 649, A1, \dodoi{10.1051/0004-6361/202039657}

\bibitem[{{Garn} {et~al.}(2007){Garn}, {Green}, {Hales}, {Riley}, \&
  {Alexander}}]{2007MNRAS.376.1251G}
{Garn}, T., {Green}, D.~A., {Hales}, S. E.~G., {Riley}, J.~M., \& {Alexander},
  P. 2007, \mnras, 376, 1251, \dodoi{10.1111/j.1365-2966.2007.11514.x}

\bibitem[{{Ghirlanda} {et~al.}(2019){Ghirlanda}, {Salafia}, {Paragi},
  {Giroletti}, {Yang}, {Marcote}, {Blanchard}, {Agudo}, {An}, {Bernardini},
  {Beswick}, {Branchesi}, {Campana}, {Casadio}, {Chassande-Mottin}, {Colpi},
  {Covino}, {D'Avanzo}, {D'Elia}, {Frey}, {Gawronski}, {Ghisellini}, {Gurvits},
  {Jonker}, {van Langevelde}, {Melandri}, {Moldon}, {Nava}, {Perego},
  {Perez-Torres}, {Reynolds}, {Salvaterra}, {Tagliaferri}, {Venturi},
  {Vergani}, \& {Zhang}}]{2019Sci...363..968G}
{Ghirlanda}, G., {Salafia}, O.~S., {Paragi}, Z., {et~al.} 2019, Science, 363,
  968, \dodoi{10.1126/science.aau8815}

\bibitem[{{Ghisellini} {et~al.}(2015){Ghisellini}, {Haardt}, {Ciardi},
  {Sbarrato}, {Gallo}, {Tavecchio}, \& {Celotti}}]{2015MNRAS.452.3457G}
{Ghisellini}, G., {Haardt}, F., {Ciardi}, B., {et~al.} 2015, \mnras, 452, 3457,
  \dodoi{10.1093/mnras/stv1541}

\bibitem[{{Ghisellini} \& {Sbarrato}(2016)}]{2016MNRAS.461L..21G}
{Ghisellini}, G., \& {Sbarrato}, T. 2016, \mnras, 461, L21,
  \dodoi{10.1093/mnrasl/slw089}

\bibitem[{{Gordon} {et~al.}(2016){Gordon}, {Jacobs}, {Beasley}, {Peck},
  {Gaume}, {Charlot}, {Fey}, {Ma}, {Titov}, \& {Boboltz}}]{2016AJ....151..154G}
{Gordon}, D., {Jacobs}, C., {Beasley}, A., {et~al.} 2016, \aj, 151, 154,
  \dodoi{10.3847/0004-6256/151/6/154}

\bibitem[{{Gordon} {et~al.}(2020){Gordon}, {Boyce}, {O'Dea}, {Rudnick},
  {Andernach}, {Vantyghem}, {Baum}, {Bui}, \&
  {Dionyssiou}}]{2020RNAAS...4..175G}
{Gordon}, Y.~A., {Boyce}, M.~M., {O'Dea}, C.~P., {et~al.} 2020, Research Notes
  of the American Astronomical Society, 4, 175,
  \dodoi{10.3847/2515-5172/abbe23}

\bibitem[{{Gregory} \& {Condon}(1991)}]{1991ApJS...75.1011G}
{Gregory}, P.~C., \& {Condon}, J.~J. 1991, \apjs, 75, 1011,
  \dodoi{10.1086/191559}

\bibitem[{{Greisen}(2003)}]{2003ASSL..285..109G}
{Greisen}, E.~W. 2003, in Astrophysics and Space Science Library, Vol. 285,
  Information Handling in Astronomy - Historical Vistas, ed. A.~{Heck}, 109,
  \dodoi{10.1007/0-306-48080-8_7}

\bibitem[{{Gurvits}(2000)}]{2000pras.conf..183G}
{Gurvits}, L.~I. 2000, in Perspectives on Radio Astronomy: Science with Large
  Antenna Arrays, ed. M.~P. {van Haarlem}, 183

\bibitem[{{Gurvits} {et~al.}(2015){Gurvits}, {Frey}, \&
  {Paragi}}]{2015IAUS..313..327G}
{Gurvits}, L.~I., {Frey}, S., \& {Paragi}, Z. 2015, in Extragalactic Jets from
  Every Angle, ed. F.~{Massaro}, C.~C. {Cheung}, E.~{Lopez}, \&
  A.~{Siemiginowska}, Vol. 313, 327--328, \dodoi{10.1017/S1743921315002434}

\bibitem[{{Gurvits} {et~al.}(1999){Gurvits}, {Kellermann}, \&
  {Frey}}]{1999A&A...342..378G}
{Gurvits}, L.~I., {Kellermann}, K.~I., \& {Frey}, S. 1999, \aap, 342, 378.
\newblock \doarXiv{astro-ph/9812018}

\bibitem[{{Haiman} {et~al.}(2004){Haiman}, {Quataert}, \&
  {Bower}}]{2004ApJ...612..698H}
{Haiman}, Z., {Quataert}, E., \& {Bower}, G.~C. 2004, \apj, 612, 698,
  \dodoi{10.1086/422834}

\bibitem[{{Helmboldt} {et~al.}(2007){Helmboldt}, {Taylor}, {Tremblay},
  {Fassnacht}, {Walker}, {Myers}, {Sjouwerman}, {Pearson}, {Readhead},
  {Weintraub}, {Gehrels}, {Romani}, {Healey}, {Michelson}, {Blandford}, \&
  {Cotter}}]{2007ApJ...658..203H}
{Helmboldt}, J.~F., {Taylor}, G.~B., {Tremblay}, S., {et~al.} 2007, \apj, 658,
  203, \dodoi{10.1086/511005}

\bibitem[{{H{\"o}gbom}(1974)}]{1974A&AS...15..417H}
{H{\"o}gbom}, J.~A. 1974, \aaps, 15, 417

\bibitem[{{Hogg} {et~al.}(2002){Hogg}, {Baldry}, {Blanton}, \&
  {Eisenstein}}]{2002astro.ph.10394H}
{Hogg}, D.~W., {Baldry}, I.~K., {Blanton}, M.~R., \& {Eisenstein}, D.~J. 2002,
  arXiv e-prints, astro.
\newblock \doarXiv{astro-ph/0210394}

\bibitem[{{Holt} {et~al.}(2004){Holt}, {Benn}, {Vigotti}, {Pedani}, {Carballo},
  {Gonz{\'a}lez-Serrano}, {Mack}, \& {Garc{\'\i}a}}]{2004MNRAS.348..857H}
{Holt}, J., {Benn}, C.~R., {Vigotti}, M., {et~al.} 2004, \mnras, 348, 857,
  \dodoi{10.1111/j.1365-2966.2004.07423.x}

\bibitem[{{Hunt} {et~al.}(2021){Hunt}, {Johnson}, {Cigan}, {Gordon}, \&
  {Spitzak}}]{2021AJ....162..121H}
{Hunt}, L.~R., {Johnson}, M.~C., {Cigan}, P.~J., {Gordon}, D., \& {Spitzak}, J.
  2021, \aj, 162, 121, \dodoi{10.3847/1538-3881/ac135d}

\bibitem[{{Hunter}(2007)}]{2007CSE.....9...90H}
{Hunter}, J.~D. 2007, Computing in Science and Engineering, 9, 90,
  \dodoi{10.1109/MCSE.2007.55}

\bibitem[{{Ighina} {et~al.}(2019){Ighina}, {Caccianiga}, {Moretti},
  {Belladitta}, {Della Ceca}, {Ballo}, \& {Dallacasa}}]{2019MNRAS.489.2732I}
{Ighina}, L., {Caccianiga}, A., {Moretti}, A., {et~al.} 2019, \mnras, 489,
  2732, \dodoi{10.1093/mnras/stz2340}

\bibitem[{{Ighina} {et~al.}(2021){Ighina}, {Caccianiga}, {Moretti},
  {Belladitta}, {Della Ceca}, \& {Diana}}]{2021MNRAS.505.4120I}
---. 2021, \mnras, 505, 4120, \dodoi{10.1093/mnras/stab1612}

\bibitem[{{Intema} {et~al.}(2017){Intema}, {Jagannathan}, {Mooley}, \&
  {Frail}}]{2017AA...598A..78I}
{Intema}, H.~T., {Jagannathan}, P., {Mooley}, K.~P., \& {Frail}, D.~A. 2017,
  \aap, 598, A78, \dodoi{10.1051/0004-6361/201628536}

\bibitem[{{Ivezi{\'c}} {et~al.}(2002){Ivezi{\'c}}, {Menou}, {Knapp}, {Strauss},
  {Lupton}, {Vanden Berk}, {Richards}, {Tremonti}, {Weinstein}, {Anderson},
  {Bahcall}, {Becker}, {Bernardi}, {Blanton}, {Eisenstein}, {Fan},
  {Finkbeiner}, {Finlator}, {Frieman}, {Gunn}, {Hall}, {Kim}, {Kinkhabwala},
  {Narayanan}, {Rockosi}, {Schlegel}, {Schneider}, {Strateva}, {SubbaRao},
  {Thakar}, {Voges}, {White}, {Yanny}, {Brinkmann}, {Doi}, {Fukugita},
  {Hennessy}, {Munn}, {Nichol}, \& {York}}]{2002AJ....124.2364I}
{Ivezi{\'c}}, {\v{Z}}., {Menou}, K., {Knapp}, G.~R., {et~al.} 2002, \aj, 124,
  2364, \dodoi{10.1086/344069}

\bibitem[{{Jackson} {et~al.}(2007){Jackson}, {Battye}, {Browne}, {Joshi},
  {Muxlow}, \& {Wilkinson}}]{2007MNRAS.376..371J}
{Jackson}, N., {Battye}, R.~A., {Browne}, I.~W.~A., {et~al.} 2007, \mnras, 376,
  371, \dodoi{10.1111/j.1365-2966.2007.11442.x}

\bibitem[{{Keimpema} {et~al.}(2015){Keimpema}, {Kettenis}, {Pogrebenko},
  {Campbell}, {Cim{\'o}}, {Duev}, {Eldering}, {Kruithof}, {van Langevelde},
  {Marchal}, {Molera Calv{\'e}s}, {Ozdemir}, {Paragi}, {Pidopryhora},
  {Szomoru}, \& {Yang}}]{2015ExA....39..259K}
{Keimpema}, A., {Kettenis}, M.~M., {Pogrebenko}, S.~V., {et~al.} 2015,
  Experimental Astronomy, 39, 259, \dodoi{10.1007/s10686-015-9446-1}

\bibitem[{{Kellermann} {et~al.}(1999){Kellermann}, {Vermeulen}, {Zensus},
  {Cohen}, \& {West}}]{1999NewAR..43..757K}
{Kellermann}, K.~I., {Vermeulen}, R.~C., {Zensus}, J.~A., {Cohen}, M.~H., \&
  {West}, A. 1999, \nar, 43, 757, \dodoi{10.1016/S1387-6473(99)00093-7}

\bibitem[{{Kewley} {et~al.}(2000){Kewley}, {Heisler}, {Dopita}, {Sutherland},
  {Norris}, {Reynolds}, \& {Lumsden}}]{2000ApJ...530..704K}
{Kewley}, L.~J., {Heisler}, C.~A., {Dopita}, M.~A., {et~al.} 2000, \apj, 530,
  704, \dodoi{10.1086/308397}

\bibitem[{{Kovalev} {et~al.}(2017){Kovalev}, {Petrov}, \&
  {Plavin}}]{2017A&A...598L...1K}
{Kovalev}, Y.~Y., {Petrov}, L., \& {Plavin}, A.~V. 2017, \aap, 598, L1,
  \dodoi{10.1051/0004-6361/201630031}

\bibitem[{{Kovalev} {et~al.}(2005){Kovalev}, {Kellermann}, {Lister}, {Homan},
  {Vermeulen}, {Cohen}, {Ros}, {Kadler}, {Lobanov}, {Zensus}, {Kardashev},
  {Gurvits}, {Aller}, \& {Aller}}]{2005AJ....130.2473K}
{Kovalev}, Y.~Y., {Kellermann}, K.~I., {Lister}, M.~L., {et~al.} 2005, \aj,
  130, 2473, \dodoi{10.1086/497430}

\bibitem[{{Krezinger} {et~al.}(2020){Krezinger}, {Frey}, {An}, {Jaiswal}, \&
  {Zhang}}]{2020MNRAS.496.1811K}
{Krezinger}, M., {Frey}, S., {An}, T., {Jaiswal}, S., \& {Zhang}, Y. 2020,
  \mnras, 496, 1811, \dodoi{10.1093/mnras/staa1669}

\bibitem[{{Lacy} {et~al.}(2020){Lacy}, {Baum}, {Chandler}, {Chatterjee},
  {Clarke}, {Deustua}, {English}, {Farnes}, {Gaensler}, {Gugliucci},
  {Hallinan}, {Kent}, {Kimball}, {Law}, {Lazio}, {Marvil}, {Mao}, {Medlin},
  {Mooley}, {Murphy}, {Myers}, {Osten}, {Richards}, {Rosolowsky}, {Rudnick},
  {Schinzel}, {Sivakoff}, {Sjouwerman}, {Taylor}, {White}, {Wrobel},
  {Andernach}, {Beasley}, {Berger}, {Bhatnager}, {Birkinshaw}, {Bower},
  {Brandt}, {Brown}, {Burke-Spolaor}, {Butler}, {Comerford}, {Demorest}, {Fu},
  {Giacintucci}, {Golap}, {G{\"u}th}, {Hales}, {Hiriart}, {Hodge}, {Horesh},
  {Ivezi{\'c}}, {Jarvis}, {Kamble}, {Kassim}, {Liu}, {Loinard}, {Lyons},
  {Masters}, {Mezcua}, {Moellenbrock}, {Mroczkowski}, {Nyland}, {O'Dea},
  {O'Sullivan}, {Peters}, {Radford}, {Rao}, {Robnett}, {Salcido}, {Shen},
  {Sobotka}, {Witz}, {Vaccari}, {van Weeren}, {Vargas}, {Williams}, \&
  {Yoon}}]{2020PASP..132c5001L}
{Lacy}, M., {Baum}, S.~A., {Chandler}, C.~J., {et~al.} 2020, \pasp, 132,
  035001, \dodoi{10.1088/1538-3873/ab63eb}

\bibitem[{{Lee} {et~al.}(2017){Lee}, {Sohn}, {Jung}, {Byun}, \&
  {Lee}}]{2017ApJS..228...22L}
{Lee}, J.~A., {Sohn}, B.~W., {Jung}, T., {Byun}, D.-Y., \& {Lee}, J.~W. 2017,
  \apjs, 228, 22, \dodoi{10.3847/1538-4365/228/2/22}

\bibitem[{{Lister} {et~al.}(2019){Lister}, {Homan}, {Hovatta}, {Kellermann},
  {Kiehlmann}, {Kovalev}, {Max-Moerbeck}, {Pushkarev}, {Readhead}, {Ros}, \&
  {Savolainen}}]{2019ApJ...874...43L}
{Lister}, M.~L., {Homan}, D.~C., {Hovatta}, T., {et~al.} 2019, \apj, 874, 43,
  \dodoi{10.3847/1538-4357/ab08ee}

\bibitem[{{Liu} {et~al.}(2022){Liu}, {Wang}, {Momjian}, {Wagg}, {Yang}, {An},
  {Shao}, {Carilli}, {Wu}, {Fan}, {Walter}, {Jiang}, {Li}, {Li}, {Fei}, \&
  {Xu}}]{2022arXiv220302922L}
{Liu}, Y., {Wang}, R., {Momjian}, E., {et~al.} 2022, arXiv e-prints,
  arXiv:2203.02922.
\newblock \doarXiv{2203.02922}

\bibitem[{{Magliocchetti} {et~al.}(2014){Magliocchetti}, {Lutz}, {Rosario},
  {Berta}, {Le Floc'h}, {Magnelli}, {Pozzi}, {Riguccini}, \&
  {Santini}}]{2014MNRAS.442..682M}
{Magliocchetti}, M., {Lutz}, D., {Rosario}, D., {et~al.} 2014, \mnras, 442,
  682, \dodoi{10.1093/mnras/stu863}

\bibitem[{{Mao} {et~al.}(2017){Mao}, {Urry}, {Marchesini}, {Landoni},
  {Massaro}, \& {Ajello}}]{2017ApJ...842...87M}
{Mao}, P., {Urry}, C.~M., {Marchesini}, E., {et~al.} 2017, \apj, 842, 87,
  \dodoi{10.3847/1538-4357/aa74b8}

\bibitem[{{Mart{\'\i}-Vidal} {et~al.}(2010){Mart{\'\i}-Vidal}, {Ros},
  {P{\'e}rez-Torres}, {Guirado}, {Jim{\'e}nez-Monferrer}, \&
  {Marcaide}}]{2010A&A...515A..53M}
{Mart{\'\i}-Vidal}, I., {Ros}, E., {P{\'e}rez-Torres}, M.~A., {et~al.} 2010,
  \aap, 515, A53, \dodoi{10.1051/0004-6361/201014203}

\bibitem[{{Massaro} {et~al.}(2015){Massaro}, {Maselli}, {Leto}, {Marchegiani},
  {Perri}, {Giommi}, \& {Piranomonte}}]{2015Ap&SS.357...75M}
{Massaro}, E., {Maselli}, A., {Leto}, C., {et~al.} 2015, \apss, 357, 75,
  \dodoi{10.1007/s10509-015-2254-2}

\bibitem[{{Middelberg} {et~al.}(2011){Middelberg}, {Deller}, {Morgan},
  {Rottmann}, {Alef}, {Tingay}, {Norris}, {Bach}, {Brisken}, \&
  {Lenc}}]{2011A&A...526A..74M}
{Middelberg}, E., {Deller}, A., {Morgan}, J., {et~al.} 2011, \aap, 526, A74,
  \dodoi{10.1051/0004-6361/201015406}

\bibitem[{{Momjian} {et~al.}(2021){Momjian}, {Ba{\~n}ados}, {Carilli},
  {Walter}, \& {Mazzucchelli}}]{2021AJ....161..207M}
{Momjian}, E., {Ba{\~n}ados}, E., {Carilli}, C.~L., {Walter}, F., \&
  {Mazzucchelli}, C. 2021, \aj, 161, 207, \dodoi{10.3847/1538-3881/abe6ae}

\bibitem[{{Momjian} {et~al.}(2018){Momjian}, {Carilli}, {Ba{\~n}ados},
  {Walter}, \& {Venemans}}]{2018ApJ...861...86M}
{Momjian}, E., {Carilli}, C.~L., {Ba{\~n}ados}, E., {Walter}, F., \&
  {Venemans}, B.~P. 2018, \apj, 861, 86, \dodoi{10.3847/1538-4357/aac76f}

\bibitem[{{Momjian} {et~al.}(2008){Momjian}, {Carilli}, \&
  {McGreer}}]{2008AJ....136..344M}
{Momjian}, E., {Carilli}, C.~L., \& {McGreer}, I.~D. 2008, \aj, 136, 344,
  \dodoi{10.1088/0004-6256/136/1/344}

\bibitem[{{Momjian} {et~al.}(2005){Momjian}, {Carilli}, \&
  {Petric}}]{2005AJ....129.1809M}
{Momjian}, E., {Carilli}, C.~L., \& {Petric}, A.~O. 2005, \aj, 129, 1809,
  \dodoi{10.1086/428598}

\bibitem[{{Momjian} {et~al.}(2004){Momjian}, {Petric}, \&
  {Carilli}}]{2004AJ....127..587M}
{Momjian}, E., {Petric}, A.~O., \& {Carilli}, C.~L. 2004, \aj, 127, 587,
  \dodoi{10.1086/381300}

\bibitem[{{Morganti} {et~al.}(2013){Morganti}, {Fogasy}, {Paragi}, {Oosterloo},
  \& {Orienti}}]{2013Sci...341.1082M}
{Morganti}, R., {Fogasy}, J., {Paragi}, Z., {Oosterloo}, T., \& {Orienti}, M.
  2013, Science, 341, 1082, \dodoi{10.1126/science.1240436}

\bibitem[{{Mosoni} {et~al.}(2006){Mosoni}, {Frey}, {Gurvits}, {Garrett},
  {Garrington}, \& {Tsvetanov}}]{2006A&A...445..413M}
{Mosoni}, L., {Frey}, S., {Gurvits}, L.~I., {et~al.} 2006, \aap, 445, 413,
  \dodoi{10.1051/0004-6361:20053473}

\bibitem[{{Myers} {et~al.}(2003){Myers}, {Jackson}, {Browne}, {de Bruyn},
  {Pearson}, {Readhead}, {Wilkinson}, {Biggs}, {Blandford}, {Fassnacht},
  {Koopmans}, {Marlow}, {McKean}, {Norbury}, {Phillips}, {Rusin}, {Shepherd},
  \& {Sykes}}]{2003MNRAS.341....1M}
{Myers}, S.~T., {Jackson}, N.~J., {Browne}, I.~W.~A., {et~al.} 2003, \mnras,
  341, 1, \dodoi{10.1046/j.1365-8711.2003.06256.x}

\bibitem[{{Ochsenbein} {et~al.}(2000){Ochsenbein}, {Bauer}, \&
  {Marcout}}]{2000A&AS..143...23O}
{Ochsenbein}, F., {Bauer}, P., \& {Marcout}, J. 2000, \aaps, 143, 23,
  \dodoi{10.1051/aas:2000169}

\bibitem[{{O'Dea} \& {Saikia}(2021)}]{2021AARv..29....3O}
{O'Dea}, C.~P., \& {Saikia}, D.~J. 2021, \aapr, 29, 3,
  \dodoi{10.1007/s00159-021-00131-w}

\bibitem[{{O'Sullivan} {et~al.}(2011){O'Sullivan}, {Gabuzda}, \&
  {Gurvits}}]{2011MNRAS.415.3049O}
{O'Sullivan}, S.~P., {Gabuzda}, D.~C., \& {Gurvits}, L.~I. 2011, \mnras, 415,
  3049, \dodoi{10.1111/j.1365-2966.2011.18915.x}

\bibitem[{{Page} {et~al.}(2014){Page}, {Simpson}, {Mortlock}, {Warren},
  {Hewett}, {Venemans}, \& {McMahon}}]{2014MNRAS.440L..91P}
{Page}, M.~J., {Simpson}, C., {Mortlock}, D.~J., {et~al.} 2014, \mnras, 440,
  L91, \dodoi{10.1093/mnrasl/slu022}

\bibitem[{{Paragi} {et~al.}(1999){Paragi}, {Frey}, {Gurvits}, {Kellermann},
  {Schilizzi}, {McMahon}, {Hook}, \& {Pauliny-Toth}}]{1999AA...344...51P}
{Paragi}, Z., {Frey}, S., {Gurvits}, L.~I., {et~al.} 1999, \aap, 344, 51.
\newblock \doarXiv{astro-ph/9901396}

\bibitem[{{Parijskij} {et~al.}(2014){Parijskij}, {Thomasson}, {Kopylov},
  {Zhelenkova}, {Muxlow}, {Beswick}, {Soboleva}, {Temirova}, \&
  {Verkhodanov}}]{2014MNRAS.439.2314P}
{Parijskij}, Y.~N., {Thomasson}, P., {Kopylov}, A.~I., {et~al.} 2014, \mnras,
  439, 2314, \dodoi{10.1093/mnras/stu047}

\bibitem[{{Pearson}(1995)}]{1995ASPC...82..267P}
{Pearson}, T.~J. 1995, in Astronomical Society of the Pacific Conference
  Series, Vol.~82, Very Long Baseline Interferometry and the VLBA, ed. J.~A.
  {Zensus}, P.~J. {Diamond}, \& P.~J. {Napier}, 267

\bibitem[{{Perger} {et~al.}(2017){Perger}, {Frey}, {Gab{\'a}nyi}, \&
  {T{\'o}th}}]{2017FrASS...4....9P}
{Perger}, K., {Frey}, S., {Gab{\'a}nyi}, K.~{\'E}., \& {T{\'o}th}, L.~V. 2017,
  Frontiers in Astronomy and Space Sciences, 4, 9,
  \dodoi{10.3389/fspas.2017.00009}

\bibitem[{{Perger} {et~al.}(2018){Perger}, {Frey}, {Gab{\'a}nyi}, {An},
  {Britzen}, {Cao}, {Cseh}, {Dennett-Thorpe}, {Gurvits}, {Hong}, {Hook},
  {Paragi}, {Schilizzi}, {Yang}, \& {Zhang}}]{2018MNRAS.477.1065P}
{Perger}, K., {Frey}, S., {Gab{\'a}nyi}, K.~{\'E}., {et~al.} 2018, \mnras, 477,
  1065, \dodoi{10.1093/mnras/sty837}

\bibitem[{{Petrov}(2013)}]{2013AJ....146....5P}
{Petrov}, L. 2013, \aj, 146, 5, \dodoi{10.1088/0004-6256/146/1/5}

\bibitem[{{Petrov}(2021)}]{2021AJ....161...14P}
---. 2021, \aj, 161, 14, \dodoi{10.3847/1538-3881/abc4e1}

\bibitem[{{Petrov} {et~al.}(2012{\natexlab{a}}){Petrov}, {Honma}, \&
  {Shibata}}]{2012AJ....143...35P}
{Petrov}, L., {Honma}, M., \& {Shibata}, S.~M. 2012{\natexlab{a}}, \aj, 143,
  35, \dodoi{10.1088/0004-6256/143/2/35}

\bibitem[{{Petrov} {et~al.}(2005){Petrov}, {Kovalev}, {Fomalont}, \&
  {Gordon}}]{2005AJ....129.1163P}
{Petrov}, L., {Kovalev}, Y.~Y., {Fomalont}, E., \& {Gordon}, D. 2005, \aj, 129,
  1163, \dodoi{10.1086/426920}

\bibitem[{{Petrov} {et~al.}(2006){Petrov}, {Kovalev}, {Fomalont}, \&
  {Gordon}}]{2006AJ....131.1872P}
{Petrov}, L., {Kovalev}, Y.~Y., {Fomalont}, E.~B., \& {Gordon}, D. 2006, \aj,
  131, 1872, \dodoi{10.1086/499947}

\bibitem[{{Petrov} {et~al.}(2008){Petrov}, {Kovalev}, {Fomalont}, \&
  {Gordon}}]{2008AJ....136..580P}
---. 2008, \aj, 136, 580, \dodoi{10.1088/0004-6256/136/2/580}

\bibitem[{{Petrov} {et~al.}(2012{\natexlab{b}}){Petrov}, {Lee}, {Kim}, {Jung},
  {Oh}, {Sohn}, {Byun}, {Chung}, {Je}, {Wi}, {Song}, {Kang}, {Han}, {Lee},
  {Kim}, {Chung}, \& {Kim}}]{2012AJ....144..150P}
{Petrov}, L., {Lee}, S.-S., {Kim}, J., {et~al.} 2012{\natexlab{b}}, \aj, 144,
  150, \dodoi{10.1088/0004-6256/144/5/150}

\bibitem[{{Plavin} {et~al.}(2019){Plavin}, {Kovalev}, \&
  {Petrov}}]{2019ApJ...871..143P}
{Plavin}, A.~V., {Kovalev}, Y.~Y., \& {Petrov}, L.~Y. 2019, \apj, 871, 143,
  \dodoi{10.3847/1538-4357/aaf650}

\bibitem[{{Plotkin} {et~al.}(2008){Plotkin}, {Anderson}, {Hall}, {Margon},
  {Voges}, {Schneider}, {Stinson}, \& {York}}]{2008AJ....135.2453P}
{Plotkin}, R.~M., {Anderson}, S.~F., {Hall}, P.~B., {et~al.} 2008, \aj, 135,
  2453, \dodoi{10.1088/0004-6256/135/6/2453}

\bibitem[{{Pushkarev} \& {Kovalev}(2012)}]{2012AA...544A..34P}
{Pushkarev}, A.~B., \& {Kovalev}, Y.~Y. 2012, \aap, 544, A34,
  \dodoi{10.1051/0004-6361/201219352}

\bibitem[{{Readhead}(1994)}]{1994ApJ...426...51R}
{Readhead}, A. C.~S. 1994, \apj, 426, 51, \dodoi{10.1086/174038}

\bibitem[{{Romani} {et~al.}(2004){Romani}, {Sowards-Emmerd}, {Greenhill}, \&
  {Michelson}}]{2004ApJ...610L...9R}
{Romani}, R.~W., {Sowards-Emmerd}, D., {Greenhill}, L., \& {Michelson}, P.
  2004, \apjl, 610, L9, \dodoi{10.1086/423201}

\bibitem[{{Saxton} {et~al.}(2008){Saxton}, {Read}, {Esquej}, {Freyberg},
  {Altieri}, \& {Bermejo}}]{2008AA...480..611S}
{Saxton}, R.~D., {Read}, A.~M., {Esquej}, P., {et~al.} 2008, \aap, 480, 611,
  \dodoi{10.1051/0004-6361:20079193}

\bibitem[{{Sbarrato}(2021)}]{2021Galax...9...23S}
{Sbarrato}, T. 2021, Galaxies, 9, 23, \dodoi{10.3390/galaxies9020023}

\bibitem[{{Sbarrato} {et~al.}(2013){Sbarrato}, {Ghisellini}, {Nardini},
  {Tagliaferri}, {Greiner}, {Rau}, \& {Schady}}]{2013MNRAS.433.2182S}
{Sbarrato}, T., {Ghisellini}, G., {Nardini}, M., {et~al.} 2013, \mnras, 433,
  2182, \dodoi{10.1093/mnras/stt882}

\bibitem[{{Schinzel} {et~al.}(2017){Schinzel}, {Petrov}, {Taylor}, \&
  {Edwards}}]{2017ApJ...838..139S}
{Schinzel}, F.~K., {Petrov}, L., {Taylor}, G.~B., \& {Edwards}, P.~G. 2017,
  \apj, 838, 139, \dodoi{10.3847/1538-4357/aa6439}

\bibitem[{{Schneider} {et~al.}(2010){Schneider}, {Richards}, {Hall}, {Strauss},
  {Anderson}, {Boroson}, {Ross}, {Shen}, {Brandt}, {Fan}, {Inada}, {Jester},
  {Knapp}, {Krawczyk}, {Thakar}, {Vanden Berk}, {Voges}, {Yanny}, {York},
  {Bahcall}, {Bizyaev}, {Blanton}, {Brewington}, {Brinkmann}, {Eisenstein},
  {Frieman}, {Fukugita}, {Gray}, {Gunn}, {Hibon}, {Ivezi{\'c}}, {Kent}, {Kron},
  {Lee}, {Lupton}, {Malanushenko}, {Malanushenko}, {Oravetz}, {Pan}, {Pier},
  {Price}, {Saxe}, {Schlegel}, {Simmons}, {Snedden}, {SubbaRao}, {Szalay}, \&
  {Weinberg}}]{2010AJ....139.2360S}
{Schneider}, D.~P., {Richards}, G.~T., {Hall}, P.~B., {et~al.} 2010, \aj, 139,
  2360, \dodoi{10.1088/0004-6256/139/6/2360}

\bibitem[{{Schwab} \& {Cotton}(1983)}]{1983AJ.....88..688S}
{Schwab}, F.~R., \& {Cotton}, W.~D. 1983, \aj, 88, 688, \dodoi{10.1086/113360}

\bibitem[{{Shao} {et~al.}(2022){Shao}, {Wagg}, {Wang}, {Momjian}, {Carilli},
  {Walter}, {Riechers}, {Intema}, {Weiss}, {Brunthaler}, \&
  {Menten}}]{2022A&A...659A.159S}
{Shao}, Y., {Wagg}, J., {Wang}, R., {et~al.} 2022, \aap, 659, A159,
  \dodoi{10.1051/0004-6361/202142489}

\bibitem[{{Shen} {et~al.}(2011){Shen}, {Richards}, {Strauss}, {Hall},
  {Schneider}, {Snedden}, {Bizyaev}, {Brewington}, {Malanushenko},
  {Malanushenko}, {Oravetz}, {Pan}, \& {Simmons}}]{2011ApJS..194...45S}
{Shen}, Y., {Richards}, G.~T., {Strauss}, M.~A., {et~al.} 2011, \apjs, 194, 45,
  \dodoi{10.1088/0067-0049/194/2/45}

\bibitem[{{Shepherd}(1997)}]{1997ASPC..125...77S}
{Shepherd}, M.~C. 1997, in Astronomical Society of the Pacific Conference
  Series, Vol. 125, Astronomical Data Analysis Software and Systems VI, ed.
  G.~{Hunt} \& H.~{Payne}, 77

\bibitem[{{Shepherd} {et~al.}(1994){Shepherd}, {Pearson}, \&
  {Taylor}}]{1994BAAS...26..987S}
{Shepherd}, M.~C., {Pearson}, T.~J., \& {Taylor}, G.~B. 1994, in Bulletin of
  the American Astronomical Society, Vol.~26, 987--989

\bibitem[{{Shimwell} {et~al.}(2017){Shimwell}, {R{\"o}ttgering}, {Best},
  {Williams}, {Dijkema}, {de Gasperin}, {Hardcastle}, {Heald}, {Hoang},
  {Horneffer}, {Intema}, {Mahony}, {Mandal}, {Mechev}, {Morabito}, {Oonk},
  {Rafferty}, {Retana-Montenegro}, {Sabater}, {Tasse}, {van Weeren},
  {Br{\"u}ggen}, {Brunetti}, {Chy{\.z}y}, {Conway}, {Haverkorn}, {Jackson},
  {Jarvis}, {McKean}, {Miley}, {Morganti}, {White}, {Wise}, {van Bemmel},
  {Beck}, {Brienza}, {Bonafede}, {Calistro Rivera}, {Cassano}, {Clarke},
  {Cseh}, {Deller}, {Drabent}, {van Driel}, {Engels}, {Falcke}, {Ferrari},
  {Fr{\"o}hlich}, {Garrett}, {Harwood}, {Heesen}, {Hoeft}, {Horellou},
  {Israel}, {Kapi{\'n}ska}, {Kunert-Bajraszewska}, {McKay}, {Mohan},
  {Orr{\'u}}, {Pizzo}, {Prandoni}, {Schwarz}, {Shulevski}, {Sipior}, {Smith},
  {Sridhar}, {Steinmetz}, {Stroe}, {Varenius}, {van der Werf}, {Zensus}, \&
  {Zwart}}]{2017AA...598A.104S}
{Shimwell}, T.~W., {R{\"o}ttgering}, H.~J.~A., {Best}, P.~N., {et~al.} 2017,
  \aap, 598, A104, \dodoi{10.1051/0004-6361/201629313}

\bibitem[{{Sotnikova} {et~al.}(2021){Sotnikova}, {Mikhailov}, {Mufakharov},
  {Mingaliev}, {Bursov}, {Semenova}, {Stolyarov}, {Udovitskiy}, {Kudryashova},
  \& {Erkenov}}]{2021MNRAS.508.2798S}
{Sotnikova}, Y., {Mikhailov}, A., {Mufakharov}, T., {et~al.} 2021, \mnras, 508,
  2798, \dodoi{10.1093/mnras/stab2114}

\bibitem[{{Spingola} {et~al.}(2020){Spingola}, {Dallacasa}, {Belladitta},
  {Caccianiga}, {Giroletti}, {Moretti}, \& {Orienti}}]{2020AA...643L..12S}
{Spingola}, C., {Dallacasa}, D., {Belladitta}, S., {et~al.} 2020, \aap, 643,
  L12, \dodoi{10.1051/0004-6361/202039458}

\bibitem[{{Szomoru} {et~al.}(2004){Szomoru}, {Biggs}, {Garrett}, {van
  Langevelde}, {Olnon}, {Paragi}, {Parsley}, {Pogrebenko}, \&
  {Reynolds}}]{2004evn..conf..257S}
{Szomoru}, A., {Biggs}, A., {Garrett}, M., {et~al.} 2004, in European VLBI
  Network on New Developments in VLBI Science and Technology, 257--260.
\newblock \doarXiv{astro-ph/0412686}

\bibitem[{{Tremblay} {et~al.}(2016){Tremblay}, {Taylor}, {Ortiz}, {Tremblay},
  {Helmboldt}, \& {Romani}}]{2016MNRAS.459..820T}
{Tremblay}, S.~E., {Taylor}, G.~B., {Ortiz}, A.~A., {et~al.} 2016, \mnras, 459,
  820, \dodoi{10.1093/mnras/stw592}

\bibitem[{{Urry} \& {Padovani}(1995)}]{1995PASP..107..803U}
{Urry}, C.~M., \& {Padovani}, P. 1995, \pasp, 107, 803, \dodoi{10.1086/133630}

\bibitem[{{Veres} {et~al.}(2010){Veres}, {Frey}, {Paragi}, \&
  {Gurvits}}]{2010AA...521A...6V}
{Veres}, P., {Frey}, S., {Paragi}, Z., \& {Gurvits}, L.~I. 2010, \aap, 521, A6,
  \dodoi{10.1051/0004-6361/201014957}

\bibitem[{{Vignali} {et~al.}(2005){Vignali}, {Brandt}, {Schneider}, \&
  {Kaspi}}]{2005AJ....129.2519V}
{Vignali}, C., {Brandt}, W.~N., {Schneider}, D.~P., \& {Kaspi}, S. 2005, \aj,
  129, 2519, \dodoi{10.1086/430217}

\bibitem[{{Volonteri} {et~al.}(2011){Volonteri}, {Haardt}, {Ghisellini}, \&
  {Della Ceca}}]{2011MNRAS.416..216V}
{Volonteri}, M., {Haardt}, F., {Ghisellini}, G., \& {Della Ceca}, R. 2011,
  \mnras, 416, 216, \dodoi{10.1111/j.1365-2966.2011.19024.x}

\bibitem[{{Wang} {et~al.}(2017){Wang}, {Momjian}, {Carilli}, {Wu}, {Fan},
  {Walter}, {Strauss}, {Wang}, \& {Jiang}}]{2017ApJ...835L..20W}
{Wang}, R., {Momjian}, E., {Carilli}, C.~L., {et~al.} 2017, \apjl, 835, L20,
  \dodoi{10.3847/2041-8213/835/2/L20}

\bibitem[{{Weisskopf} {et~al.}(2002){Weisskopf}, {Brinkman}, {Canizares},
  {Garmire}, {Murray}, \& {Van Speybroeck}}]{2002PASP..114....1W}
{Weisskopf}, M.~C., {Brinkman}, B., {Canizares}, C., {et~al.} 2002, \pasp, 114,
  1, \dodoi{10.1086/338108}

\bibitem[{{White} {et~al.}(1997){White}, {Becker}, {Helfand}, \&
  {Gregg}}]{1997ApJ...475..479W}
{White}, R.~L., {Becker}, R.~H., {Helfand}, D.~J., \& {Gregg}, M.~D. 1997,
  \apj, 475, 479, \dodoi{10.1086/303564}

\bibitem[{{Wright}(2006)}]{2006PASP..118.1711W}
{Wright}, E.~L. 2006, \pasp, 118, 1711, \dodoi{10.1086/510102}

\bibitem[{{Wu} {et~al.}(2013{\natexlab{a}}){Wu}, {An}, {Baan}, {Hong},
  {Stanghellini}, {Frey}, {Xu}, {Liu}, \& {Wang}}]{2013A&A...550A.113W}
{Wu}, F., {An}, T., {Baan}, W.~A., {et~al.} 2013{\natexlab{a}}, \aap, 550,
  A113, \dodoi{10.1051/0004-6361/201219700}

\bibitem[{{Wu} {et~al.}(2013{\natexlab{b}}){Wu}, {Brandt}, {Miller}, {Garmire},
  {Schneider}, \& {Vignali}}]{2013ApJ...763..109W}
{Wu}, J., {Brandt}, W.~N., {Miller}, B.~P., {et~al.} 2013{\natexlab{b}}, \apj,
  763, 109, \dodoi{10.1088/0004-637X/763/2/109}

\bibitem[{{Wyithe} \& {Loeb}(2012)}]{2012MNRAS.425.2892W}
{Wyithe}, J. S.~B., \& {Loeb}, A. 2012, \mnras, 425, 2892,
  \dodoi{10.1111/j.1365-2966.2012.21127.x}

\bibitem[{{Zhang} {et~al.}(2020){Zhang}, {An}, \& {Frey}}]{2020SciBu..65..525Z}
{Zhang}, Y., {An}, T., \& {Frey}, S. 2020, Science Bulletin, 65, 525,
  \dodoi{10.1016/j.scib.2020.01.008}

\bibitem[{{Zhang} {et~al.}(2017){Zhang}, {An}, {Frey}, {Gab{\'a}nyi}, {Paragi},
  {Gurvits}, {Sohn}, {Jung}, {Kino}, {Lao}, {Lu}, \&
  {Mohan}}]{2017MNRAS.468...69Z}
{Zhang}, Y., {An}, T., {Frey}, S., {et~al.} 2017, \mnras, 468, 69,
  \dodoi{10.1093/mnras/stx392}

\bibitem[{{Zhu} {et~al.}(2019){Zhu}, {Brandt}, {Wu}, {Garmire}, \&
  {Miller}}]{2019MNRAS.482.2016Z}
{Zhu}, S.~F., {Brandt}, W.~N., {Wu}, J., {Garmire}, G.~P., \& {Miller}, B.~P.
  2019, \mnras, 482, 2016, \dodoi{10.1093/mnras/sty2832}

\end{thebibliography}
\bibliographystyle{aasjournal}



\end{document}